\documentclass[journal=jacsat,manuscript=article,layout=twocolumn]{achemso}
\usepackage[T1]{fontenc}                        
\usepackage{lmodern}                            
\usepackage[fontsize=11pt]{fontsize}            
\usepackage[version=3]{mhchem}
\usepackage{verbatim}                           
\usepackage{siunitx}                            
\usepackage{color}                              
\usepackage{listings}
\usepackage{booktabs}
\usepackage[normalem]{ulem}
\usepackage{tikz}                               
\usepackage{colortbl}
\usepackage{stfloats}                           
\usepackage{caption}
\DeclareSIUnit\permille{\text{\textperthousand}}

\graphicspath {{Figures/}}

\definecolor{grey}{rgb}{0.95,0.95,0.95}
\definecolor{mint}{rgb}{0.92,1.0,0.92}
\definecolor{rose}{rgb}{1.0,0.92,0.92}
\definecolor{dRed}{rgb}{0.62,0.0,0.0}
\definecolor{dGreen}{rgb}{0.0,0.62,0.0}
\definecolor{dBlue}{rgb}{0.0,0.0,0.62}
\definecolor{rose}{rgb}{1.0,0.88,0.88}
\newcommand*{\mc}[3]{\multicolumn{#1}{#2}{#3}}  
\newcommand*{\A}{$^{(a)}$}                       
\newcommand*{\B}{$^{(b)}$}

\newcommand*{\cm}{\,cm$^{-1}$\ }
\newcommand*{\mum}{\SI{}{\micro\metre}}
\newcommand*{\mus}{\SI{}{\micro\second}}
\newcommand*{\muJ}{\SI{}{\micro\joule}}

\def\permille{\ensuremath{{}^\text{o}\mkern-5mu/\mkern-3mu_\text{oo}}}
\makeatletter
\newcommand*\bigcdot{\mathpalette\bigcdot@{.5}}
\newcommand*\bigcdot@[2]{\mathbin{\vcenter{\hbox{\scalebox{#2}{$\m@th#1\bullet$}}}}}
\makeatother

\newcommand*\circledd[1]{\tikz[baseline=(char.base)]{
            \node[shape=circle,draw,inner sep=1pt] (char) {\small{#1}};}}
\newcommand*\circleds[1]{\tikz[baseline=(char.base)]{
            \node[shape=circle,draw,inner sep=1pt] (char) {\footnotesize{#1}};}}

\SectionNumbersOn

\author{Thomas Schultz}
\email{schultz@unist.ac.kr}
\phone{+82-52-217-5425}
\fax{+82-52-217-5409}
\affiliation[]
{Department of Chemistry, Ulsan National Institute of Science and Technology (UNIST)}

\title{Review\\
 Correlated Rotational Alignment Spectroscopy: A New Tool for High-Resolution Spectroscopy and the Analysis of Heterogeneous Samples}

\abbreviations{CRASY, FWHM, RCS, FT, SNR}

\begin{document}

\sffamily                                           
\renewcommand{\familydefault}{\sfdefault}           
\renewcommand{\textfraction}{0.10}                  
\renewcommand{\topfraction}{0.95}                   
\renewcommand{\bottomfraction}{0.90}
\renewcommand{\floatpagefraction}{0.20}
\renewcommand*{\thefootnote}{\fnsymbol{footnote}}   

\setlength{\textfloatsep}{16pt plus 4.0pt minus 4.0pt}      
\setlength{\dbltextfloatsep}{16pt plus 4.0pt minus 4.0pt}   
\setlength{\floatsep}{12pt plus 4.0pt minus 4.0pt}          
\setlength{\dblfloatsep}{12pt plus 4.0pt minus 4.0pt}       
\setlength{\intextsep}{16pt plus 4.0pt minus 4.0pt}         

\tableofcontents

\begin{abstract}

Correlated rotational alignment spectroscopy correlates observables of ultrafast gas-phase spectroscopy with high-resolution, broad-band rotational Raman spectra. This article reviews the measurement principle of CRASY, existing implementations for mass-correlated measurements, and the potential for future developments. New spectroscopic capabilities are discussed in detail: Signals for individual sample components can be separated even in highly heterogeneous samples. Isotopologue rotational spectra can be observed at natural isotope abundance. Fragmentation channels are readily assigned in molecular and cluster mass spectra. And finally, rotational Raman spectra can be measured with sub-MHz resolution, an improvement of several orders-of-magnitude as compared to preceding experiments.
\end{abstract}


\section{Introduction}
Scientific progress is based on the observation of the natural world, the description of observed properties with qualitative or quantitative models and, finally, the utilization of our model-based understanding for theoretical or practical advances. Spectroscopy extends our observations to the realm of quantized matter and is of fundamental importance for all molecular sciences. Spectroscopic tools allow us to characterize the composition of samples, to identify molecular structure, and to observe a large range of interesting molecular properties. New spectroscopic capabilities therefore enable the progress of science: The ability to observe new types of samples and to subsequently manipulate and control their properties are an indispensable pre-requisite for the development of new science and technology. Conversely, our inability to characterize particular samples will leave us blind to their potential utility.

The development of correlated rotational alignment spectroscopy (CRASY) aims to remove a blind spot in the observation of heterogeneous, or impure, molecular samples. Most molecular samples are impure and chemists routinely purify their compounds of interest from a natural sample or a synthetic mixture. Purification is often difficult, e.g., for rare isotopologues and energetic species. In some cases purification is impossible, e.g., for molecular tautomers or instable molecules. Spectra for impure samples show averaged spectroscopic signals and weak sample components cannot, usually, be resolved. Would it not be nice if we could mark each component in a heterogeneous sample and, subsequently, sort the spectroscopic results to assign observed properties to each sample component?

CRASY marks spectroscopic data by correlation to high-resolution rotational Raman spectra. Correlation is achieved by observing the interference of coherent rotational states, as they are probed by a separate spectroscopic measurement. The probed spectroscopic observable can, in principle, be any signal generated by the interaction of a gas-phase molecular sample with femto or picosecond laser pulses. The correlated rotational spectra can serve as molecular fingerprints and used to assign the probed spectroscopic results. Beyond serving as a fingerprint, the rotational spectra can be analyzed to characterize the structure of detected molecules. Currently, CRASY experimental data only exists for photoionization mass and electron spectroscopy and the former is the main topic of this review. Section \ref{sec:Outlook: Future CRASY Experiments} offers a perspective on other possible CRASY experiments.

\begin{figure}[htb]
\centering
  \includegraphics[width=8.3cm]{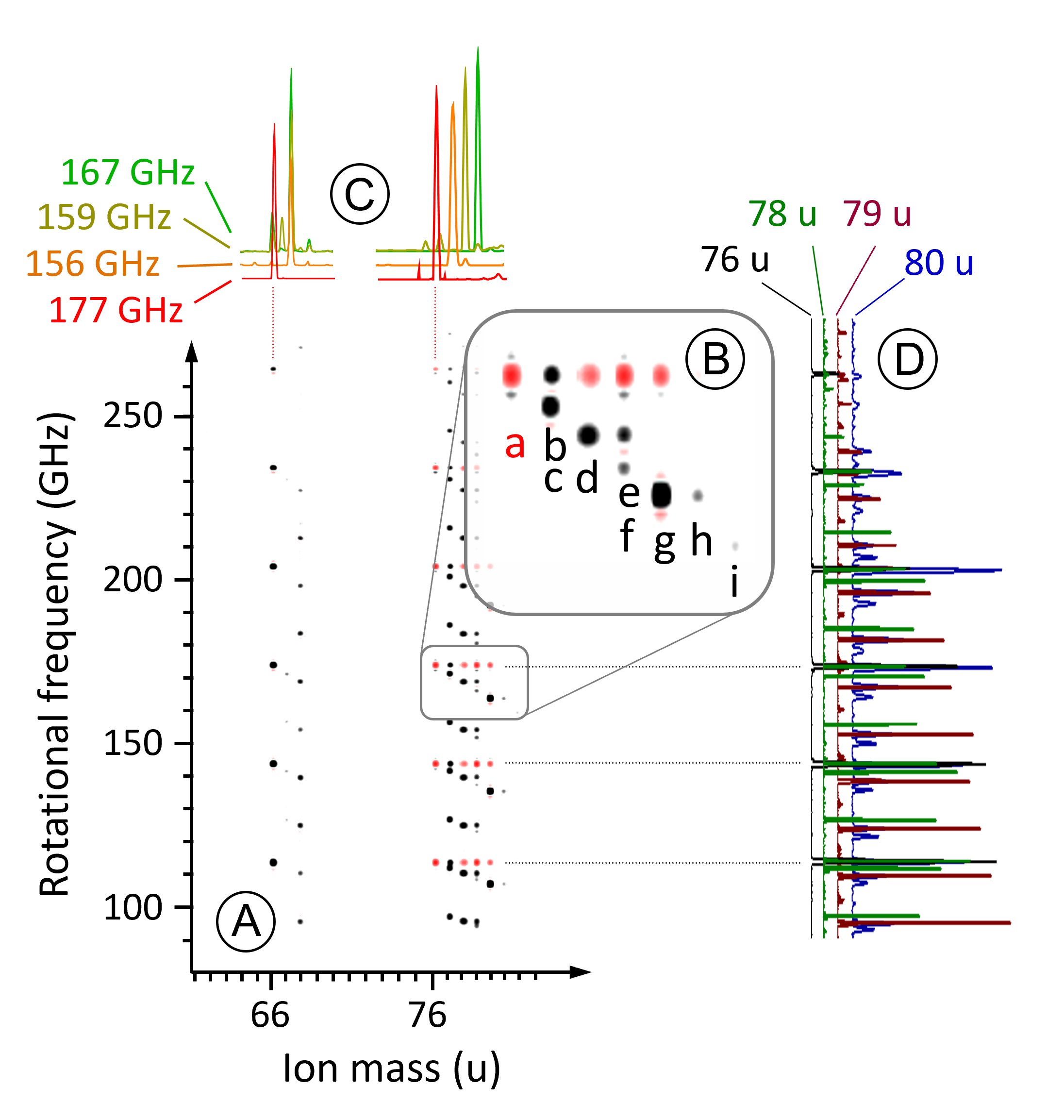}
    \caption[Caption for Fig1_Correlated_Spectroscopy]{\small \circleds{A} Two dimensional mass-CRASY data for \ce{CS2} shows correlated information of molecular mass and rotational frequency. Red signals show an inverted phase due to signal saturation. \circleds{B} An enlarged section illustrates the assignment of 9 naturally occurring isotopologue signals (a--i), containing \ce{^{12}C}, \ce{^{13}C}, \ce{^{32}S}, \ce{^{33}S}, \ce{^{34}S}, and \ce{^{36}S} isotopes. \circleds{C} One dimensional cuts at selected rotational transition frequencies resolve mass spectra correlated with a single isotopologue and reveal the correlation between parent and fragment species. \circleds{D} One dimensional cuts at selected masses separate isotopologue rotational spectra. }
  \label{Fig1_Correlated_Spectroscopy}
\end{figure}

It is a truism in the physical sciences that any order-of magnitude improvement of a measurement will lead to new and interesting insights. The same, certainly, holds true in chemistry, but the relevant technological developments are usually hidden in the internals of commercial devices that are used for chemical sample characterization. The current implementation of mass-CRASY delivered order-of-magnitude increases in spectroscopic information via two separate avenues. First, the correlated spectroscopic information corresponds to the product, not the sum, of the corresponding uncorrelated spectra, greatly increasing the quantity and quality of spectroscopic information. This is readily visualized by considering the two dimensional mass-CRASY data plotted in Fig.\ \ref{Fig1_Correlated_Spectroscopy}, where the information is spread across an area corresponding to the product of a mass axis and a rotational frequency axis. The correlated rotation-mass information facilitates the analysis of heterogeneous samples, as described in Section \ref{sec:Applications of Mass-CRASY}. Second, progress in the CRASY measurement technology increased the resolution and accuracy of rotational Raman spectra by several orders-of-magnitude. Relevant developments are described in Section \ref{sec:High-Resolution, High-Accuracy Rotational Raman Spectra}. But before delving into the concepts and capabilities of CRASY, we should give a brief review of earlier work that laid the foundation for the development of CRASY.



\section{Background}
\label{sec:Related_Techniques}
\subsection{Inception of CRASY}
The CRASY measurement concept was inspired by developments in the field of nuclear magnetic resonance spectroscopy (NMR) in the 1960s to 1980s.\cite{Ernst1992} Time-domain, Fourier-transform NMR was initially developed to increase the available spectroscopic signal bandwidth and contrast. The resulting understanding of coherent spin-state superpositions and the development of radio wave technology to excite and probe these state superpositions led to the development of correlated, multi-dimensional NMR methods and the 1991 Chemistry Nobel Prize for Richard Ernst.\cite{Ernst1992} The resolution of NMR spectra is inherently limited by spin lifetimes and available magnetic field strengths. The information content of uncorrelated NMR spectra was therefore only sufficient to resolve the structure of small molecules. Multi-dimensional NMR spectra, on the other hand, resolved spin coupling in congested spectra and allowed to resolve the structure of macromolecules and even proteins, leading to the 2002 Chemistry Nobel Prize for Kurt W\"uthrich.\cite{Wuthrich2003}

\begin{figure}[htb]
\centering
  \includegraphics[width=8.3cm]{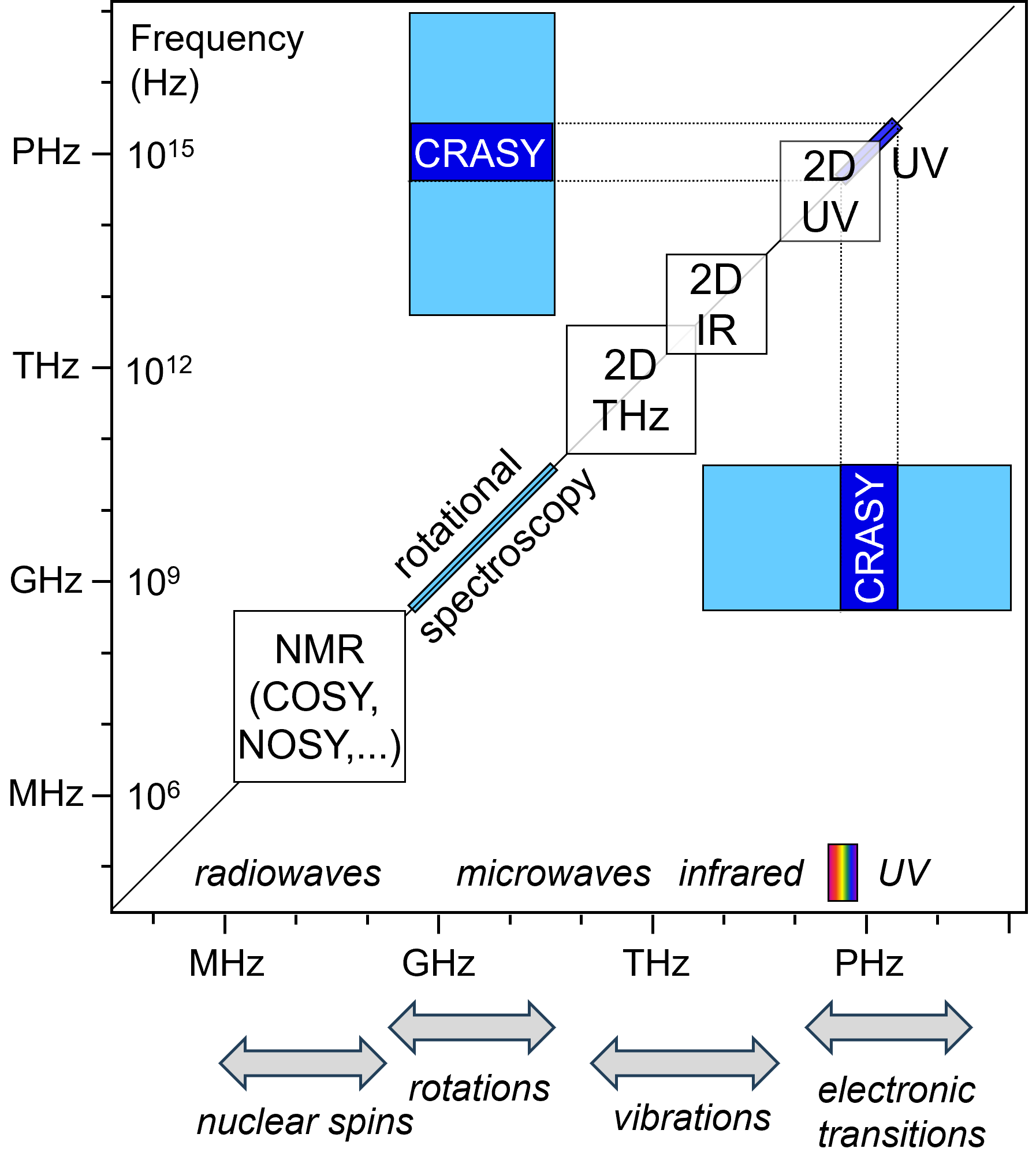}
  \caption{Spectroscopic regime of current CRASY experiments (blue) and possible future CRASY experiments (light blue), as compared to multi-dimensional NMR, teraherz (THz), infrared (IR), and ultraviolet (UV) spectroscopy. CRASY correlates dissimilar observables, e.g., molecular rotation (GHz regime) with ultraviolet photoexcitation and ionization. [Adapted from Ref.\ \citenum{Schultz2015}.]}
  \label{Fig2_regime_of_CRASY}
\end{figure}

The idea to perform correlated laser spectroscopy, akin to NMR experiments, emerged in 2004, motivated by the search for a method to separate spectroscopic signals from biomolecular tautomers and clusters.\cite{SFB450} Just as time-domain NMR experiments excite and probe a superposition of spin states with radio wave pulses, a superposition of rotational states can be excited and probed by ultrafast laser pulses. As reviewed in the context of femtochemistry\cite{Zewail2000,Zewail2000a}, ultrafast laser spectroscopy can resolve a plethora of molecular observables, hence rotational spectra can be correlated with a wide number of other observables. CRASY therefore correlates fundamentally different spectroscopic regimes, as illustrated in Fig.\ \ref{Fig2_regime_of_CRASY}. This is different from most established two-dimensional techniques, such as 2D\,NMR, 2D\,THz spectroscopy\cite{Reimann2021}, or 2D\,vibrational spectroscopy \cite{Hamm2011,Jonas2003a}, which correlate information from two identical spectroscopic measurements.

\subsection{Related Measurement Techniques}
The fundamental concepts of quantum state superpositions, coherence, and their effect on measured signals was understood in the early days of quantum physics. The experimental observation of coherent rotational states in dipolar molecules emerged with the development of strong microwave sources and forms the basis of modern Fourier-transform microwave spectroscopy (FTMW).\cite{Flygare1976,Grabow2011,Shipman2011,Park2016} Impulsive excitation of rotational states became possible with the advent of ultrafast lasers and was first observed by Heritage in 1975.\cite{Heritage1975} Baskin, Felker, and Zewail recognized the utility of rotational wavepacket observation for molecular spectroscopy and used rotational coherence spectroscopy (RCS) to characterize a number of chromophores.\cite{Baskin1986,Felker1986,Baskin1987,Felker1987} The state of the field and numerous spectroscopic implementations were reviewed by Felker\cite{Felker1992} and by Frey et al.\cite{Frey2011}

Multiple groups developed their own variants of RCS. Riehn et al.\ measured mass-selected RCS traces \cite{Riehn2000} and implemented four-wave-mixing schemes \cite{Riehn2002}. Frey et al.\ started a systematic investigation of ground-state rotational constants in non-dipolar molecules \cite{Frey2004} and pushed the boundaries in terms of measurement accuracy.\cite{Frey2011} In most RCS implementations, the analysis of rotational properties was performed in time domain data, i.e., by analyzing observed signal modulations in time domain traces. Rouz\'{e}e et al.\ Fourier-transformed signal modulations observed in Raman-induced polarization spectroscopy, a variant of RCS, and fitted the spectroscopic results in the frequency domain.\cite{Rouzee2006}

When molecules with highly anisotropic polarizability interact with strong laser pulses, the interaction can lead to significant molecular alignment. Molecular alignment spectroscopy characterized the resulting molecular orientation with the spatially resolved detection of fragment ions (ion imaging) after photoionization and fragmentation, often at high laser intensities that induced Coulomb explosion. Methods to maximize non-adiabatic alignment, which occurs after interaction with a short and intense laser pulse, were explored in-depth in the experimental and theoretical work of Stapelfeldt, Seideman, and colleagues.\cite{Stapelfeldt2003,Peronne2003,Peronne2004,Poulsen2004,Hamilton2005,Holmegaard2007,Viftrup2007}.

In recent years, the fields of RCS and molecular alignment spectroscopy converged and ion imaging experiments were used for the characterization of rotational spectra in atomic and molecular clusters. Wu et al.\cite{Wu2011}, Veltheim et al.\cite{Veltheim2014} and Mizuse et al.\cite{Mizuse2022} investigated the spectra of nobel gas dimers with increasing fidelity. Galinis et al.\cite{Galinis2014} investigated the spectrum of the acetylene-helium cluster. Chatterley et al.\cite{Chatterley2020} Fourier-transformed alignment traces for carbon disulfide dimer and determined rotational constants for this molecular cluster. Several groups investigated alignment and rotational coherence for molecular clusters in helium droplets.\cite{Chatterley2019,Chatterley2020a,Schouder2021} A recent review from Schouder at al.\ summarized the state of the field.\cite{Schouder2022} We disambiguate RCS and molecular alignment experiments from CRASY because the correlation to additional observables is absent or plays a subordinate role in the former but is a central aspect in the latter.

The advent of ultrafast lasers also spurred the development of correlated multi-dimensional spectroscopy for condensed phase samples. This field gave rise to a rich variety of experimental implementations for 2D\,THz,\cite{Reimann2021} 2D\,IR,\cite{Cho2008,Fritzsch2020} and 2D\,electronic spectroscopy.\cite{Jonas2003a,Tollerud2017} Condensed phase experiments are performed with high sample densities and therefore deliver better signal contrast than gas phase experiments. But the strong coupling of rotational and vibrational motion to the condensed environment leads to short coherence times and to inherently low spectroscopic resolution. The high-resolution gas-phase experimental approaches discussed here are therefore not in competition to condensed phase multi-dimensional spectroscopy.

Hole-burning spectroscopy offers an alternative tool for the separation of spectroscopic signals in heterogeneous samples. This method was used extensively for the analysis of biomolecular building blocks in the gas phase, as reviewed by deVries and Hobza.\cite{deVries2007} In hole-burning experiments, an optical transition of one sample component is saturated (burned out) and thereby removed from a subsequent spectroscopic measurement. Where correlated spectroscopy reveals all signal correlations within the probed spectral range, hole-burning isolates only the signal for one selected species at a time.

The Chemists' traditional approach to handle heterogeneous samples is to purify each sample component before characterization. However, purification of minor sample components is difficult and this approach may leave us blind to whole classes of molecules, e.g., short-lived reactive species or molecular tautomers. A particularly powerful approach to characterize compositionally heterogeneous samples is the direct combination of a separation technique with a subsequent spectroscopic analysis, e.g., as implemented in gas-chromatography coupled to mass spectrometry (GC-MS, see Ref.\ \citenum{Fang2015_GCMS} and references therein) or in MS$^n$ approaches\cite{Glish2008,Agthoven2019}. To appreciate the respective utility of such methods as compared to correlated spectroscopy, it is important to recognize the inherent technological limitations for each of these approaches. E.g., the signal origin in GC-MS may be obscured by a propensity for molecular fragmentation in the chromatographic column and MS$^n$ experiments are only feasible if molecules form stable ionic species.

With the correlation of gas-phase rotational Raman spectra to other spectroscopic observables, CRASY offers a new and complementary approach to characterize heterogeneous samples. Gas-phase rotational spectroscopy can be performed with extraordinary resolution\cite{Grabow2011,Shipman2011,Frey2011} and the separation of signal contributions in CRASY data is straightforward. The choice of correlated observables is only limited by the spectral range and power of available short-pulse light sources. With modern high-harmonic sources and free-electron lasers,\cite{Schultz2014} the accessible spectroscopic range can, in principle, reach into the XUV and X-ray regime.

\section{Experimental Principle}
\label{sec:Experimental_Concept}
\subsection{Exciting and Probing Rotational Wave Packets}
\begin{figure}[htb]
\centering
  \includegraphics[width=8.3cm]{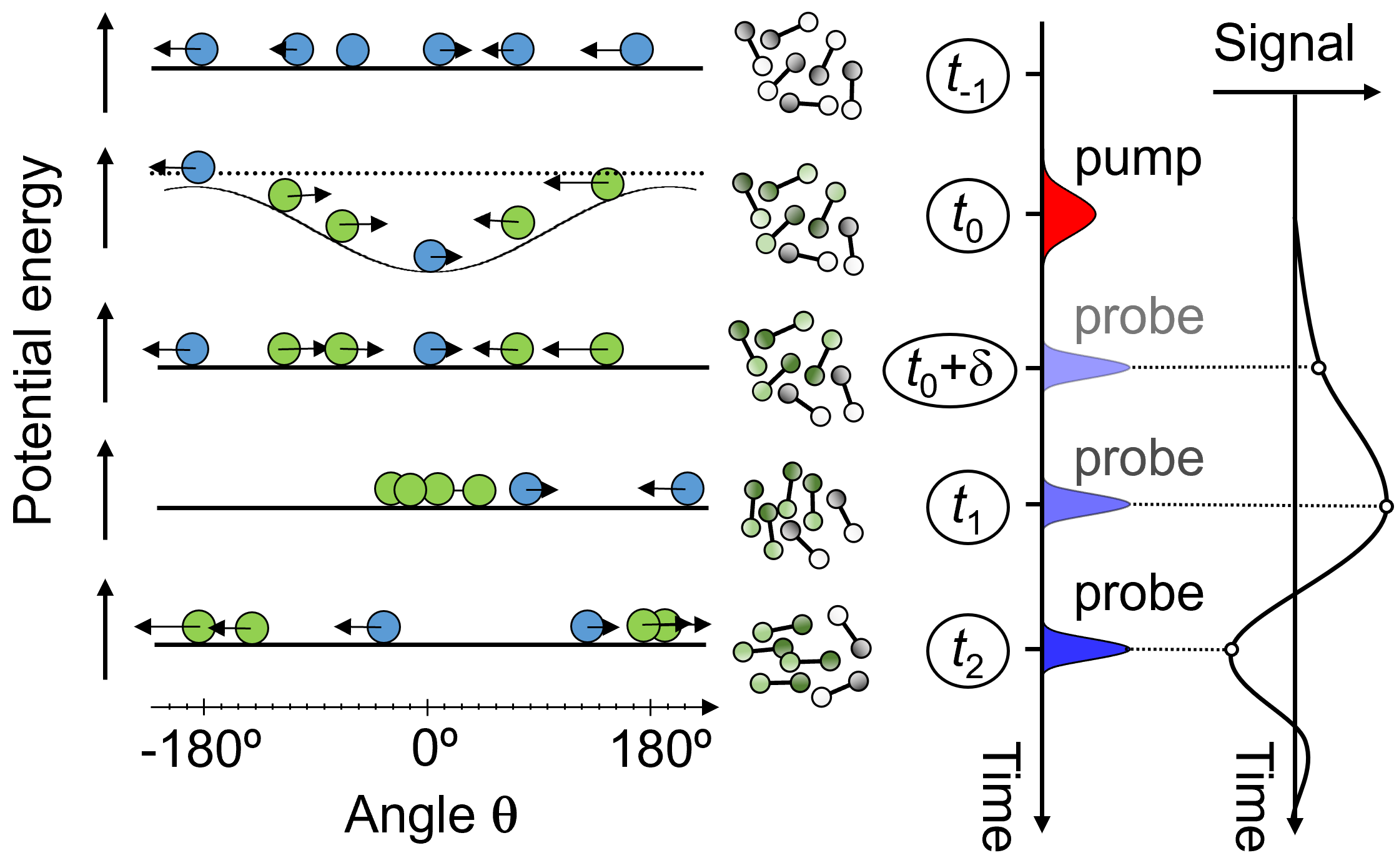}
  \caption{Pseudo-classical representation of the CRASY pump-probe scheme. ($t_{-1}$) A molecular ensemble consists of randomly oriented and freely rotating molecules. Angular momentum is represented by horizontal arrows and $\theta$ denotes the angle between the axis of highest molecular polarizability and the laser electric field (polarization axis). ($t_{0}$) Interaction of the molecular polarizability and laser electric field distorts the angular potential. ($t\mathrm{_0+\delta}$) After the interaction, molecules rotate with a new angular momentum, turning their most polarizable axis into the pump polarization axis. This leads to transient molecular alignment ($t_{1}$) and anti-alignment ($t_{2}$). The magnitude of a probed signal (right hand side) grows and shrinks as the molecular transition moments rotate into and out-of the polarization axis of the probe laser.}
  \label{Fig3_CRASY_Classical_Scheme}
\end{figure}

The CRASY measurement is based on the coherent excitation of a rotational state superposition (wave packet) and the probing thereof over an extended period of time. Fig.\ \ref{Fig3_CRASY_Classical_Scheme} depicts a pseudo-classical model of the CRASY pump-probe scheme and illustrates the role of molecular alignment. The electric field component of a short but strong pump laser interacts with the anisotropic polarizability of a molecule and pulls the most polarizable molecular axis towards the electric field axis. This imparts an angular impulse onto the molecules, causing them to rotate into the laser polarization axis. After the brief interaction period, molecules rotate freely and move through moments of transient alignment in the lab frame. This alignment is observed by probing the molecular ensemble with a second laser pulse. If the probed transition dipoles $\mu_\mathrm{T}$ have a well-defined orientation within the molecular frame, the probed spectroscopic signals will increase and decrease as the probed transition moments rotate into and out-of the polarization axis of the probing laser. The observed modulation of the probe signal reveals the quantized rotational periods of the molecules.

\begin{figure}[htb]
\centering
  \includegraphics[width=8.3cm]{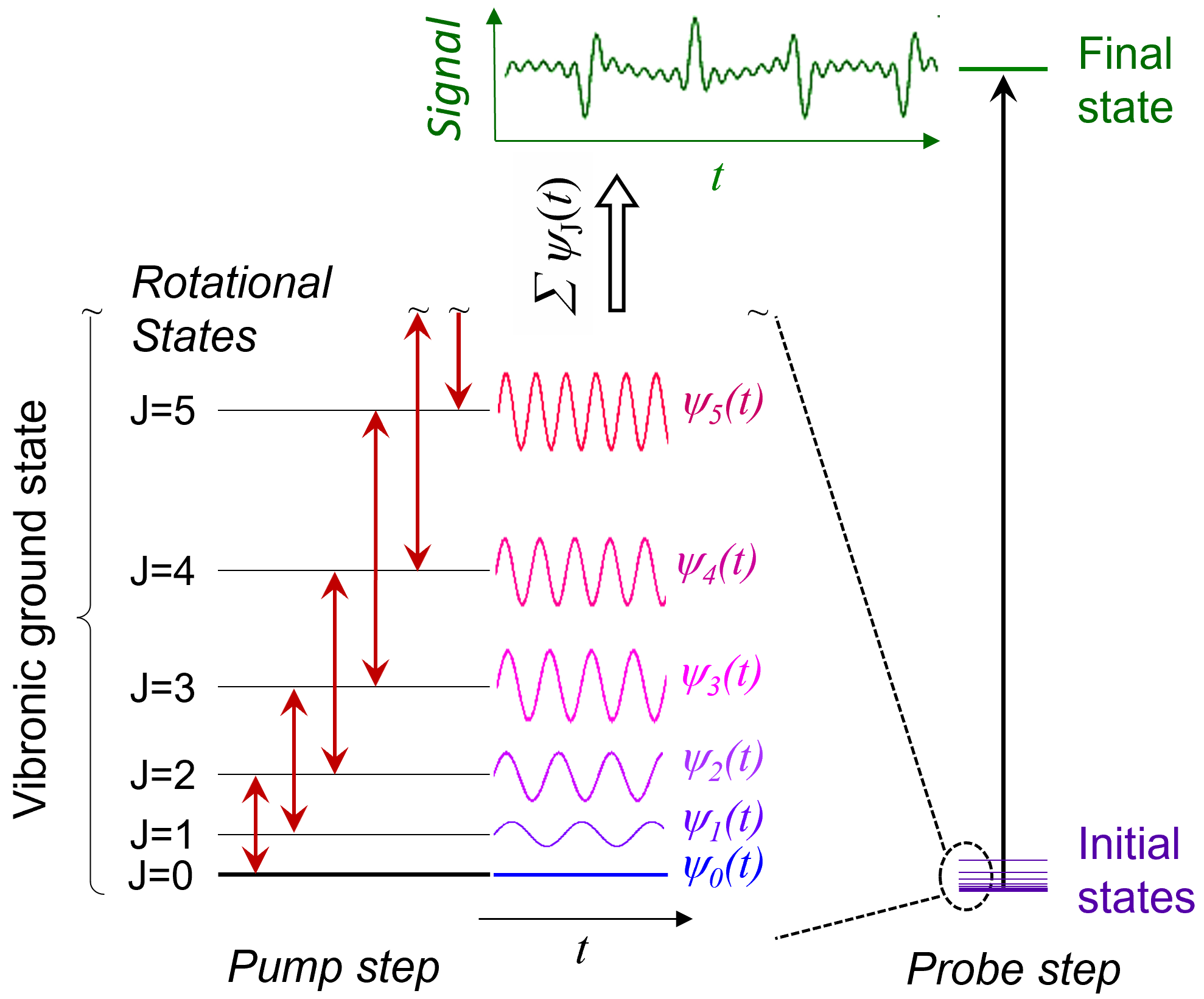}
  \caption[Caption for Fig4_CRASY_QM_Scheme]{\small Quantum mechanical scheme to illustrate the measurement principle of CRASY. In the pump step (left), a spectrally broad laser pulse Raman excites a coherent superposition of rotational states within the vibronic ground state. The coherence is reflected in the fixed phase-relation between the time-dependent state wave functions $\psi_\mathrm{J}(t)$. In the probe step (right), a second laser pulse drives transitions from the rotational states $\psi_\mathrm{J}(t)$ into a common final state. The probed signal is proportional to the coherent sum $\sum{\psi_\mathrm{J}(t)}$ of the rotational wave functions and therefore shows time-dependent signal modulation.}
  \label{Fig4_CRASY_QM_Scheme}
\end{figure}

Fig.\ \ref{Fig4_CRASY_QM_Scheme} depicts a more rigorous quantum mechanical scheme, which considers the interference of pumped rotational states when they are probed into a common set of final states. The impulsive rotational Raman excitation, driven by the pump pulse, creates a coherent superposition of time-dependent rotational states $\psi_\mathrm{J}(t) = c_\mathrm{J}\psi_\mathrm{J}(0) \cdot e^{-i (\frac{E_\mathrm{J}}{\hbar} t + \phi_\mathrm{J})}$,
with a fixed relation between the phases $\phi_\mathrm{J}$. The transition probability $P_{\psi_\mathrm{f} \leftarrow \sum \psi_\mathrm{J}} = \langle \psi_\mathrm{f} | \mu_\mathrm{T} | \sum{\psi_\mathrm{J}(t)} \rangle$ for probing this superposition of states into a final state $\psi_\mathrm{f}$ depends on the coherent sum over all pumped rotational states $\sum{\psi_\mathrm{J}(t)}$ and the transition moment $\mu_\mathrm{T}$. With a suitable transition moment, the coherent sum oscillates through moments of constructive and destructive interference. Note that time-dependent wavefunctions $\psi_\mathrm{J}(t)$ drawn in Fig.\ \ref{Fig4_CRASY_QM_Scheme} only show the real-valued wavefunction component.

Probed time-domain signal oscillations encode spectroscopic information about the transition energies $E_\mathrm{J'} \leftarrow E_\mathrm{J}$ of all coherently excited states. A Fourier transformation (FT) converts this time-domain information into a complex-valued frequency domain spectrum. The FT yields spectral amplitudes and phases, the latter encode the relative orientation of transition dipoles in the pump and probe steps. This phase information is not available from frequency domain data, unless heterodyne detection methods are employed.

\begin{figure}[htb]
\centering
  \includegraphics[width=8.3cm]{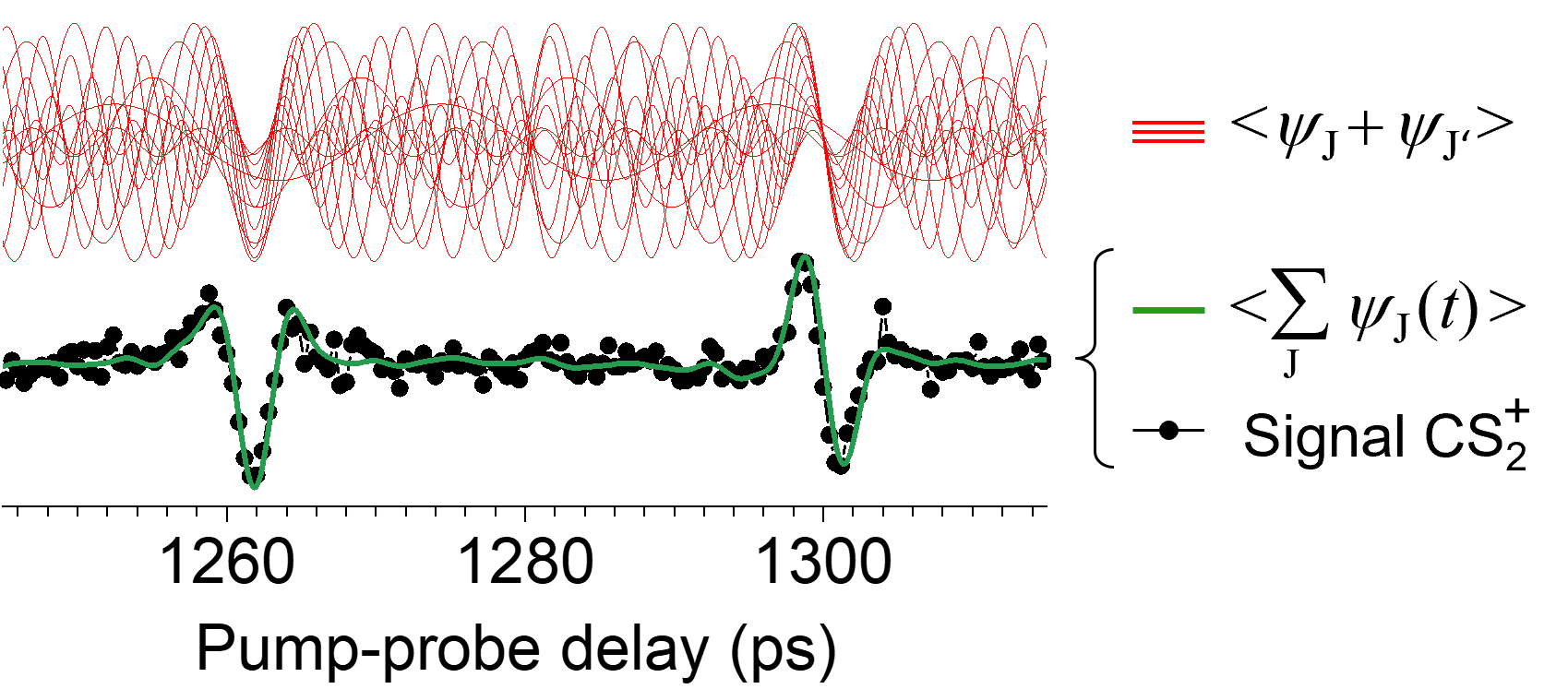}
  \caption{ Coherent rotational wavepacket observed for \ce{CS2}. Detected ion signals (black dots) show pronounced delay-dependent signal modulation. The signal can be simulated with the coherent sum (green line) of rotational wave functions $\psi_\mathrm{J}(t)$ and can be decomposed in to the pairwise sum of initial and final states (red lines) that are excited in the pump step. [Adapted from Ref.\ \citenum{Schroter2011}.]}
  \label{Fig5_Wavepacket}
\end{figure}

Fig.\ \ref{Fig5_Wavepacket} illustrates quantum mechanical wave interference based on experimentally observed ion signals for a coherent superposition of carbon disulfide (\ce{CS2}) rotational states. The delay-dependent signal modulation shows moments of strong alignment (positive signals) and anti-alignment (negative signals) and can be reproduced by the sum of multiple rotational wave functions $\psi_\mathrm{J}(t)$, with characteristic amplitudes $\psi^0 $, frequencies $\omega$, and phases $\phi_{0}$. Note that the time-dependent wave functions are complex-valued and the observed signal oscillations reflect the frequency differences $\Delta\omega = \frac{(E_\mathrm{J'} - E_\mathrm{J})}{\hbar} \, t$ between states that were coherently excited with a pump laser pulse.

Pronounced interference maxima or minima in the quantum mechanical picture correspond to moments of maximal alignment or anti-alignment. Strong alignment occurs only for a specific relation between initial phases $\phi_\mathrm{J}$, but temporal signal oscillations can be observed and interpreted independent of these initial phases. Moments of significant alignment, despite their prominence in measured signal traces, are not required to characterize rotational spectra in the time domain.

The observation of quantum state interference in form of temporal signal oscillations is expected for most sets of coherent intermediate and final states, not just rotational states. However, there can be unfavorable cases where the probed transition moments are insensitive to the pumped excitations. This would be the case, e.g., for a spherical transition moment or when the pumped and probed transition moments are at a magic angle. Another unfavorable situation can arise when observed signals involve multiple final states with different transition moments, washing out the coherent signal oscillations. A brief consideration of the transition moments used for excitation and probing is sufficient to determine the suitability of a particular pump-probe scheme. The choice of probed states should therefore be a matter of experimental convenience: The pump-probe scheme should facilitate the observation of the coherent wavepacket with maximal contrast and, in the case of CRASY, carry meaningful correlated information.

\subsection{Experimental Implementation}
\label{sec:Experimental Implementation}
Current CRASY experiments probe rotational wave packets with resonant 2-photon photoionization. The resonant excited state, with a well-oriented electronic transition dipole, can be considered as an angular filter state, which ensures that rotational state interference can be observed with high contrast. The transition dipoles for electronic excitation are often large and the detection of ions and electrons in mass- and electron-CRASY is possible with high quantum yields, facilitating the signal collection. Correlated mass and electron spectra contribute valuable correlated information about molecular mass and electron binding energies. The value of correlating other observables should be readily evident and we hope that a broader field of CRASY experiments will be developed in due course (see Section \ref{sec:Outlook: Future CRASY Experiments}). This is the reason why Fig.\ \ref{Fig2_regime_of_CRASY} plots a large regime of possible CRASY experiments, across a frequency regime from microwaves to X-rays.

\begin{figure}[htb]
\centering
  \includegraphics[width=8.3cm]{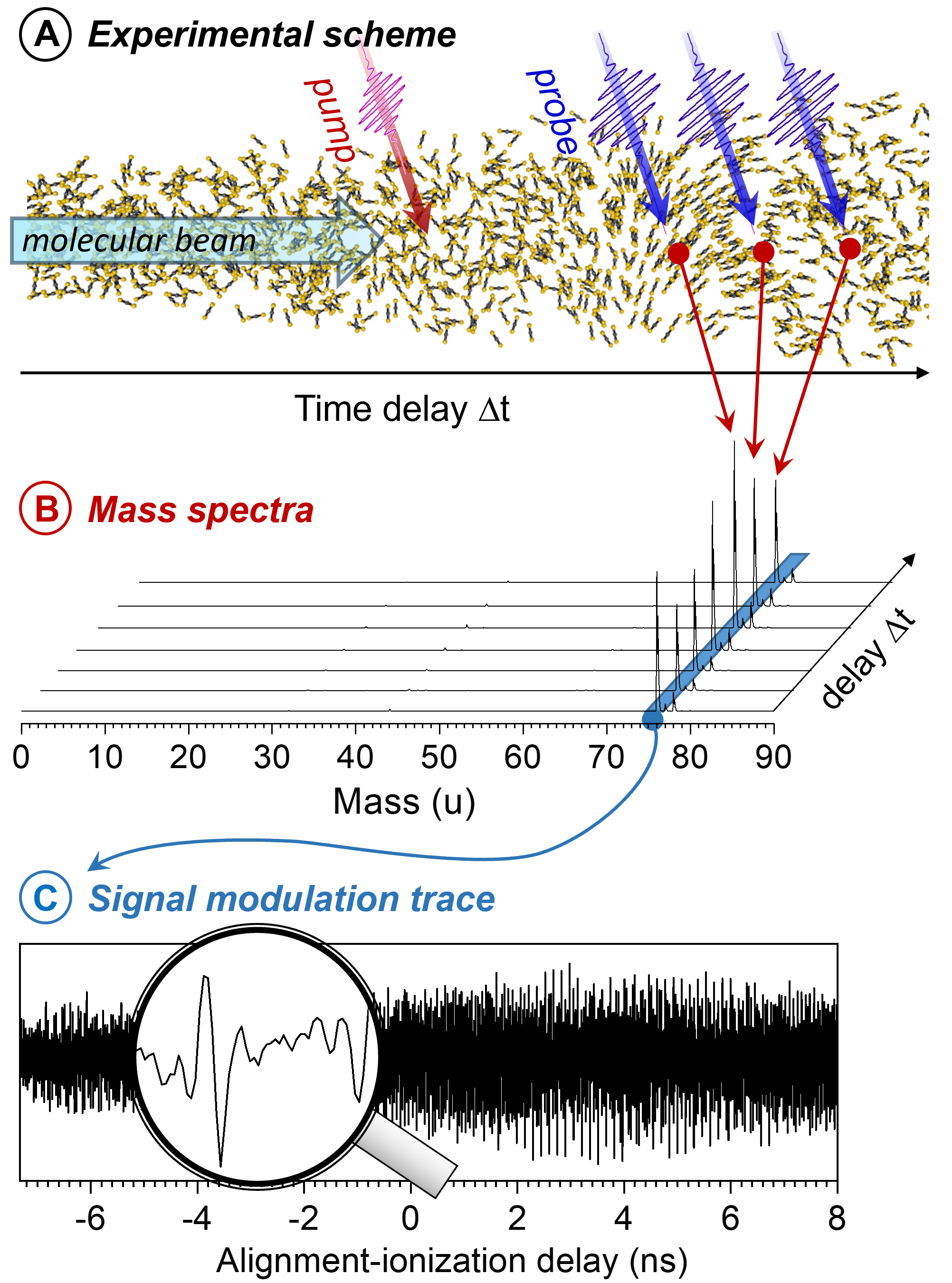}
  \caption[Fig6_mass-CRASY_implementation]{\circleds{A} Experimental implementation of mass-CRASY. A cold molecular beam in vacuo interacts with a short alignment laser pulse (pump), exciting a coherent rotational wavepacket. Molecular alignment is probed by photoionization after time delay $\Delta t$ and ions are mass-analyzed. \circleds{B} Resulting mass spectra show a modulation of ion signal amplitudes as function of $\Delta t$, due to molecular alignment. \circleds{C}  A signal modulation trace, can be plotted for each mass signal.}
  \label{Fig6_mass-CRASY_implementation}
\end{figure}

Fig.\ \ref{Fig6_mass-CRASY_implementation} illustrates the data acquisition process for a mass-CRASY experiment. In a collision-free molecular beam, a pump pulse excites a coherent superposition of rotational states, which is subsequently probed by resonant photoionization (Fig.\ \ref{Fig6_mass-CRASY_implementation}\,\circledd{A}). Laser pulses originated from an amplified Ti:Sa laser and typically 100\,\muJ, 800\,nm, 1\,ps pulses were used for rotational excitation and single-\muJ, 200\,nm or 266\,nm, 50\,fs pulses for photoionization. Ions were detected in a mass spectrometer and a mass spectrum can be plotted for each pump-probe time delay (Fig.\ \ref{Fig6_mass-CRASY_implementation}\,\circledd{B}). The experiment is repeated for multiple time delays between pump and probe laser pulses. Ion signals for each observed mass show delay-dependent signal modulation due to rotational state interference. The modulation trace (Fig.\ \ref{Fig6_mass-CRASY_implementation}\,\circledd{C}) for each ion mass therefore contains correlated information about the rotational periods of the neutral ground state molecule. The FT of the modulation trace yields the correlated rotational Raman spectrum for the selected ion mass.

Throughout this manuscript, rotational Raman signals are plotted as power spectra, based on the real and imaginary components of the Fourier-transformed signals ($S_\mathrm{power} = S_\mathrm{FT,real}^2 + S_\mathrm{FT,imag}^2$). Traces were zero-padded before Fourier transformation.\footnote{Zero padding extends the signal trace with an array of zeros. This interpolates points in the FT spectrum.} Zero padding by at least a factor 2 is used to recover the full spectroscopic information into the power spectrum.\cite{Bartholdi1973} The analysis of frequency domain spectra is then performed as established in the literature\cite{Gordy1984} and using spectroscopic analysis software, such as Pgopher.\cite{Western2010,Western2017}

A proper sampling of the time domain signal is required to characterize the full molecular spectrum. With a discrete sampling step size $t_\mathrm{step}$, only frequencies up to the Nyquist critical frequency $f_\mathrm{c} = (2\cdot t_\mathrm{step})^{-1}$ are resolved.\cite{NumericalRecipes} Higher spectral frequencies $f_2 > f_\mathrm{c}$ cause spurious signals due to aliasing, i.e., will appear at a sampled frequency $f_1 < f_\mathrm{c}$, where $f_1$ and $f_2$ differ by a multiple of $2\cdot$$f_c$. If transition frequencies are expected outside the sampled frequency regime, aliasing may be avoided by either \emph{(i)} restricting the excitation bandwidth to the sampled frequency regime, i.e., using an impulsive excitation pulse duration equal or larger than the sampling step size, or \emph{(ii)} continuous sampling with a subsequent discretization of sampled delays, or \emph{(iii)} irregular sampling, i.e., the sampling of data at delays $\mathrm{n} \cdot t_\mathrm{step} + r$  with random values for $r$ in the range $\pm \frac{1}{2} \cdot t_\mathrm{step}$.

As described in more detail in Section \ref{sec:High-Resolution, High-Accuracy Rotational Raman Spectra}, the rotational resolution scales proportional to the scanned delay range. A large number of delay-dependent mass spectra must therefore be measured to obtain a good resolution across a large spectral range. Mass-CRASY experiments are typically based on the measurement of more than 10\,000 mass spectra. Each mass spectrum contains more than 100\,000 data points and a typical measurement creates gigabyte data sets. Because mass spectra are highly discrete, in-memory compression algorithms\cite{ZLIB} are very effective to reduce data quantity and data transfer times. The information content of mass-CRASY data is quite unique and tailored data analysis routines were programmed in the Python programming language, utilizing sparse data formats and efficient algorithms to limit memory consumption and computation times. The analysis software is under continuous development to handle the evolving data types obtained by CRASY. A documented set of analysis scripts and mass-CRASY sample data is available at a data repository.\cite{Figshare_CrasyDataAnalysis}

\section{Applications of Mass-CRASY}
\label{sec:Applications of Mass-CRASY}
\subsection{Rotational Spectra of Molecular Isotopologues}
\label{sec:Rotational Spectra of Molecular Isotopologues}
Most elements have multiple naturally occurring isotopes,\cite{Michael2011} but the isotope distribution is often dominated by a single isotopic species, e.g., with an abundance ratio of 98.9\,:\,1.1 for \ce{^{12}C} to \ce{^{13}C}, of 99.8\,:\,0.20 for \ce{^{16}O} to \ce{^{18}O}, of 99.6\,:\,0.37 for \ce{^{14}N} to \ce{^{15}N}, and of 95.0\,:\,0.7\,:\,4.2 for \ce{^{32}S} to \ce{^{33}S} and \ce{^{34}S}. The size of isotopologue spectroscopic signals is proportional to the corresponding isotope abundance and it is difficult to resolve rare isotopologue signals in uncorrelated spectra. Nevertheless, isotopologue-selective spectroscopy is an important tool for the study of molecular structure and reactivity and measurements are commonly performed with synthesized isotopologues, an expensive and time-consuming process. In CRASY data, isotopologue rotational spectra can be separated by their correlated mass and rare isotopologue signals may be observed without interference. This facilitates the assignment of spectra for isotopologues in samples with heterogeneous isotopic composition and at natural isotope abundance.

With exceptions for some rather exotic molecules,\cite{Fleming2014} isotopic substitution does not significantly affect chemical bonding but can induce isotope effects that reveal whether a substituted isotope is involved in a rate-determining reaction step.\cite{Wolfsberg1969} Isotopic substitution also allows to mark and track the position of a particular atom throughout the course of a chemical reaction. Section \ref{sec:Analysis of Cationic Fragmentation} gives examples for isotopic tracking in cationic fragmentation reactions. The effect of isotopic substitution on vibrational and rotational transition frequencies yields valuable information about molecular structure. Section \ref{sec:De-Novo Structure Determination} gives examples for \emph{de novo} structure analysis based on CRASY experimental data.

\begin{figure*}[htb]
\centering
  \includegraphics[width=12.cm]{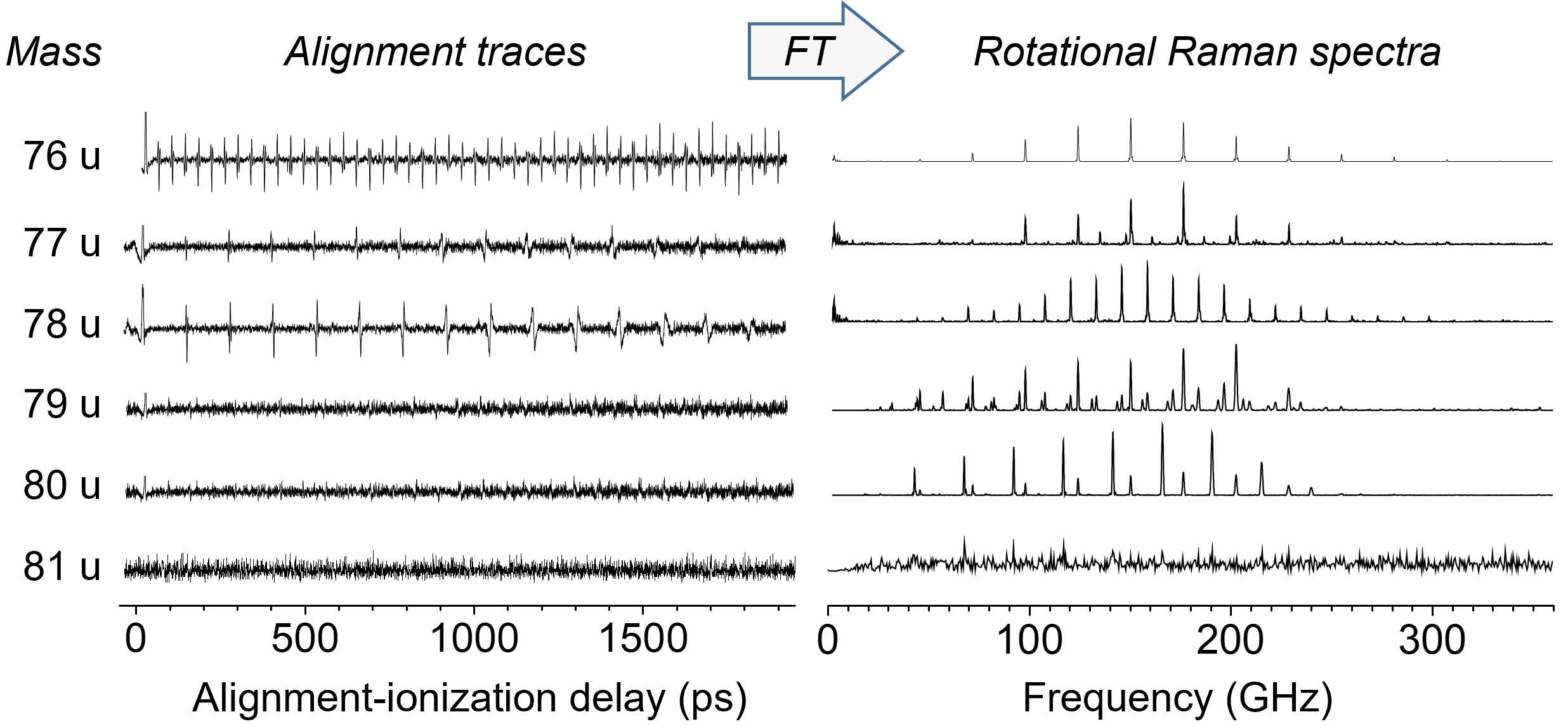}
  \caption{ (Left) Delay dependent ion signal modulation traces for \ce{CS2} isotopologues in mass channels 76\,u to 81\,u. (Right) Power spectra obtained by FT of the traces yields mass-correlated rotational Raman spectra. [Adapted from Ref.\ \citenum{Schroter2011}.]}
  \label{Fig7_FFT}
\end{figure*}

The first experimental implementation for mass- and electron-CRASY experiments was published in 2011 and analyzed rotational spectra of \ce{CS2} isotopologues.\cite{Schroter2011} \ce{CS2} is a simple linear molecule with 20 naturally occurring isotopologues, due to the presence of stable \ce{^{12}C}, \ce{^{13}C}, \ce{^{32}S}, \ce{^{33}S}, \ce{^{34}S}, and \ce{^{36}S} isotopes. Mass spectra and a signal modulation trace for the main \ce{CS2} isotopologue are plotted in Fig.\ \ref{Fig6_mass-CRASY_implementation}. Signal modulation traces in six \ce{CS2} isotopologue mass channels  were Fourier-transformed to obtain mass-correlated rotational Raman spectra, as shown in Fig.\ \ref{Fig7_FFT}. Rotational spectra remained badly resolved for isotopologues with low abundance. This problem was overcome by measuring with higher ion count rates, albeit at the cost of saturation artifacts.

\begin{figure}[htb]
\centering
  \includegraphics[width=8.3cm]{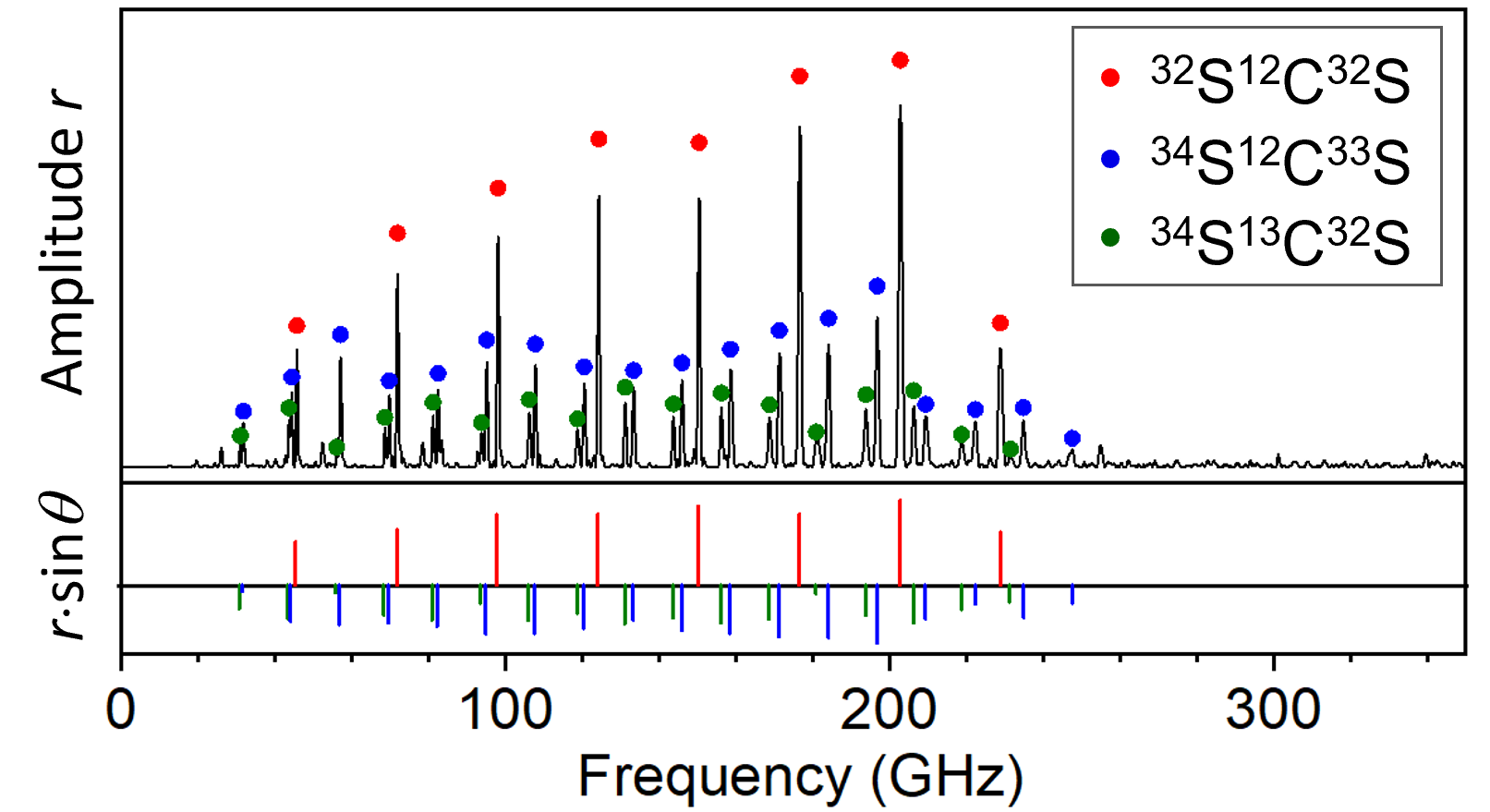}
  \caption{ (Top) A spectrum correlated to mass 79\,u shows signals for the \ce{^{34}S^{12}C^{33}S} and \ce{^{34}S^{13}C^{32}S} isotopologues, but also saturation artifacts for the most abundant \ce{^{32}S^{12}C^{32}S} isotopologue. (Bottom) Multiplication of the spectral amplitudes with the sine of their phases reveals an inverted phase for saturation artifacts. [Adapted from Ref.\ \citenum{Schroter2011}.]}
  \label{Fig8_FFT_Amp_and_Phase}
\end{figure}

Saturation artifacts occurred because the detector sensitivity was suppressed after the detection of large ion signals. Signal modulation frequencies from the most abundant isotopologue therefore appeared with an inverted sign in heavier isotopologue mass channels. The inverted sign of signal modulations corresponds to a 180$^\circ$ phase shift at corresponding signal frequencies, which are readily identified in FT spectra. Fig.\ \ref{Fig8_FFT_Amp_and_Phase} shows a spectrum for mass 79\,u, which contains the expected signals for the \ce{^{34}S^{12}C^{33}S} and \ce{^{34}S^{13}C^{32}S} isotopologues, but also saturation artifacts for the \ce{^{32}S^{12}C^{32}S} main isotopologue. The bottom inset shows a phase-scaled spectrum and reveals the opposite phase of the saturation artifact signals. Phase scaling was also used in the 2D data shown in Fig.\ \ref{Fig1_Correlated_Spectroscopy}\,\circledd{A} and \circledd{B} to distinguish saturation artifacts (red color) from ordinary signal (black color).

The facile assignment of individual isotopologue signals is illustrated in Fig.\ \ref{Fig1_Correlated_Spectroscopy}\,\circledd{B}. Note that some isotopologues are readily distinguished by their mass but appear at identical rotational transition frequencies (e.g., \ce{^{32}S^{12}C^{32}S} at mass 32\,u versus \ce{^{32}S^{13}C^{32}S} at mass 33\,u), while others are readily distinguished by their rotational transition frequencies but appear at identical mass (e.g., \ce{^{32}S^{13}C^{32}S} at 177\,GHz versus \ce{^{32}S^{12}C^{33}S} at 159\,GHz). I.e., neither a mass spectrum alone, nor a rotational spectrum alone could resolve all isotopologue-specific signals for the sample.

A single CRASY measurement resolved rotational Raman signals for 10 different isotopologues at natural abundances down to $10^{-6}$.\cite{Schroter2011} To appreciate the scope of this achievement, these results should be contrasted with the preceding literature for isotope-specific \ce{CS2} rotational constants: Decades of infrared spectroscopy for natural and isotopically enriched samples (see Refs.\ \citenum{Cheng1996,Ahonen1997,Winther1988,Horneman2005} and references therein) only resolved rotational constants for 7 distinct isotopologues. The extraordinary sensitivity of CRASY was achieved because ion counting is highly efficient and nearly background-free and because rare isotopologue signals appear in separate mass channels, largely unaffected by noise from the the abundant isotopologue signals.

Since the initial work on \ce{CS2}, CRASY resolved rotational Raman spectra for many molecular isotopologues containing \ce{^{13}C}, \ce{^{18}O}, and \ce{^{33}S}, \ce{^{34}S}, and \ce{^{36}S} isotopes at natural abundance, e.g., in \ce{CS2}\cite{Schroter2018}, trans-butadiene,\cite{Schroter2015,Ozer2020} benzene,\cite{Heo2022a,Heo2022b} furan,\cite{Ozer2024} acetylacetone, and thiophene. We should note that the high resolution and signal contrast in modern FTMW experiments also allows the assignment of rare isotopologue signals, even though mass-correlation is not possible. Examples for FTMW data resolving \ce{^{13}C} and \ce{^{15}N} isotopologue spectra at natural abundance can be found in Refs.\ \citenum{Shipman2011} and \citenum{Zinn2015,Gougoula2019,Perez2020}.


As discussed in the following sections, the ability to assign the position of rare isotopes in molecular isotopologues creates a new capability for the analysis of molecular structure and fragmentation pathways. When natural isotopes have insufficient abundance for mass-CRASY analysis, unspecific isotopic labeling should be sufficient to resolve and assign additional isotopologue spectra. We expect that this capability of CRASY will greatly reduce the difficulty and cost of isotopologue studies.

\subsection{De-Novo Structure Determination}
\label{sec:De-Novo Structure Determination}
The facile characterization of isotopologue rotational spectra with CRASY opened a new avenue for high-resolution gas-phase structure determination. Gas phase structures are unperturbed by environmental effects and provide excellent reference data for theory. A critical review of structure determination methods in gas and condensed phase, if a bit dated, can be found in Ref.\ \citenum{Lide1974}. Missing from the review are modern NMR\cite{Wuthrich1990,Wuthrich2003} and 3D electron diffraction methods,\cite{Gruene2021} but neither of these methods can deliver high resolution molecular structures with m\AA\ accuracy.

The Gordy\&Cook textbook\cite{Gordy1984} offers an exhaustive description of molecular structure analysis based on rotational spectroscopy. The analysis of rovibrational corrections was further improved based on the fitting of experimental data\cite{Watson1999} or ab initio calculations.\cite{Piccardo2015} The following paragraphs aim to give a concise summary of the relevant concepts and mathematical equations, sufficient to fit molecular structure parameters from experimentally determined rotational constants.

In first-order approximation, isotopic substitution does not affect molecular bond lengths. The effective distance $r\mathrm{_{0,i}^{a,b,c}}$ of an atom $\mathrm{i}$ to a rotational axis $\mathrm{a,b,c}$ is therefore directly related to the atomic mass $m_\mathrm{i}$ and the measured molecular inertial moment $I\mathrm{_0^{a,b,c}}$, or the corresponding rotational constant $A_\mathrm{0}, B_\mathrm{0}$, or $C_\mathrm{0}$:
\begin{equation} \label{eq:Inertial_Moments}
\begin{split}
I\mathrm{_0^{a,b,c}} &= \sum_{\mathrm{i=1}}^{N} m_\mathrm{i} \cdot <\!(r\mathrm{_{0,i}^{a,b,c}})^2\!>\mathrm{, with} \\
A_\mathrm{0},B_\mathrm{0},C_\mathrm{0} &= \frac{\hbar^2}{2 \cdot I\mathrm{_0^{a,b,c}}}
\end{split}
\end{equation}
Angle brackets are used to indicate that only the expectation values for squared distances $<\!r^2\!>$ are accessible. If rotational constants are known for multiple isotopologues, then the resulting set of eqs.\ (\ref{eq:Inertial_Moments}) is readily solved to obtain the effective positions $r\mathrm{_0}$ of each isotopically substituted atom with respect to the rotational axes $\mathrm{a,b,c}$. Bond lengths and angles are then calculated by a coordinate transformation into internal molecular coordinates.

Vibrational motion, including zero-point vibration, leads to a slight isotope-dependence of the bond lengths and the calculated effective bond lengths do not correspond to the exact bond length in any specific isotopologue. Additional terms $\epsilon^\mathrm{a,b,c}$ can be introduced to account for rovibrational corrections along the rotational axes: $I\mathrm{_0^{a,b,c}} =  I\mathrm{^{a,b,c}} + \epsilon^\mathrm{a,b,c}$. Kraitchmann proposed the analysis of inertial moment differences between pairs of isotopologues.\cite{Kraitchman1953} This removes the rovibrational correction terms $\epsilon^\mathrm{a,b,c}$ under the assumption that these corrections can be described by constant, isotope-independent factors. The resulting substitution geometry parameters, $r\mathrm{_s}$, are expected to fall between effective ($r_\mathrm{0}$) and equilibrium ($r_\mathrm{e}$) bond lengths.

Costain estimated the expected uncertainty bounds that should include $r_\mathrm{0}$, $r_\mathrm{s}$ and $r_\mathrm{e}$ bond lengths.\cite{Costain1958} Gordy\&Cook (chapter 13.8 in Ref.\ \citenum{Gordy1984}) give the Costain uncertainty as $\delta r = (0.0015\,\mathrm{A^2})/r_\mathrm{s}$, but a range of estimates for this uncertainty can be found in the literature.\cite{Eijck1982,Watson1999,Demaison2002} Larger inaccuracies are expected for $r_\mathrm{s}$ values calculated based on \ce{H},\ce{D} substitution or when atomic positions are close to a rotational axis.\cite{Demaison2002}

Laurie proposed an explicit mass-dependent correction term for the large \ce{H},\ce{D},\ce{T} isotope corrections.\cite{Laurie1962} The mass-corrected bond length $r\mathrm{_m}$ for each hydrogen atom is corrected based on the molecular mass $M$, the H-isotope mass $m_\mathrm{x}$ ($\mathrm{x}$ = \ce{H},\ce{D},\ce{T}) and the Laurie correction factor $\delta_\mathrm{H}$:
\begin{equation} \label{eq:Laurie_Corrections}
r\mathrm{_{x}^L} = r\mathrm{_{m}} + \delta_\mathrm{H}(M/[(m_\mathrm{x})(M-m_\mathrm{x})])^{1/2}
\end{equation}
Note that, unlike parameter $r_\mathrm{0}$ in eq.\ \ref{eq:Inertial_Moments}, $r\mathrm{_m}$ denotes an actual \ce{X-H} bond length. The determination of $\delta_\mathrm{H}$ parameters from experimental data often fails, but for molecules with sufficiently large molecular mass, the parameter remains close to $\delta_\mathrm{H} \approx$ 10\,u$^{1/2}$\AA\ and may be held at this value.\cite{Watson1999}

Watson derived generalized mass-dependent rovibrational correction terms $\epsilon^\mathrm{a,b,c}$ that approximate harmonic and anharmonic rovibrational corrections for all atoms:\footnote{A harmonic correction is required because the width of the vibrational wave function is $\propto m^{-1/2}$. This width leads to a mass-dependent divergence between $<\!r\!>^2$ and $<\!r^2\!>$.}
\begin{equation} \label{eq:Watson}
I_\mathrm{0,\alpha} = I_\mathrm{m,\alpha} \;+\;
\mathrm{c_\alpha} \cdot \sqrt{I_\mathrm{m,\alpha}} \;+\;
\mathrm{d_{\alpha}} \cdot \left(\frac{\displaystyle \prod_{\mathrm{i}=1}^\mathrm{N}{m_\mathrm{i}}} {\displaystyle\sum_{\mathrm{i}=1}^\mathrm{N}{m_\mathrm{i}}}\right)^{1/(2\mathrm{N}-2)}
\end{equation}
The $\mathrm{c_\alpha}$ and $\mathrm{d_{\alpha}}$ terms represent the approximate harmonic and anharmonic corrections with respect to the $\mathrm{\alpha = a,b,c}$ rotational axes. Resulting geometry parameters are labeled $r_\mathrm{m}^{I}$ (only $\mathrm{c}$-terms) and $r_\mathrm{m}^{II}$ ($\mathrm{c}$ and $\mathrm{d}$-terms) and provide a purely experimental estimate for the equilibrium geometry with typical bond length accuracies $\ll$\,1\,m\AA\ for heavy atoms and close to 1\,m\AA\ for \ce{H} and \ce{D}. Parameters $\mathrm{c_\alpha}$ and $\mathrm{d_{\alpha}}$ must be fitted from experimental data and solving eqs.\ \ref{eq:Watson} is only feasible if rotational constants are available for a large number of isotopologues.

The fitting of $r\mathrm{_0}$, $r\mathrm{_s}$, and $r_\mathrm{m}$ geometry parameters to experimental constants is straightforward and was implemented by Kisiel in the STRFIT program.\cite{STRFIT,Kisiel2003} We created a Python script with a graphical user interface for this purpose\cite{Figshare_FitMOI}, including tools for the facile exploitation of molecular symmetry and for the Monte-Carlo propagation of experimental errors. It should be noted that semi-experimental methods can give even better estimates for equilibrium bond lengths by combining experimental data with ab initio estimates for rovibrational corrections.\cite{Demaison2011,Piccardo2015} The calculation of anharmonic force fields is required to obtain best estimates for the rovibrational correction terms.

As described in Section \ref{sec:Rotational Spectra of Molecular Isotopologues}, CRASY data can resolve rotational spectra and rotational constants for multiple isotopologue species in a single measurement. This greatly reduces the experimental effort for obtaining isotopologue rotational constants as compared to the traditional approach of synthesizing and characterizing individual isotopologues.

\emph{De novo} structure analysis from a single CRASY data set was first presented for the carbon backbone of butadiene, based on the observation of \ce{^{13}C} isotopologues at natural abundance.\cite{Ozer2020} Rotational spectra for the main isotopologue and both \ce{^{13}C} isotopologues were obtained with a full-width at half-maximum (FWHM) resolution <\,10\,MHz and fitted rotational constants showed relative uncertainties $\Delta \nu / \nu$ in the $10^{-5}$ to $10^{-7}$ regime. Rotational constants agreed well with literature values obtained by high-resolution FTIR spectroscopy of synthesized isotopologues.\cite{Craig2004,Craig2004a,Craig2006,Craig2006a} A butadiene substitution structure for the carbon frame was calculated based on Kraitchman's formalism and is given in Table \ref{tab:ButadieneGeometry}. The resulting bond lengths and angles for the carbon frame of the molecule agreed within the Costain errors with $r_\mathrm{0}$ literature values obtained from high-resolution IR, FTMW, and electron diffraction data.\cite{Craig2006,Kveseth1980}

\begin{table}[htb]
\small
   \caption{ Bond lengths $r$ (in \AA) and bond angle $\alpha$ (in degrees) for the carbon backbone of butadiene. Numbers in brackets give the 1-$\sigma$ standard deviation in the corresponding last digits.}
   \begin{tabular}{llcc}
    \toprule
    & \multicolumn{1}{c}{CRASY\cite{Ozer2020}} & \multicolumn{2}{c}{Craig\cite{Craig2006}}  \\
      \cmidrule(lr){2-2}             \cmidrule(lr){3-4}
    & \multicolumn{1}{c}{$r_s$}    & $r_s$     & $r_0$    \\
    \midrule
    r(C--C)           & 1.466(7)   & 1.451(4)  & 1.458(3) \\
    r(C=C)            & 1.342(14)  & 1.344(6)  & 1.346(3) \\
    $\alpha$(C=C--C)  & 123.1(1.4) & 123.6(6)  & 123.4(1) \\
    \bottomrule
   \end{tabular}
  \label{tab:ButadieneGeometry}%
\end{table}

The actual power of a measurement technique comes into focus when measurement results do not confirm our expectations, but give unexpected results that force us to reevaluate our understanding. This case occurred when CRASY data was used for the structure analysis of benzene. CRASY experiments resolved the rotational Raman spectra of 5 benzene isotopologues, \ce{C6H6}, \ce{^{13}C-C5H6}, \ce{C6D6}, \ce{^{13}C-C5D6}, and \ce{^{13}C6H6}, sufficient for a full $r_\mathrm{m}^{II}$ structure analysis.\cite{Heo2022a,Heo2022b} Table \ref{tab:BenzeneGeometryParameters} compares effective and equilibrium geometries obtained from CRASY data with literature values.\cite{Pliva1990,Kunishige2015,CCCBDB,Ceselin2021} Effective bond lengths agreed near-perfectly with literature values, but estimates for equilibrium bond lengths deviated significantly.

\begin{table}[htb]
  \centering
  \caption{ Effective ($r\mathrm{_0}$) and equilibrium ($r\mathrm{_e}$) benzene bond lengths (all values in \AA). Colors highlight significant deviations between literature values and values derived from CRASY data.}
\small
    \begin{tabular}{lll}
    \toprule
                   & {$r\mathrm{(CC)}$} & {$r\mathrm{(CH)}$}\\
    \midrule
    $r_\mathrm{0}$, Ref.~\citenum{Pliva1990}        & 1.3969    & 1.0815     \\
    $r_\mathrm{0}$, Ref.~\citenum{Kunishige2015}    & 1.3971    & 1.0805     \\
    $r_\mathrm{0}$, CRASY                           & 1.3971(0) & 1.0803(2)  \\
    \midrule
    $r_\mathrm{e}$, Ref.~\citenum{Pliva1990}        & \cellcolor{rose}1.3893    & \cellcolor{rose}1.0857     \\
    $r_\mathrm{e}$, Ref.~\citenum{Kunishige2015}    & \cellcolor{rose}1.3892    & \cellcolor{rose}1.0864     \\
    $r_\mathrm{e}$, CRASY                           & \cellcolor{mint}1.3914(2) & \cellcolor{mint}1.0814(3)  \\
    \midrule
    $r_\mathrm{e}$, calc.\A                         & 1.3920    & 1.0802     \\
    $r_\mathrm{e}$, s.e.\B                          & 1.3916    & 1.0799     \\
    \bottomrule
    \end{tabular}
    \footnotesize{ \\
    \A Reference values from the CCCBDB database \cite{CCCBDB} for a coupled-cluster (full) calculation with aug-cc-pVTZ basis. \B Semi-experimental value.\cite{Ceselin2021}.}
  \label{tab:BenzeneGeometryParameters}%
\end{table}

Preceding literature values for the benzene geometry were calculated based on FTIR and FTMW data for numerous benzene isotopologues.\cite{Pliva1982JMol,Pliva1987,Pliva1989a,Pliva1990,Pliva1991,Baba2011,Kunishige2015,Hirano2021} These geometries predicted negligible or inverted \ce{H},\ce{D} rovibrational corrections and isotope effects, i.e., $r\mathrm{_e(CH)} > r\mathrm{_0(CH)}$ and $r\mathrm{_0(CD)} \approx r\mathrm{_0(CH)}$. This contradicts the general expectation of positive rovibrational correction terms and the more specific expectation of a positive Laurie correction factor $\delta_\mathrm{H} \approx$ 10\,u$^{1/2}$\AA. The experimental geometries were later supported by theoretical calculations, which predicted a fortuitous canceling of rovibrational effects from \ce{C-H} stretching and bending modes.\cite{Hirano2021,Udagawa2023} Based on these results, unexpected and unusual isotope effects must be expected for benzene and possibly other aromatic molecules. 

An integrated analysis of all available spectroscopic data revealed why the CRASY results diverged from those in the literature.\cite{Heo2022b} Literature analyses were based almost exclusively on data for deuterated benzene isotopologues, which are readily synthesized and in some cases have sufficient dipoles for FTMW analysis. But the scarcity of carbon substituted isotopologue data made it difficult to disentangle rovibrational corrections for the \ce{C-H} and \ce{C-C} bonds: eqs.\ \ref{eq:Inertial_Moments} for deuterated isotopologues are linearly dependent to a very high degree. Data for additional deuterated isotopologues therefore contributed little information for the structure analysis and introduced a bias in the $r_\mathrm{e}$ estimates. The CRASY data contributed rotational constants for multiple \ce{^{13}C} isotopologues, which removed the bias in the $r_\mathrm{e}$ structure determination and revealed an ordinary H,D isotope effect.\cite{Heo2022b}

Structure parameters from the combined analysis and those calculated from CRASY data alone were in excellent agreement with semi-experimental\cite{Ceselin2021} and theoretical\cite{CCCBDB} estimates for equilibrium bond lengths (see Table \ref{tab:BenzeneGeometryParameters}). A subsequent comparison of semiempirical and theoretical structures, based on highest levels of theory, showed excellent agreement with the CRASY results.\cite{Esselman2023}

CRASY data for furan resolved rotational constants for the main isotopologue and \ce{^{13}C} and \ce{^{18}O} isotopologues at natural abundance.\cite{Ozer2024} Furan is a dipolar molecule and highly accurate rotational spectra were available from FTMW measurements.\cite{Bak1962,Barnum2021} The agreement between CRASY data and published rotational constants for \ce{^{18}O-furan} were unsatisfactory, but a reanalysis of literature data, including quartic distortion constants, improved the agreement.\cite{Ozer2024} Effective and equilibrium geometries were calculated based on rotational constants from CRASY data and literature values and were in good agreement with published semi-experimental geometry parameters.\cite{Demaison2011}

\subsection{Analysis of Cationic Fragmentation}
\label{sec:Analysis of Cationic Fragmentation}
Mass spectra readily resolve molecular fragmentation and thereby offer rich information about chemical bonding and thermodynamic stability, providing the basis for analytical mass spectrometry.\cite{Pretsch2000} But based on mass information alone, it can be difficult to determine the origin of fragment signals: in a heterogeneous sample, fragments may originate from different precursor molecules or from sample impurities.

Mass-CRASY can resolve this issue by correlating each cation mass signal with the rotational spectrum of the neutral precursor molecule. The rotational spectrum can be considered as a molecular fingerprint, which is shared by the parent and each of its cationic fragments. A comparison of correlated rotational spectra therefore allows to assign fragments to their parent molecules. This creates a fundamentally new capability to assign cationic fragments in mass spectra of complex, heterogeneous samples.

As outlined in Section \ref{sec:Rotational Spectra of Molecular Isotopologues}, mass-CRASY data routinely resolves isotopologue rotational Raman spectra for rare isotopologues at natural abundance. The ability to assign parent-fragment relations can be combined with the observation of isotopologues to track atomic positions from a parent molecule to respective fragments: The position of a rare isotope in the parent is readily determined by an analysis of the correlated rotational Raman spectrum and its presence in a fragment is revealed by the fragment mass. The tracking of atoms throughout a chemical reaction represents the gold standard for the characterization of chemical reaction pathways. In the past, isotopic tracking required the synthesis of isotopologues with targeted isotopic substitution, which is expensive and time consuming. Mass-CRASY can deliver comparable information based on the observation of naturally occurring isotopologues or heterogeneous isotopologue mixtures. 

\begin{figure}[htb]
\centering
  \includegraphics[width=8.3cm]{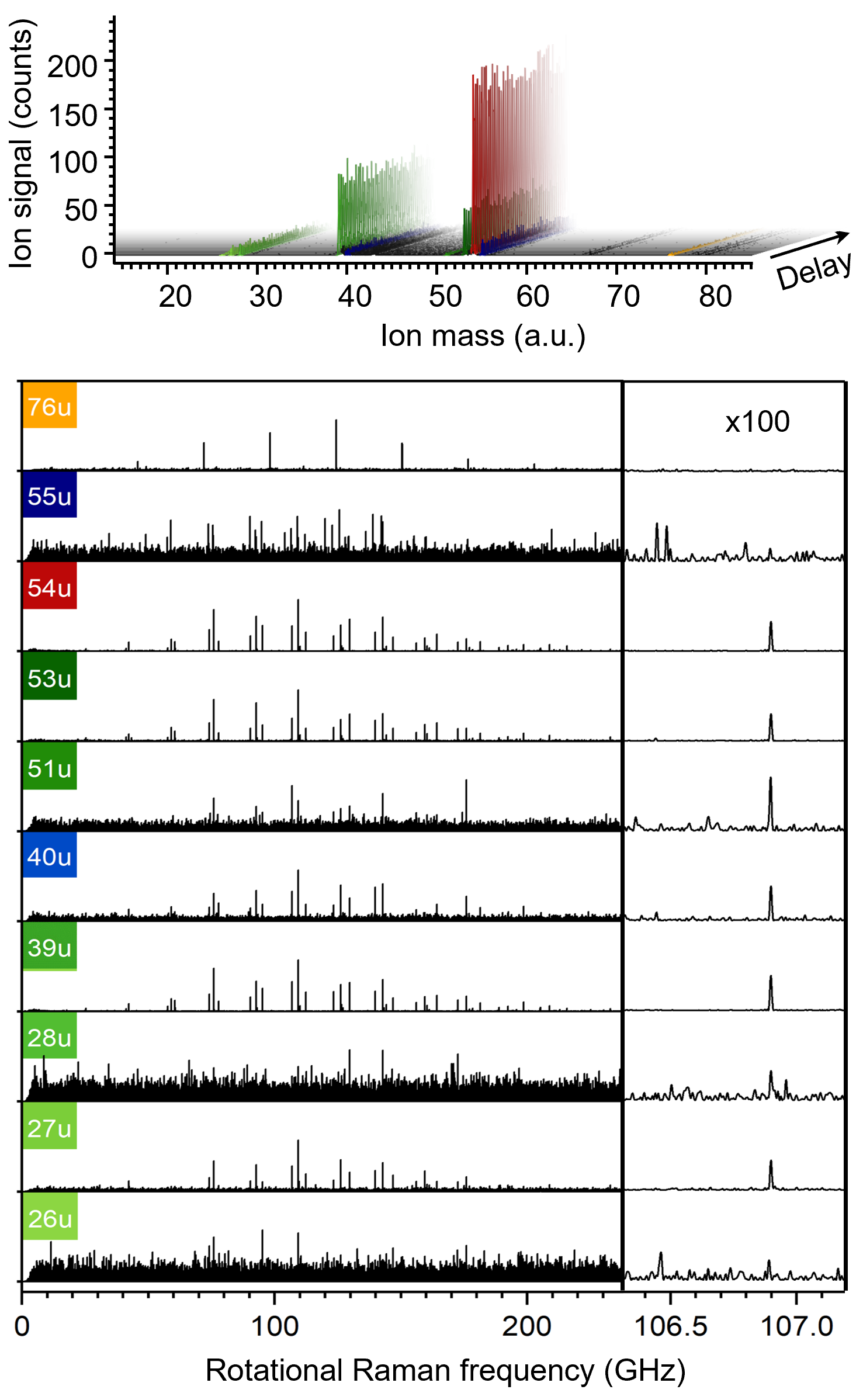}
  \caption{ (Top) Delay-dependent mass spectra show ion signals for butadiene (54\,u, red), its cationic fragments (green), and naturally occurring \ce{^{13}C-isotopologues} (blue). Signal at mass 76\,u originated from residual \ce{CS2} (orange) in the sample container and signal at 80\,u may be due to fragmentation of 4-vinylcyclohexene, a common impurity of butadiene.\cite{ButadieneProduct} (Bottom) Rotational Raman spectra correlated to the 10 largest ion signals in the mass spectrum. [Adapted from Ref.\ \citenum{Ozer2020}.]}
  \label{Fig9_Butadiene_CRASY}
\end{figure}

CRASY data for 1,3-butadiene is shown in Fig.\ \ref{Fig9_Butadiene_CRASY} and illustrates the assignment of cationic fragments to a parent molecule.\cite{Lee2019} Fig.\ \ref{Fig9_Butadiene_CRASY}, top, shows the rich fragment spectrum generated by photoionization of butadiene with a 200\,nm laser pulse. Ion signals were observed for butadiene (mass 54\,u), its naturally occurring \ce{^{13}C-isotopologues} (55\,u) and fragments at lower mass. The sample also contained impurities of 4-vinylcyclohexene (108\,u) and possibly larger condensates\cite{ButadieneProduct}, as well as traces of \ce{CS2} (76\,u). Observed fragments may therefore originate from butadiene or sample impurities.

Mass-correlated rotational Raman spectra are shown in Fig.\ \ref{Fig9_Butadiene_CRASY}, bottom, and revealed that at least six fragments were formed from the butadiene precursor: Spectra correlated with butadiene showed identical line positions as those for fragment masses 53\,u (C$_4$H$_5^{+}$), 51\,u (C$_4$H$_3^{+}$), 39\,u (C$_3$H$_3^{+}$), 28\,u (C$_2$H$_4^{\boldsymbol{\cdot}+}$), 27\,u (C$_2$H$_3^{+}$), and 26\,u (C$_2$H$_2^{\boldsymbol{\cdot}+}$). Spectra correlated with mass 55\,u (\ce{^{13}C-butadiene}) and 76\,u (\ce{CS2}) were markedly different. A heavy isotopologue fragment (\ce{^{13}C-C2H3^{+}}) was expected at mass 40\,u, originating from \ce{^{13}C-butadiene}, but the spectrum in this mass channel also showed lines correlated to the 54\,u main butadiene isotopologue. A closer inspection of this mass signal revealed that the 40\,u signal was contaminated by the much stronger signal at 39\,u: delayed fragmentation on the time-scale of ion acceleration created a broad, asymmetric shoulder in the 39\,u mass signal that overlapped with the 40\,u mass channel.

\begin{figure}[htb]
\centering
  \includegraphics[width=8.3cm]{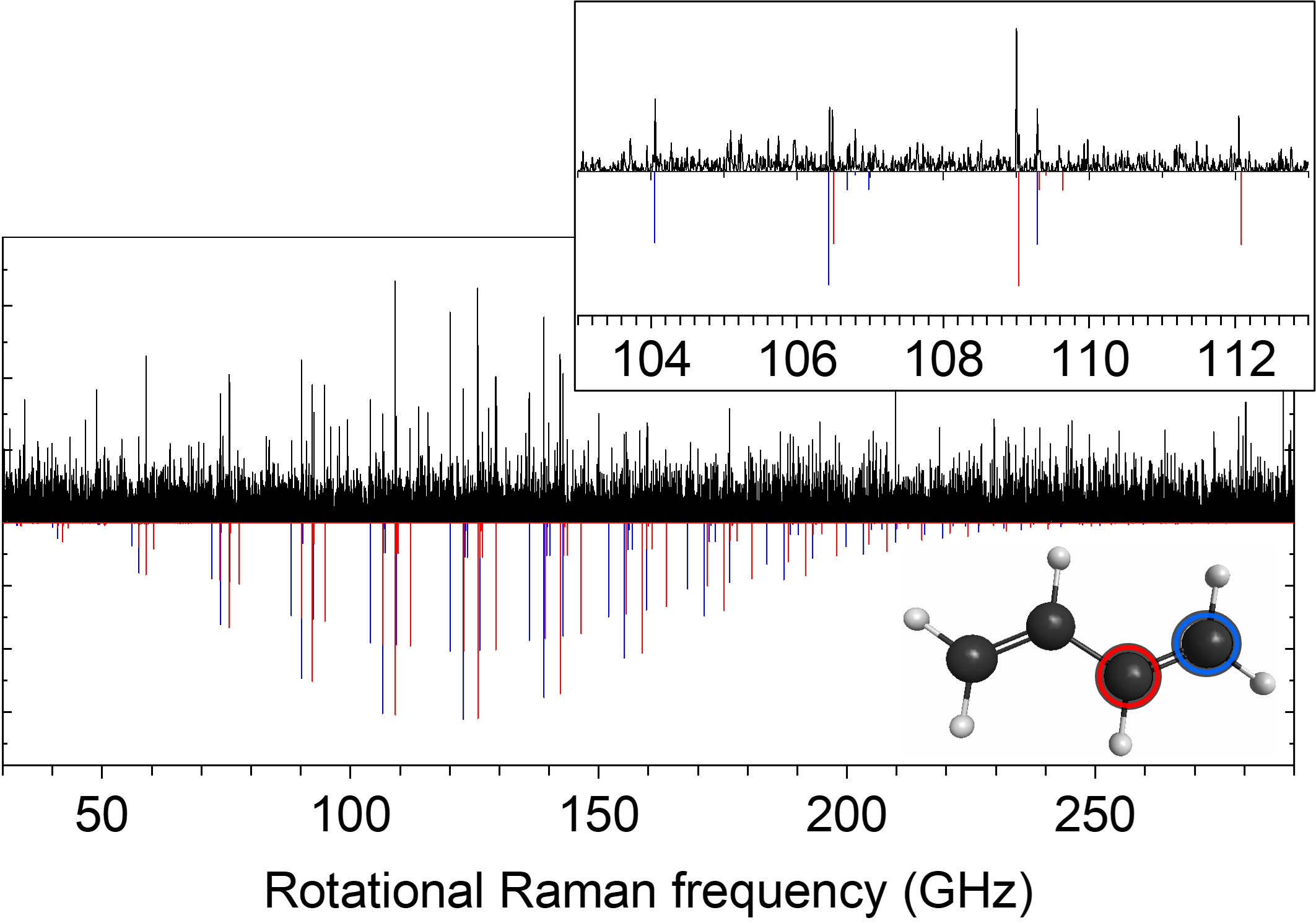}
  \caption{ (Positive trace) CRASY rotational Raman spectrum observed for a butadiene sample in mass channel 55\,u. (Negative traces) Pgopher simulation for spectra of 1-\ce{^{13}C-butadiene} (blue) and 2-\ce{^{13}C-butadiene} (red). Insets show an enlarged section of the spectrum and the molecular structure of 1,3 butadiene, marking the relevant 1-\ce{^{13}C} and 2-\ce{^{13}C} substitution positions. [Adapted from Ref.\ \citenum{Lee2019}.]}
  \label{Fig10_Butadiene_isotope_assignment}
\end{figure}

\ce{^{13}C} isotopologues occur with a natural abundance of $\approx$ \SI{1.1}{\percent}.\cite{Michael2011} The corresponding \ce{^{13}C-butadiene} isotopologues were observed at mass 55\,u. Fig.\ \ref{Fig10_Butadiene_isotope_assignment} illustrates the assignment of rotational spectra for both isotopologues, with the \ce{^{13}C-isotope} in the outer (\ce{1-^{13}C}) or inner (\ce{2-^{13}C}) carbon position. Both isotopologues appeared with identical abundance and integrated signal amplitudes were of comparable size.

\begin{figure}[htb]
\centering
  \includegraphics[width=8.3cm]{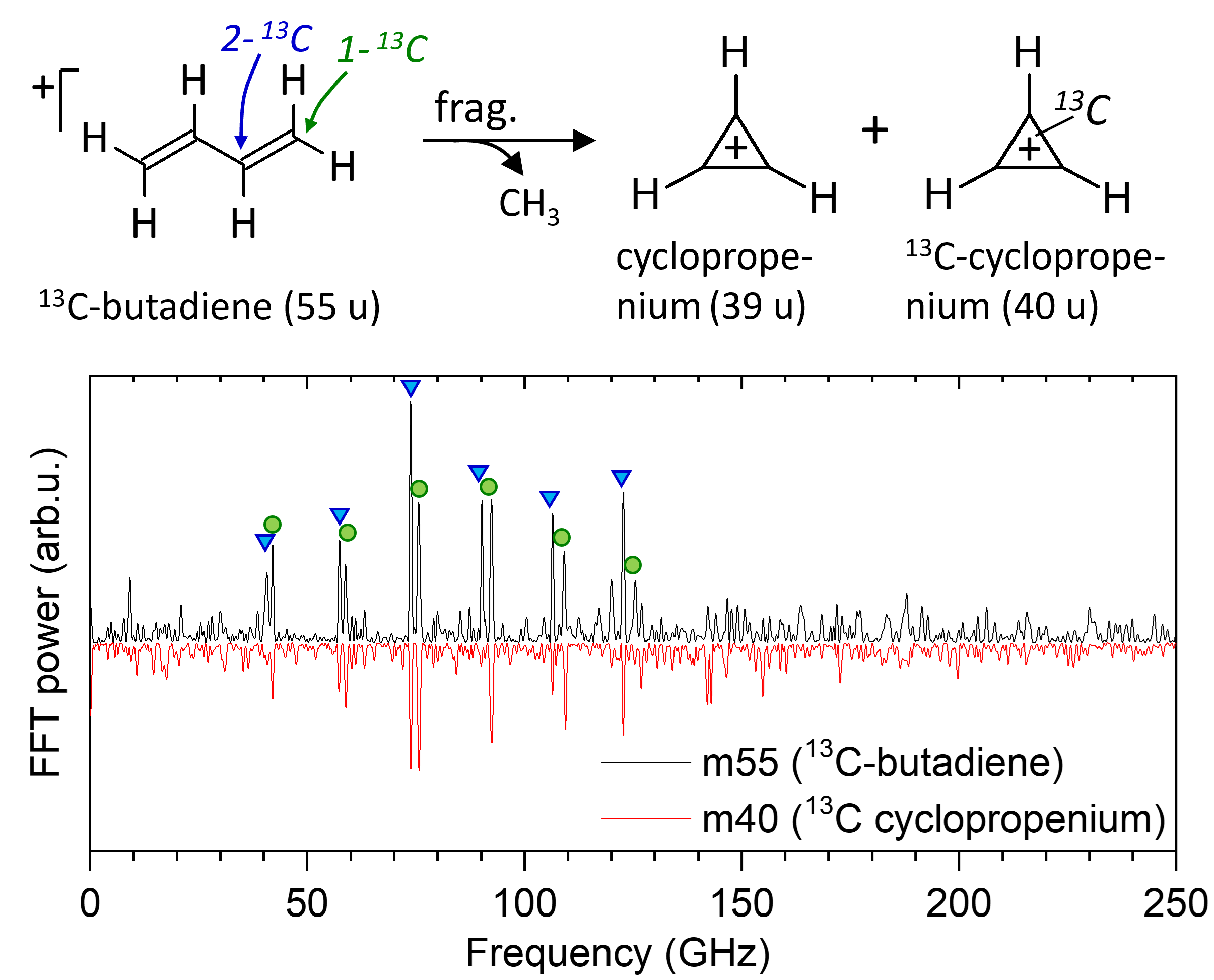}
  \caption{ (Top) The dominant fragmentation channel  of \ce{^{13}C-butadiene} is the loss of \ce{^{13}CH3} or \ce{^{12}CH3}, forming cyclopropenium and \ce{^{13}C-cyclopropenium} fragments. (Bottom, positive signal) The spectrum correlated to mass 55\,u shows transitions for 1-\ce{^{13}C-butadiene} (green bullets) and 2-\ce{^{13}C-butadiene} (blue triangles). (Bottom, negative signal) Corresponding transitions are observed correlated to mass 40\,u (\ce{^{13}C-cyclopropenium}) and reveal the fragmentation propensity for the two \ce{^{13}C} isotopologues. [Adapted from Ref.\ \citenum{Schroter2015}.]}
  \label{Fig11_13CButadiene_Fragments}
\end{figure}

The tracking of \ce{^{13}C} isotopes from butadiene to its fragments can reveal the propensity for the loss of inner versus outer carbon atoms and thereby reveal mechanistic aspects of the fragmentation process. The dominant fragmentation pathway for butadiene cation is the loss of methyl, forming a cyclopropenium cation. As illustrated in Fig.\ \ref{Fig11_13CButadiene_Fragments}, top, fragmentation of heavy \ce{^{13}C-butadiene} may form ordinary cyclopropenium with mass 39\,u (loss of \ce{^{13}CH3}) or \ce{^{13}C-cyclopropenium} with mass 40\,u (loss of \ce{^{12}CH3}). It is natural to assume that fragmentation is due to the loss of an outer carbon atom. In that case, half of 1-\ce{^{13}C-butadiene} should fragment into \ce{^{13}C-cyclopropenium} and the other half should fragment into \ce{^{12}C-cyclopropenium}, whereas 2-\ce{^{13}C-butadiene} should exclusively fragment into \ce{^{13}C-cyclopropenium}. Rotational spectra correlated to the \ce{^{13}C-cyclopropenium} fragment mass channel should then show an amplitude ratio of 1:2 for the transition lines of the two \ce{^{13}C-butadiene} precursors.

\begin{figure}[htb]
\centering
  \includegraphics[width=8.3cm]{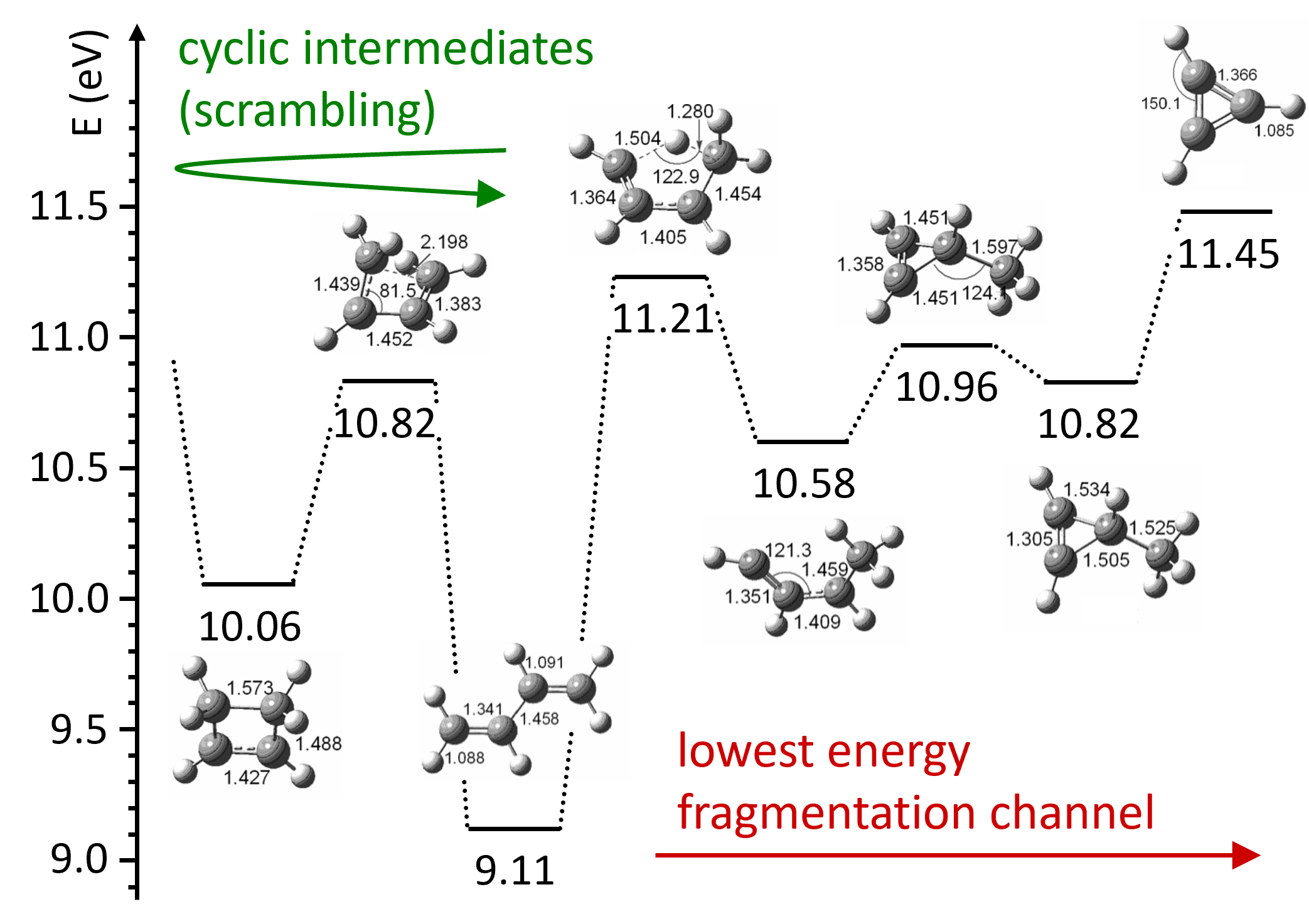}
  \caption{ Pathway for atomic scrambling in butadiene cation via theoretically predicted cyclic intermediates. [Calculated structures and energies from Ref.\ \citenum{Fang2011}.]}
  \label{Fig12_13CButadiene_scrambling}
\end{figure}

Fig.\ \ref{Fig11_13CButadiene_Fragments} (bottom) compares rotational spectra in \ce{^{13}C-butadiene} and \ce{^{13}C-cyclopropenium} mass channels in a low-resolution data set. The expected 2:1 amplitude ratio of isotopologue transition lines was not observed and a quantitative analysis revealed near-equal signal contributions from both isotopologues. Atomic scrambling of the carbon atom positions must therefore exchange inner and outer carbon atoms in butadiene cation before fragmentation occurs.\cite{Schroter2015} DFT calculations in the literature predicted low-energy cyclic intermediates on the cationic potential energy surface.\cite{Fang2011} As illustrated in Fig.\ \ref{Fig12_13CButadiene_scrambling}, transient formation of cyclic intermediates, combined with hydrogen migration, may explain the observed atomic scrambling.



\begin{figure}[htb]
\centering
  \includegraphics[width=8.3cm]{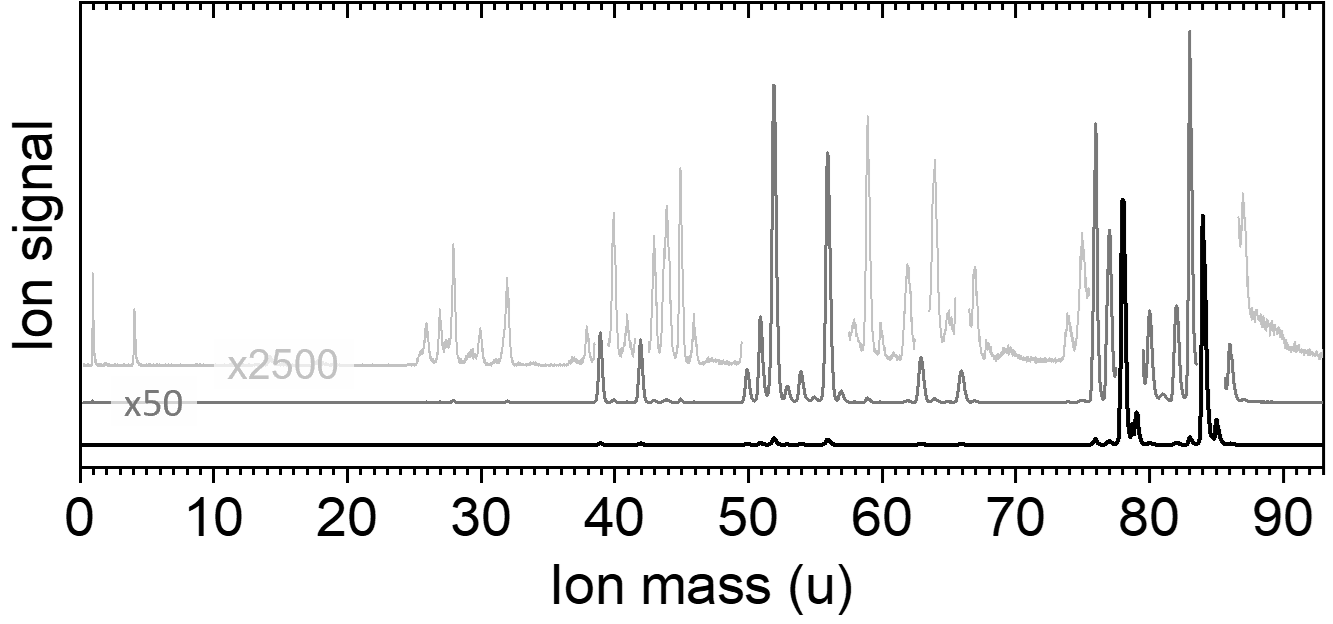}
  \caption{ Mass spectrum from a CRASY data set for benzene (78\,u), perdeuterated benzene (84\,u), \ce{CS2} (76\,u), and their natural isotopologues. Grey lines show the spectrum with vertical offset and 50-fold or 2500-fold enlarged ordinate.}
  \label{Fig13_Benzene_isotopologue_MS}
\end{figure}

The number and variety of fragments observed in UV multiphoton ionization mass spectra can be rather large. This is particularly true for photoionization with intense femtosecond laser pulses, where high pulse intensities can lead to significant contributions from above-threshold photoionization and the formation of energetic fragments. Fig.\ \ref{Fig13_Benzene_isotopologue_MS} shows the mass spectrum from a CRASY measurement of benzene and perdeuterated benzene. Rotational Raman spectra were resolved for 28 ion signals with an amplitude down to \SI{2.5}{\permille} of the largest signal. A systematic comparison should then be performed between 378 pairs of mass-correlated spectra, while accounting for the full spectral range (500\,GHz) and effective resolution (8\,MHz). A manual comparison of such data quantities is quite challenging.

\begin{figure*}[htb]
\centering
  \includegraphics[width=14.8cm]{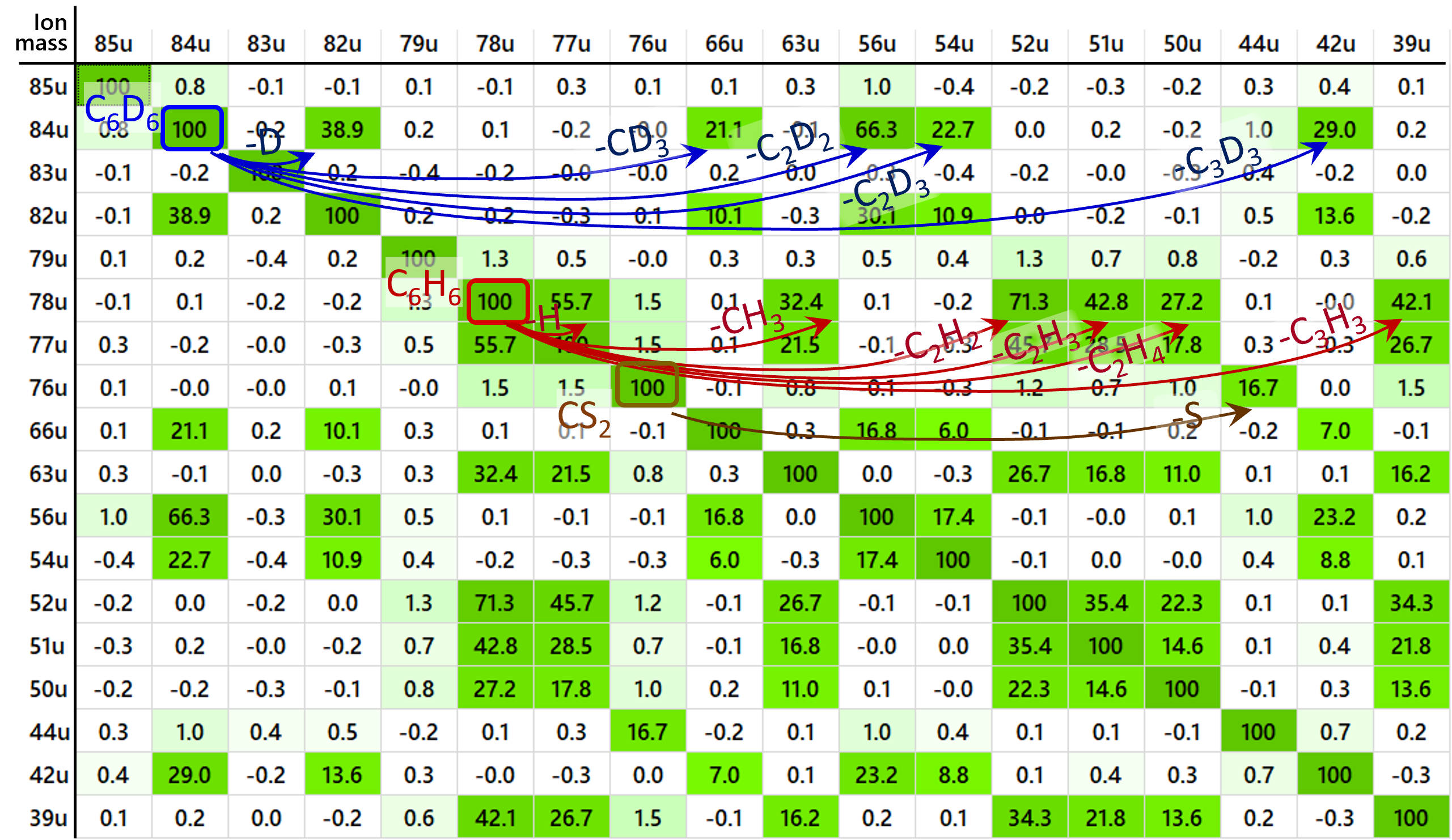}
  \caption[Pearson_Correlation_Table]{\small Pearson correlations between rotational spectra correlated with ion signals in Fig.\ \ref{Fig13_Benzene_isotopologue_MS}. Column and row headers denote the ion mass. Correlation values of 100 indicate identical spectra, 0 indicates no correlation, and negative values indicate anti-correlation. Annotations in the top half of the table denote fragmentation pathways of perdeuterated benzene (parent mass 84\,u), benzene (parent mass 78\,u) and \ce{CS2} (parent mass 76\,u).}
  \label{Fig14_Pearson_correlation_table_benzene}
\end{figure*}

A computer-generated correlation map can replace the manual comparison of spectra and helps to rapidly diagnose parent-fragment relationships in CRASY data. Fig.\ \ref{Fig14_Pearson_correlation_table_benzene} shows an annotated Pearson correlation map for rotational spectra correlated with major mass signals in Fig.\ \ref{Fig13_Benzene_isotopologue_MS}. High correlation values in off-diagonal table cells denote a strong similarity between two spectra. Negative correlation values are due to noise and carry no meaning.

Significant correlations were observed between benzene (78\,u) and six fragment channels (77\,u, 63\,u, 52\,u, 51\,u, 50\,u, and 39\,u), between perdeuterated benzene (84\,u) and five fragment channels (82\,u, 66\,u, 56\,u, 54\,u, and 42\,u), and between \ce{CS2} (76\,u) and one fragment channel (44\,u). A weak correlation between signals at mass 78\,u and 79\,u, as well as between 84\,u, 85\,u was found to be an artifact due to the limited resolution of the mass-spectrometer: Imperfect signal separation between signals for abundant \ce{^{12}C} and less abundant \ce{^{13}C-isotopologues} contaminates the latter. These artifactual correlations can be partially suppressed by shifting the integration boundaries for the respective ion channels.

\begin{figure}[htb]
\centering
  \includegraphics[width=8.3cm]{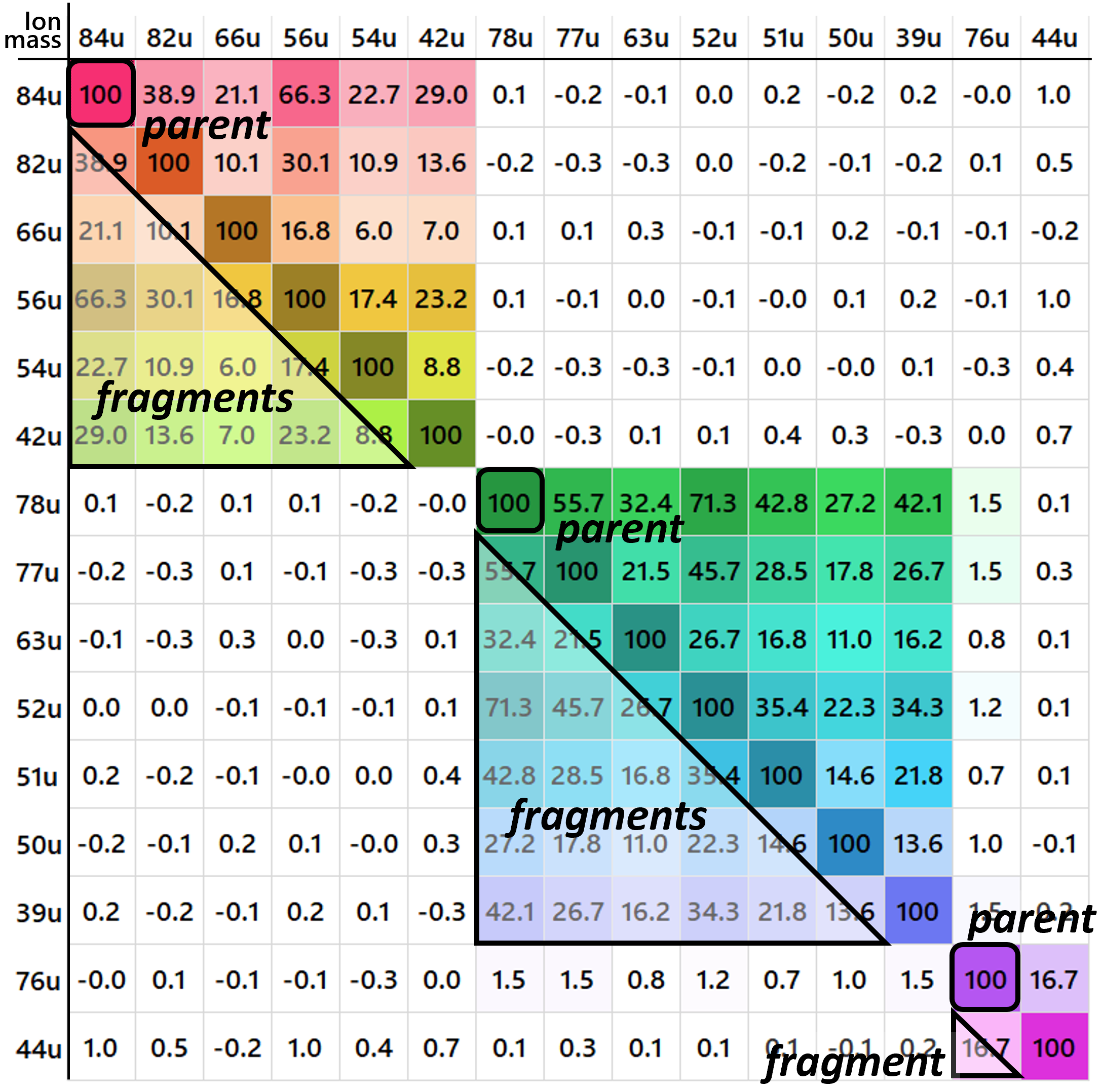}
  \caption[Pearson_Correlation_Table]{\small Pearson correlation map for ordinary and perdeuterated benzene ion signals (see Fig.\ \ref{Fig14_Pearson_correlation_table_benzene}) sorted into correlated blocks to facilitate the recognition parent-fragment correlations. Correlation values may deviate from those shown in Fig.\ \ref{Fig14_Pearson_correlation_table_benzene} due to a different choice of ion signal integration boundaries.}
  \label{Fig15_Pearson_correlation_table_Benzene_simplified}
\end{figure}

The analysis can be further simplified by sorting columns and rows in the correlation map, as shown in Fig.\ \ref{Fig15_Pearson_correlation_table_Benzene_simplified}. This representation facilitates the recognition of fragments that originate from the same parent. Columns and rows for mass channels that showed no meaningful correlations were omitted to reduce the table size. On first glance, it is curious that 6 fragmentation channels are observed for benzene, but only 5 for perdeuterated benzene: \ce{C2H4} loss is observed for benzene but the corresponding \ce{C2D4} loss channel is missing. The resulting fragment (\ce{C4D2}) should appear at mass 52\,u, and is an isobar of \ce{C4H4}. The latter is a much more abundant fragment and elevates the noise in this mass channel, thereby obscuring the presence of the former. The fact that larger signals lead to significantly elevated noise levels in mass-correlated rotational spectra is owed to the fact that we used sparse sampling to acquire this data set (see Section \ref{sec:High-Resolution Rotational Raman Spectra}).

Identified fragmentation channels are summarized in Table \ref{tab:FragmentationTable}. Col.\ 1 lists the neutral precursors, which can be identified via their mass-correlated rotational spectra. Col.\ 2 lists the observed ion masses that are correlated to a precursor species (see correlated parent-fragment blocks in Fig.\ \ref{Fig15_Pearson_correlation_table_Benzene_simplified}). Col.\ 3 suggests plausible fragmentation channels for the formation of the observed ions.

\newcommand{\cb}[2]{\color{white}\colorbox{#1}{\textbf{\textsf{#2}}}\color{black}}
\definecolor{c84}{HTML}{f62f72}
\definecolor{c82}{HTML}{db5b27}
\definecolor{c66}{HTML}{b47626}
\definecolor{c56}{HTML}{9c8026}
\definecolor{c54}{HTML}{858826}
\definecolor{c42}{HTML}{668f26}
\definecolor{c78}{HTML}{269640}
\definecolor{c77}{HTML}{28946f}
\definecolor{c63}{HTML}{299285}
\definecolor{c52}{HTML}{2a9095}
\definecolor{c51}{HTML}{2b8ea8}
\definecolor{c50}{HTML}{2e8ac6}
\definecolor{c39}{HTML}{6c77f2}
\definecolor{c76}{HTML}{b556f1}
\definecolor{c44}{HTML}{dd31dc}
\definecolor{grey}{HTML}{b1b1b1}
\begin{table}[htb]
   \small
   \caption{Assigned photoionization and fragmentation channels in a benzene data set.}
    \renewcommand{\arraystretch}{0.7}   
   \begin{tabular}{lll}
    Neutral & Ion\A        &   Fragmentation channel                  \\
   \toprule {\ce{^{13}C-C5D6}}
            & \cb{grey}{85\,u}  &\ce{^{13}CC5D6 ->[h\nu]   ^{13}CC5D6^{+}} \\
   \midrule {\ce{C6D6}}
            & \cb{c84}{84\,u}   &     \ce{C6D6  ->[h\nu]         C6D6^{+}} \\
            & \cb{c82}{82\,u}   &     \ce{C6D6  ->[h\nu] D     + C6D5^{+}} \\
            & \cb{c66}{66\,u}   &     \ce{C6D6  ->[h\nu] CD3   + C5D3^{+}} \\
            & \cb{c56}{56\,u}   &     \ce{C6D6  ->[h\nu] C2D2  + C4D4^{+}} \\
            & \cb{c54}{54\,u}   &     \ce{C6D6  ->[h\nu] C2D3  + C4D3^{+}} \\
            & \cb{c42}{42\,u}   &     \ce{C6D6  ->[h\nu] C3D3  + C3D3^{+}} \\
   \midrule {\ce{C6D5H}}
            & \cb{grey}{83\,u}  &     \ce{C6D5H ->[h\nu]        C6D5H^{+}} \\
   \midrule {\ce{^{13}C-C5H6}}
            & \cb{grey}{79\,u}  &\ce{^{13}CC5H6 ->[h\nu]   ^{13}CC5H6^{+}} \\
   \midrule {\ce{C6H6}}
            & \cb{c78}{78\,u}   &     \ce{C6H6  ->[h\nu]         C6H6^{+}} \\
            & \cb{c77}{77\,u}   &     \ce{C6H6  ->[h\nu] H     + C6H5^{+}} \\
            & \cb{c63}{63\,u}   &     \ce{C6H6  ->[h\nu] CH3   + C5H3^{+}} \\
            & \cb{c52}{52\,u}   &     \ce{C6H6  ->[h\nu] C2H2  + C4H4^{+}} \\
            & \cb{c51}{51\,u}   &     \ce{C6H6  ->[h\nu] C2H3  + C4H3^{+}} \\
            & \cb{c50}{50\,u}   &     \ce{C6H6  ->[h\nu] C2H4  + C4H2^{+}} \\
            & \cb{c39}{39\,u}   &     \ce{C6H6  ->[h\nu] C3H3  + C3H3^{+}} \\
   \midrule {\ce{(CS2)}}
            & \cb{c76} {76\,u}  & \ce{(CS2)_2   ->[h\nu] CS2   +  CS2^{+}} \\
            & \cb{c44} {44\,u}  & \ce{(CS2)_2   ->[h\nu] CS2   +   CS^{+}} \\
    \bottomrule
   \end{tabular}
  \footnotesize{ \\
    \A Ion mass; colors correspond to those in Fig.\ \ref{Fig15_Pearson_correlation_table_Benzene_simplified} or are grey for uncorrelated ion signals.}
  \label{tab:FragmentationTable}%
\end{table}



The ability to identify parent-fragment relationships in complex photoionization mass spectra is of particular interest for the study of molecular clusters and molecular radicals. Cluster sources and radical sources generally produce a heterogeneous mixture of molecular species and traditional spectroscopic methods cannot routinely assign fragments to particular parent species. First examples for the analysis of cluster fragmentation channels are presented in the following section. 


\subsection{Molecular Cluster Spectroscopy}
\label{sec:Molecular Cluster Spectroscopy}
The spectroscopic analysis of molecular clusters in the gas phase allows to study non-covalent interactions in well-defined model systems. The literature on this topic is extensive and the interested reader is referred to a JPCA Virtual Issue\cite{Senent2016} to explore the wide range of research in the field. Comparison of molecular cluster data and theory is straightforward and allows to tackle complex systems and interactions, e.g., elucidating the local structure around a hydrated proton\cite{Zeng2021}.

Neutral cluster sources usually create a heterogeneous distribution of cluster sizes and cluster geometries. A large body of research relies on photoionization and mass analysis to assign the composition of the observed clusters. But the energy released due to cluster reorganization around the newly formed cation can often overcome the cluster dissociation limit and the observed cluster mass does not reliably reveal the initially photoexcited cluster size. An illustrative example for this problem can be found in the literature for phenol-ammonia clusters, where contradictory data led to decades-long discussions about the role of excited state proton or hydrogen transfer.\cite{Gregoire2001,David2002}

As presented in Section \ref{sec:Analysis of Cationic Fragmentation}, CRASY allows to assign fragment signals to their neutral parent species. CRASY can therefore help to assign cluster spectroscopy results to a specific cluster size and structure. Indeed, the initial proposal\cite{SFB450} for CRASY experiments was formulated to facilitate the analysis of photochemistry in DNA base clusters.\cite{Schultz2004,Ritze2005,Samoylova2005,Gador2007,Samoylova2008,Smith2010a}

\begin{figure}[htb]
\centering
  \includegraphics[width=7.6cm]{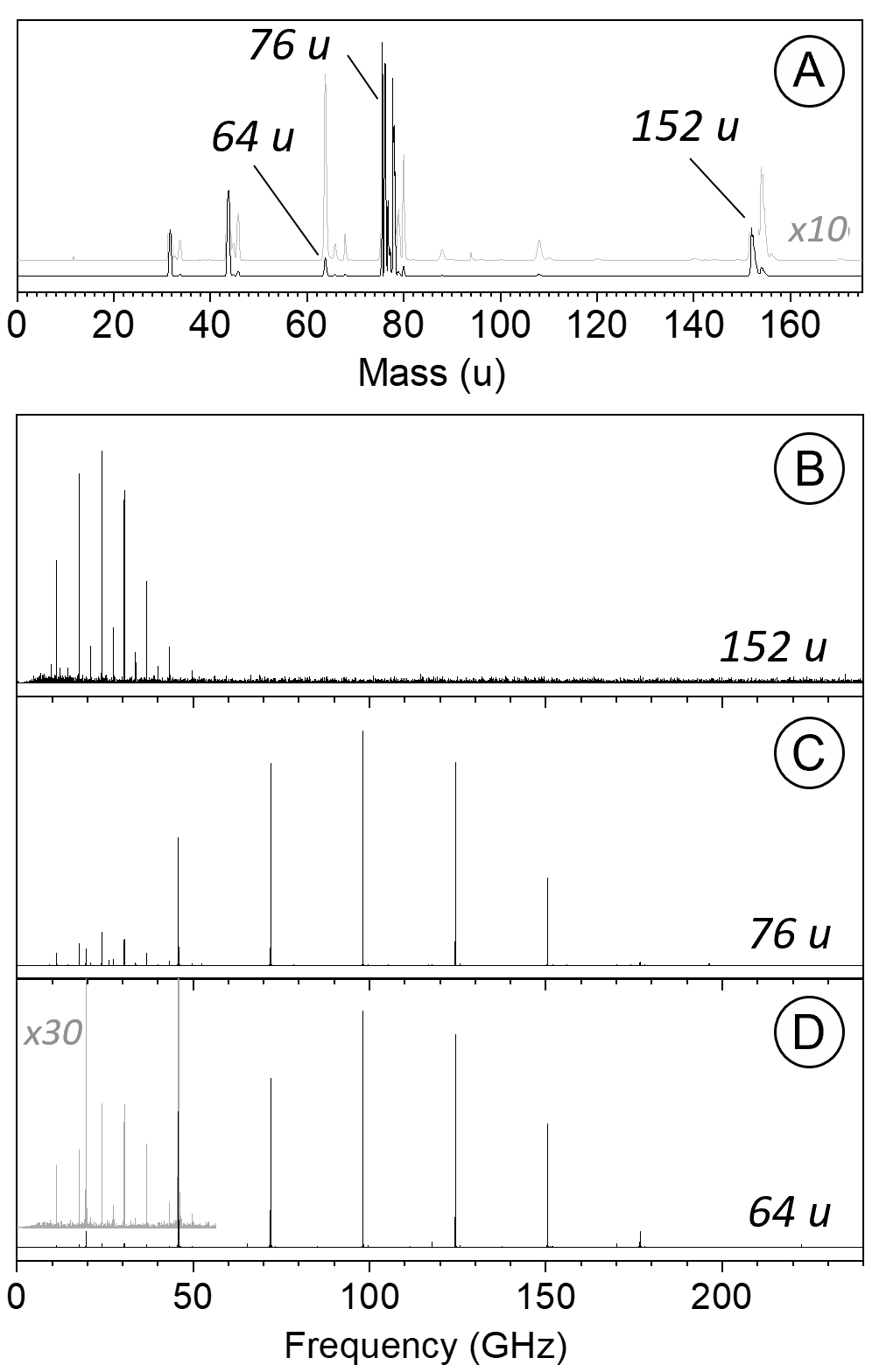}
  \caption[Caption for CS2 cluster correlation]{(A) Mass spectrum for \ce{CS2} dimer (152\,u), monomer (76\,u), isotopologues and fragments. (B-D) Rotational Raman spectra correlated to \ce{CS2} monomer and a \ce{S2} fragment (64\,u) show the \ce{CS2}-dimer transitions in the 0--50\,GHz frequency range. Grey insets shows the same data with enlarged ordinate.}
  \label{Fig16:CS2_cluster_correlation}
\end{figure}

\begin{figure}[htb]
\centering
  \includegraphics[width=7.8cm]{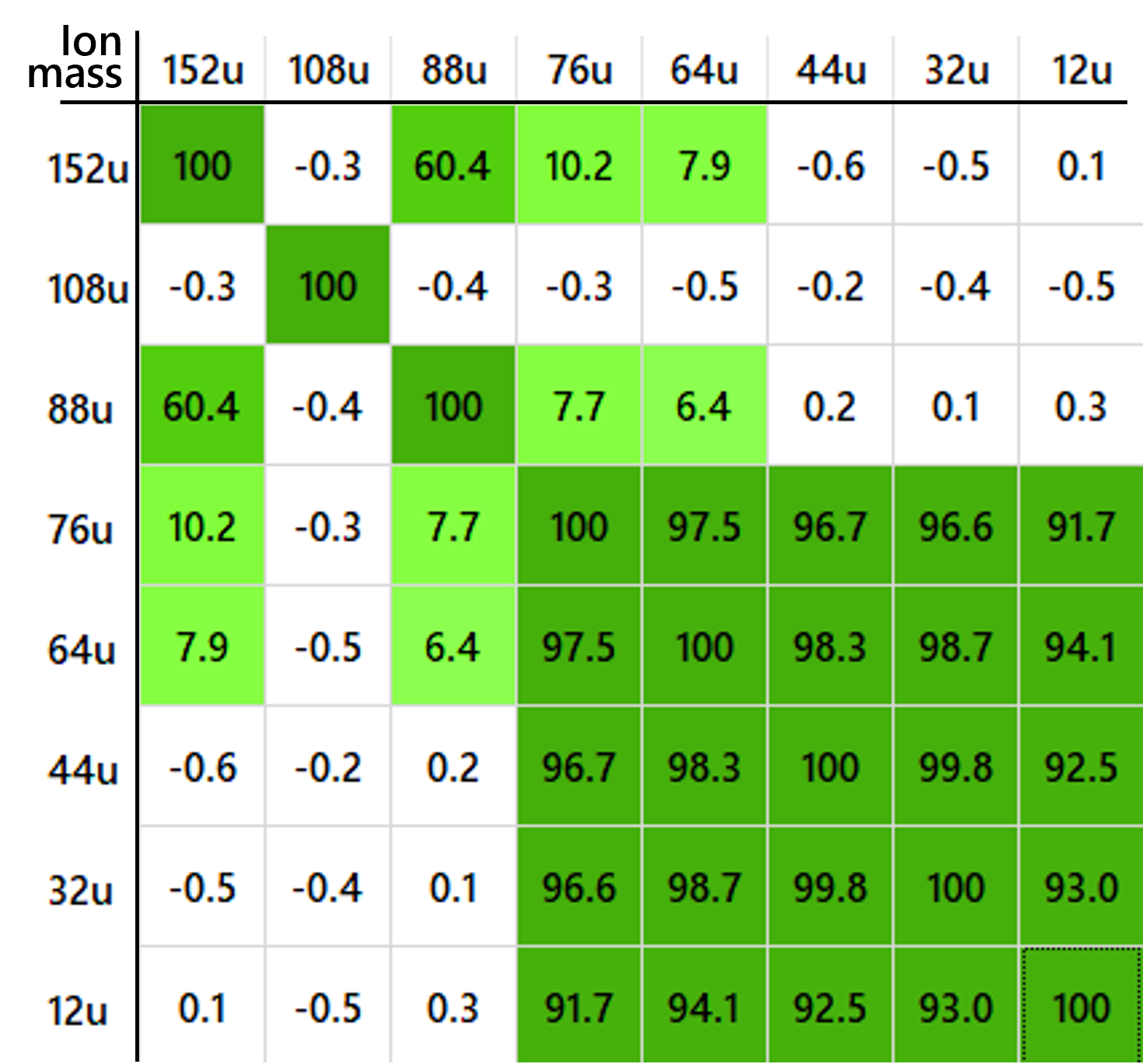}
  \caption{ Pearson correlation map for CS2 molecules and clusters.}
  \label{Fig17:CS2_dimer_correlation_map}
\end{figure}

Fig.\ \ref{Fig16:CS2_cluster_correlation} shows the mass spectrum and several mass-correlated rotational spectra from a CRASY scan with 18\, MHz FWHM resolution. Rotational transition lines for the \ce{CS2} dimer were resolved in the <\,50\,MHz spectral range. The same lines appeared in the \ce{SCCS}, \ce{CS2}, and \ce{S2} mass channels, marking them as fragmentation products of this cluster. Fig.\ \ref{Fig17:CS2_dimer_correlation_map} shows the correlation map for ion masses that might be formed by fragmentation of the dimer main isotopologue. Additional data for 20 more molecular species was omitted from the correlation map to keep it at a reasonable size. The correlation analysis revealed very selective fragmentation channels: The dimer fragmented into \ce{SCCS} (88\,u), \ce{CS2} (76\,u), and \ce{S2} (64\,u), but not into \ce{CS3} (108\,u), \ce{CS} (44\,u), \ce{S} (32\,u), or \ce{C} (12\,u). A cationic disproportionation reaction \ce{(CS2)2 -> SCCS + S2} seems to occur with a surprisingly high yield and selectivity. Signals for \ce{CS3} are a factor 2 larger than those for \ce{SCCS}, but showed no correlation to the dimer and must be formed via an unrelated pathway.

Experimental data for pyridine showed ion signals for pyridine dimer, but failed to resolve the corresponding rotational spectrum, even though signals were large.\cite{Lee2021} This failure may be due to the theoretically predicted presence of multiple stable dimer geometries,\cite{Piacenza2005,Hohenstein2009,Zhang2014,Sieranski2017} i.e., rotational lines for any single dimer species might be too small to present a recognizable spectrum.

\begin{figure}[htb]
\centering
  \includegraphics[width=7.8cm]{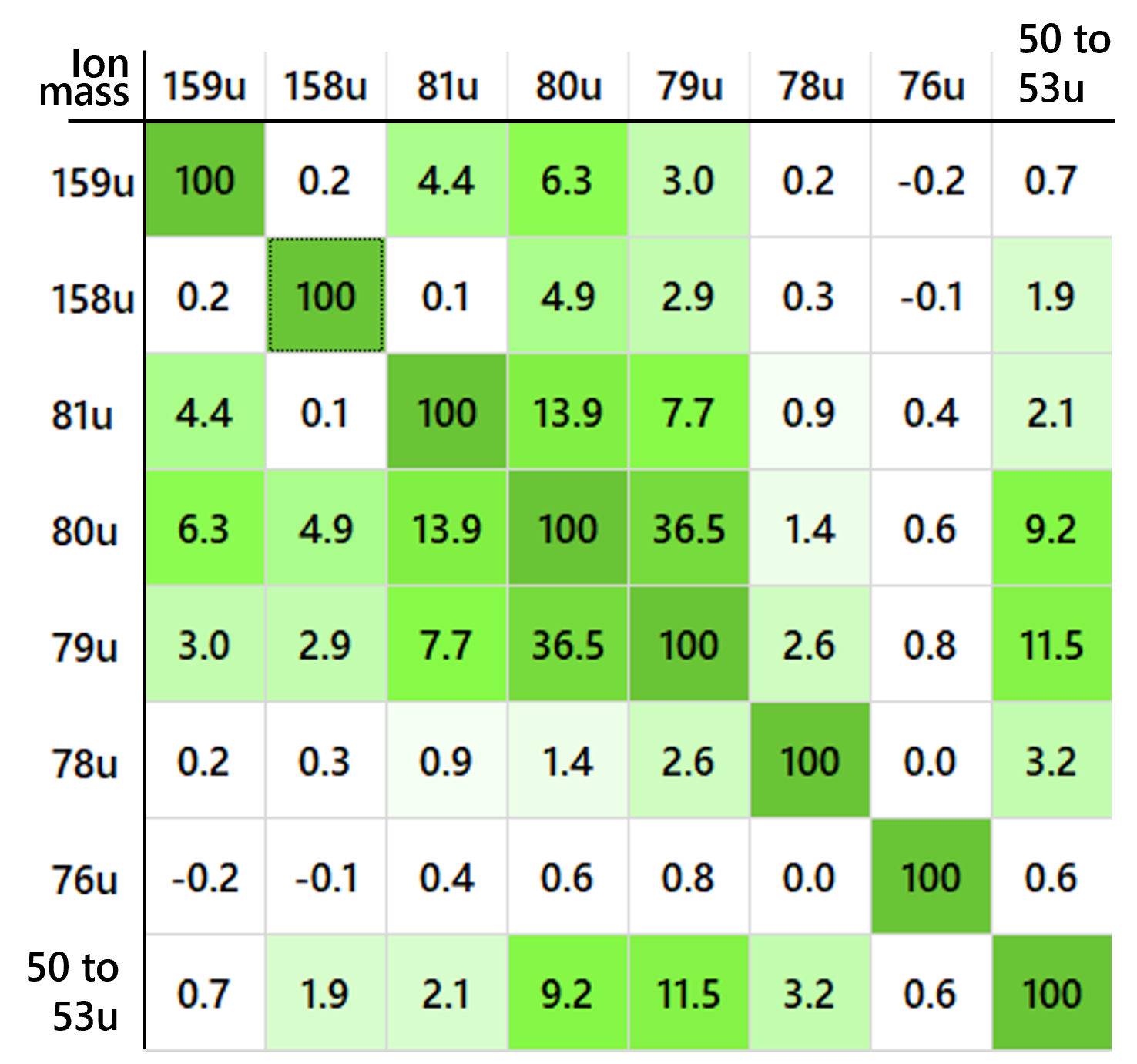}
  \caption{ Pearson correlation map for a pyridine sample. Pyridine dimer (158\,u) showed correlation with protonated (80\,u) and unprotonated (79\,u) pyridine monomer and smaller fragments (50\,u to 53\,u). }
  \label{Fig18:Pyridine_cluster_correlation}
\end{figure}

But even without a well-resolved spectrum, a correlation analysis can yield insights based on unresolved or low amplitude signals. Fig.\ \ref{Fig18:Pyridine_cluster_correlation} shows a correlation map for a CRASY scan with 200\,ns scan range, \SI{10}{\percent} sparse sampling, and 4.5\,MHz effective resolution. This measurement showed a well-resolved spectrum in the mass channel for pyridine monomer (79\,u) and \ce{CS2} (76\,u), but no assignable lines in other mass channels. Nevertheless, clear correlation was observed between pyridine, protonated pyridine (80\,u, \SI{80}{\percent} of monomer signal amplitude), pyridine dimer (158\,u, \SI{1.7}{\percent} amplitude), protonated dimer (159\,u, \SI{3.2}{\percent} amplitude), and smaller fragments (50\,u to 53\,u, \SI{6.2}{\percent} amplitude). The presence of protonated monomer and dimer signals can be explained by asymmetric fragmentation of clusters.

The observation of a large protonated pyridine signal, presumedly formed by fragmentation of the dimer, was unexpected and deserved further consideration. Subsequent ab initio calculations revealed that the lowest energy fragmentation pathway is not the initially expected symmetric fragmentation into pyridine and pyridine cation, but forms a protonated pyridine cation and an $ortho$-pyridyl radical.\cite{Lee2021}

For benzene dimer, theory predicted similar stabilities for unpolar stacked and dipolar T-shaped geometries,\cite{Sinnokrot2004} but only the dipolar geometry was observed by FTWM experiments\cite{Arunan1993,Schnell2013,Schnell2013b}. Rotational Raman spectra might additionally resolve unpolar stacked and stacked-displaced geometries, but mass-CRASY data in the dimer mass channel only revealed a dense and unresolved spectrum at low frequencies (<\,10\,GHz).

\begin{figure*}[htb]
\centering
  \includegraphics[width=17.8cm]{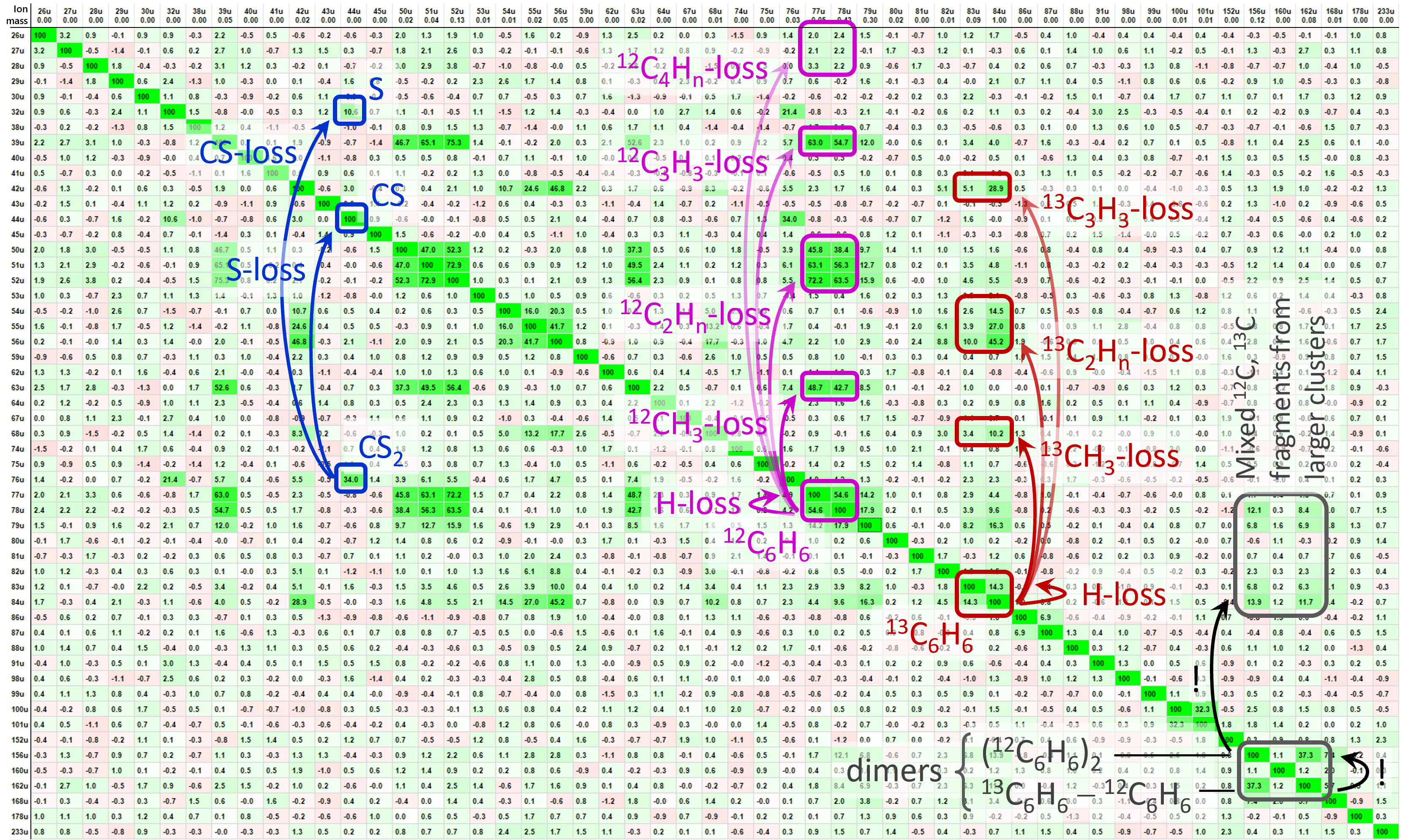}
  \caption{ Annotated Pearson correlation map for a sample of benzene, \ce{^{13}C6}-benzene, and traces of \ce{CS2}. }
  \label{Fig19:Benzene_cluster_correlation}
\end{figure*}

A CRASY correlation map for a mixed sample of benzene and \ce{^{13}C6}-benzene is shown in Fig.\ \ref{Fig19:Benzene_cluster_correlation}. Here, a full correlation map is shown and annotated to illustrate the facile assignment of fragmentation channels even in highly congested mass spectra. Monomer rotational spectra were correlated with expected fragments, e.g., \ce{CS2} (76\,u) correlated with \ce{CS} (S-loss, 44\,u) and \ce{S} (\ce{CS}-loss, 32\,u). For benzene monomers, hydrogen-loss and hydrocarbon loss was readily assigned. For the dimer species, unphysical correlations between the \ce{^{13}C6H6-^{12}C6H6} cluster (162\,u) and the \ce{^{12}C6H6} dimer (156\,u), as well as the \ce{^{12}C6H6} dimer and the \ce{^{13}C6H6} monomer (84\,u) were observed. These correlations can only be explained by the presence of larger clusters with mixed isotopic composition. This result should guide future experiments, which should be measured with significantly smaller cluster formation rate to avoid the interference from larger clusters.


\section{High-Resolution Rotational Raman Spectroscopy}
\label{sec:High-Resolution, High-Accuracy Rotational Raman Spectra}
\subsection{Spectral Range and Resolution}
\begin{figure}[htb]
\centering
  \includegraphics[width=7.9cm]{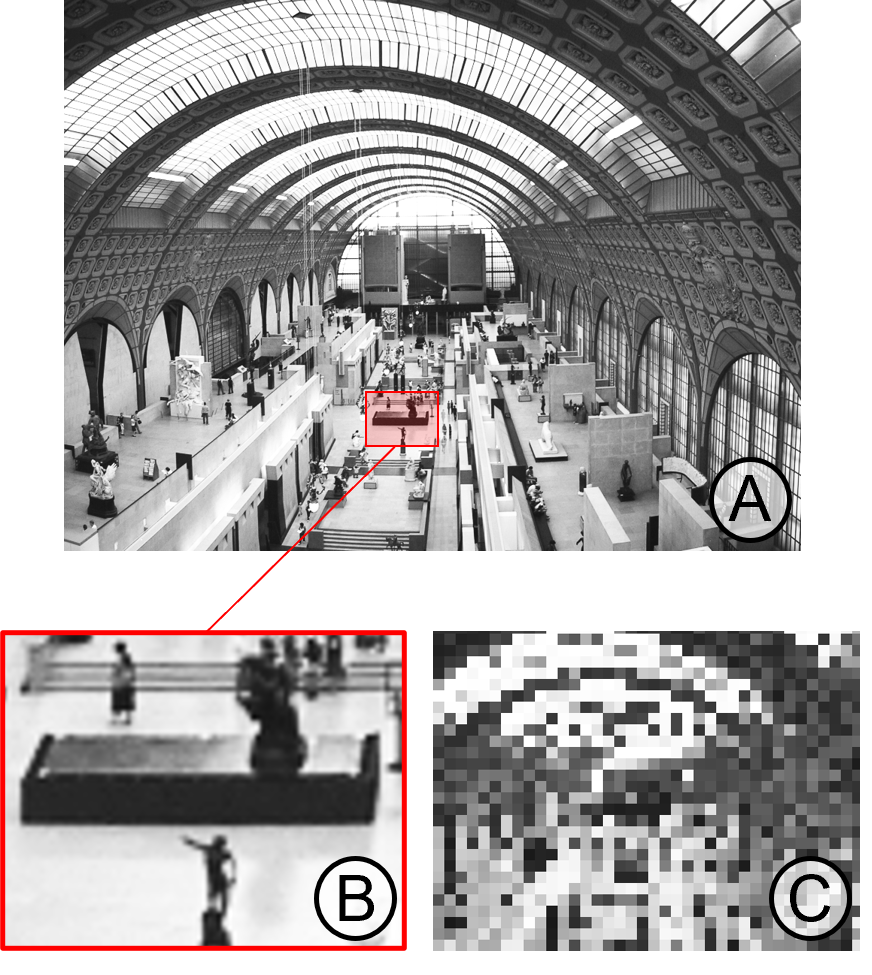}
  \caption{ (A) Photo of the Quai d'Orsay museum hall. (B,C) Photo with one order-of-magnitude reduced observation range or resolution along x and y.}
  \label{Fig20:Spectral_Range_and_Resolution}
\end{figure}

The quality of any measurement is limited by its observation range and resolution. The product of the two properties describes how many distinct data points can be resolved by a measurement. Fig.\ \ref{Fig20:Spectral_Range_and_Resolution} aims to illustrate the importance of observation range and resolution with a pictorial image. These two factors, together with a method's sensitivity and contrast, limit which aspects of the natural world we can or cannot observe. The importance of spectroscopic range and resolution is hard to overstate: order-of-magnitude improvements often lead to significant developments in associated research fields, with recent examples found in frequency comb spectroscopy (energy resolution, 2005 Physics Nobel prize), super-resolution microscopy (spatial resolution, 2014 Chemistry Nobel prize) and attosecond spectroscopy (time resolution, 2023 Physics Nobel prize).

The resolution of all spectroscopic experiments is fundamentally limited by Heisenberg's energy-time uncertainty principle: the best attainable energy resolution $\Delta E$ within an observation time $\Delta t$ is limited to $\Delta E \cdot \Delta t\geq \hbar / 2$. In frequency domain spectroscopy, the observation time is limited by the coherence time of the light source, i.e., the pulse duration when using perfectly coherent laser light. Spectroscopy with ultrafast laser pulses therefore has an inherently low resolution, e.g., a 50\,fs laser pulse has a transform-limited bandwidth of $\approx$\,300\cm and the achievable spectroscopic resolution is too low to resolve a meaningful rotational or vibrational spectrum. In many cases, the effective observation time is limited by the dephasing time, or lifetime, of the observed states. In the condensed-phase or in dense gas phase samples, collisional dephasing  severely limits the achievable resolution. As a result, high-resolution rotational spectroscopy, as described here, is only feasible in the collision-free environment of a cold molecular beam.

In time-domain spectroscopy, the observation time corresponds to the probed range of time delays $t_\textrm{range}$. The non-apodized resolution limit is given as $\Delta \nu^\textrm{FWHM} \cdot t_\textrm{range} \ge 0.61$.\cite{Albert2011}$^,$\footnote{In FTIR spectroscopy, the relation $\Delta \widetilde{\nu}^\textrm{FWHM} \ge 0.61 \cdot x^{-1}_\textrm{range}$ is formulated between the scanned distance $x_\textrm{range}$ (in cm) and the wavenumber resolution $\Delta \widetilde{\nu}$.} The term non-apodized denotes a square window for the probed time-domain data, i.e., a sudden signal onset and dropoff of the signal at the first and last probed delay. Note that the numerical value on the right-hand side of the equation depends on the apodization and the chosen measure for the signal width.

\begin{figure*}[htb]
\centering
  \includegraphics[width=17.8cm]{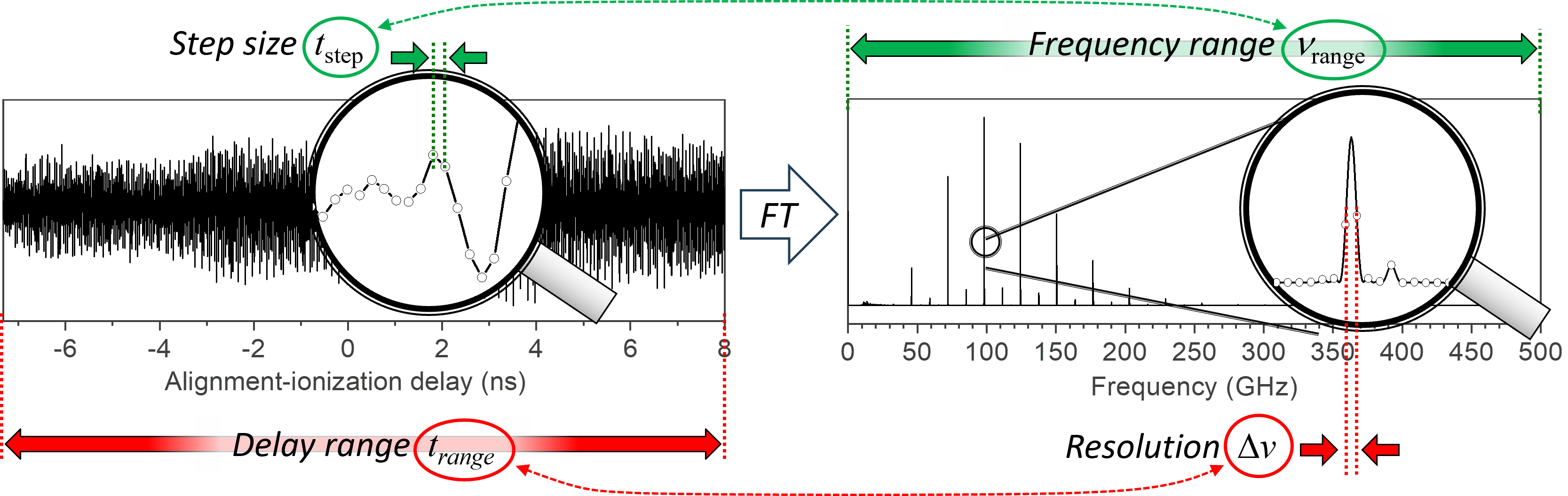}
  \caption{ Relation between spectral range and resolution in time and frequency domain data, illustrated with the delay-dependent ion signal of \ce{CS2} (left) and the corresponding FT spectrum (right). The frequency range in the spectrum is limited by the spacing of data points (step size) in the time domain. Conversely, the spacing of data points in the frequency domain (resolution) is inversely proportional to the sampled delay range. }
  \label{Fig21_FFT_Range_vs_Resolution.png}
\end{figure*}

Better spectroscopic resolution resolution can be obtained by extending the scan range, with a direct proportionality between scan range and achievable resolution. The Heisenberg uncertainty corresponds to a Fourier relation and becomes readily apparent if we consider the FT of any time-domain trace, as illustrated in Fig.\ \ref{Fig21_FFT_Range_vs_Resolution.png}. The range of sampled delays $t_\textrm{range} = (t_\textrm{max} - t_\textrm{min})$ determines the spacing of points $\Delta \nu = 1/t_\textrm{range}$ in the frequency domain. The minimal width of a signal in the frequency domain is obviously limited to a value close to $\Delta \nu$. Zero-padding of the time trace before FT can reduce the spacing of data points in the frequency domain, but does not add spectroscopic information. The accessible spectral range, on the other hand, is determined by the spacing of sampled points in the time domain: $\nu_\textrm{range} = (2 \cdot \Delta t_\textrm{step})^{-1}$. Frequencies higher that $\nu_\textrm{range}$ are not properly sampled in the time domain and appear as artifacts at lower frequencies.

A spectral range of hundreds of GHz is required to resolve the full rotational Raman spectrum in cold ($\ll$\,10\,K) molecular beams. This spectral range is readily accessible by sampling the time axis with a step-size in the single-picosecond regime, or a corresponding change of the laser beam path length with few-hundred \mum\ steps. This is achieved with the displacement of mirrors on an opto-mechanical stage.\footnote{A larger spectral range and smaller step-size is required for FTIR and modern stages can achieve reproducible step sizes $\le$\,100\,nm to give a spectral range >\,50\,000\,cm$^{-1}$.} The finite speed of light $c$ and stage length $d$ then determine the maximum range of delay times $t_\textrm{range}$ = $d/c$ and the spectroscopic resolution. The largest opto-mechanical stages, installed at national research facilities, reach path lengths in the 10\,m regime and the largest operational device seems to be a 11-chamber interferometer (Bruker ETH-SLS 2009 FTIR spectrometer prototype) installed at the Swiss Light Source, with an effective interferometer length of 11.7\,m.\cite{Albert2018} This corresponds to a delay range of 39\,ns and a FWHM resolution limit of 15.6\,MHz. Clearly, such large installations are neither practical nor affordable for modest University based research and the longest delays reached in the field of RCS were close to 5\,ns (see Ref.\ \citenum{Frey2011} and references therein).

\subsection{Extended Delay Range and Resolution}
\begin{figure}[htb]
\centering
  \includegraphics[width=8.3cm]{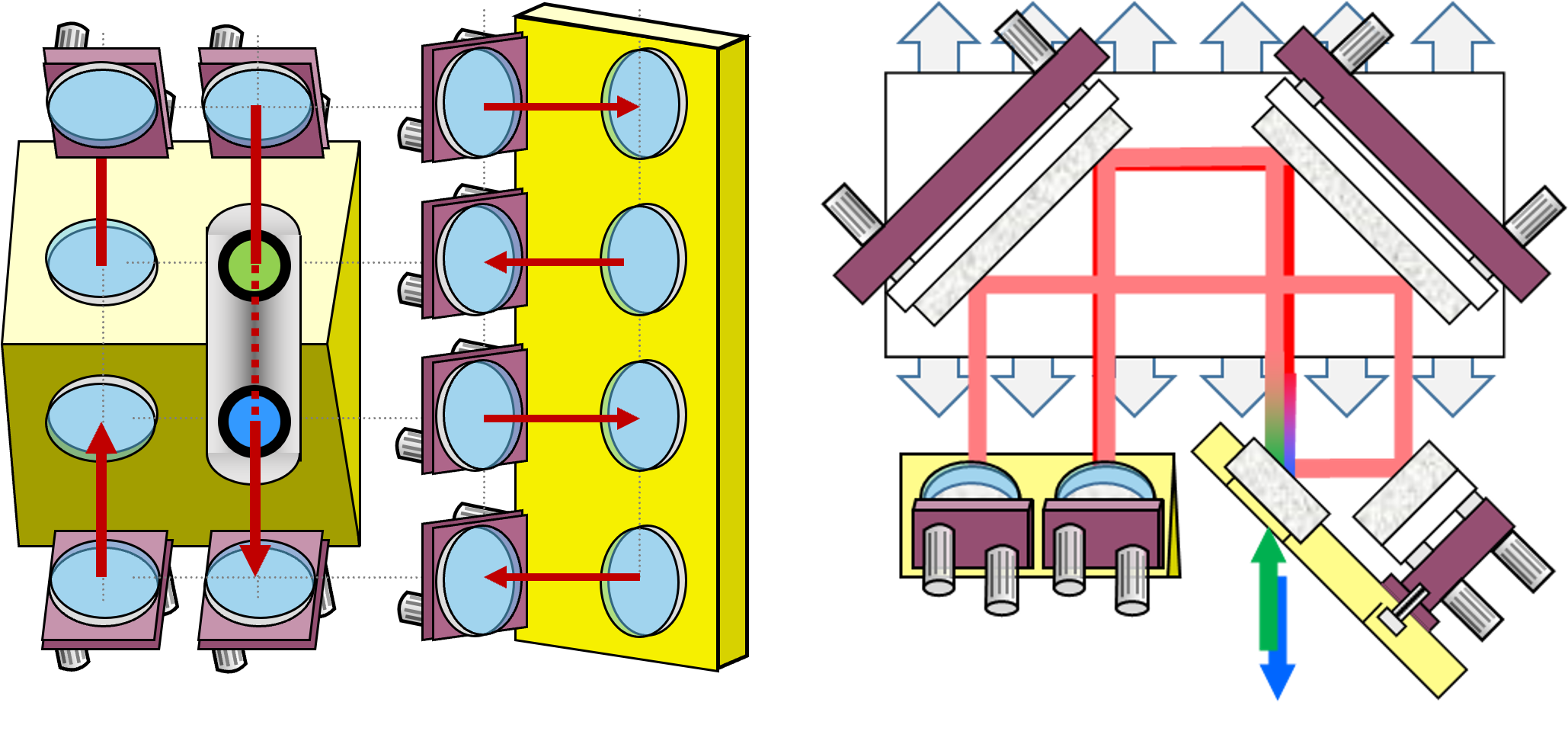}
  \caption{ Opto-mechanical set-up for 16-times folding of the optical beam path across a mechanical delay stage. (Left) Design of two opposing mirror holders, based on machined metal blocks, and 1/2-inch mirror holders. (Right) Beam path across the mirror holders and two 3-inch mirrors, mounted on the mechanical stage. Red arrows mark the beam path between the input port (green arrow or dot) and the output port (blue arrow or dot).}
  \label{Fig22_Interferometer_Folded_beam}
\end{figure}

RCS and early CRASY data was measured with fairly compact ($\ll$\,1\,m) mechanical stages. A first significant resolution increase for CRASY was obtained by extending the length of a 30\,cm opto-mechanical delay stage with a 16-times folded optical beam path. The beam folding optics were mounted on a simple metal block, as illustrated in Fig.\ \ref{Fig22_Interferometer_Folded_beam}. This extended the effective interferometer length to 4.8\,m, equivalent to a 16\,ns scan range, while maintaining a compact footprint.

\begin{figure}[htb]
\centering
  \includegraphics[width=8.3cm]{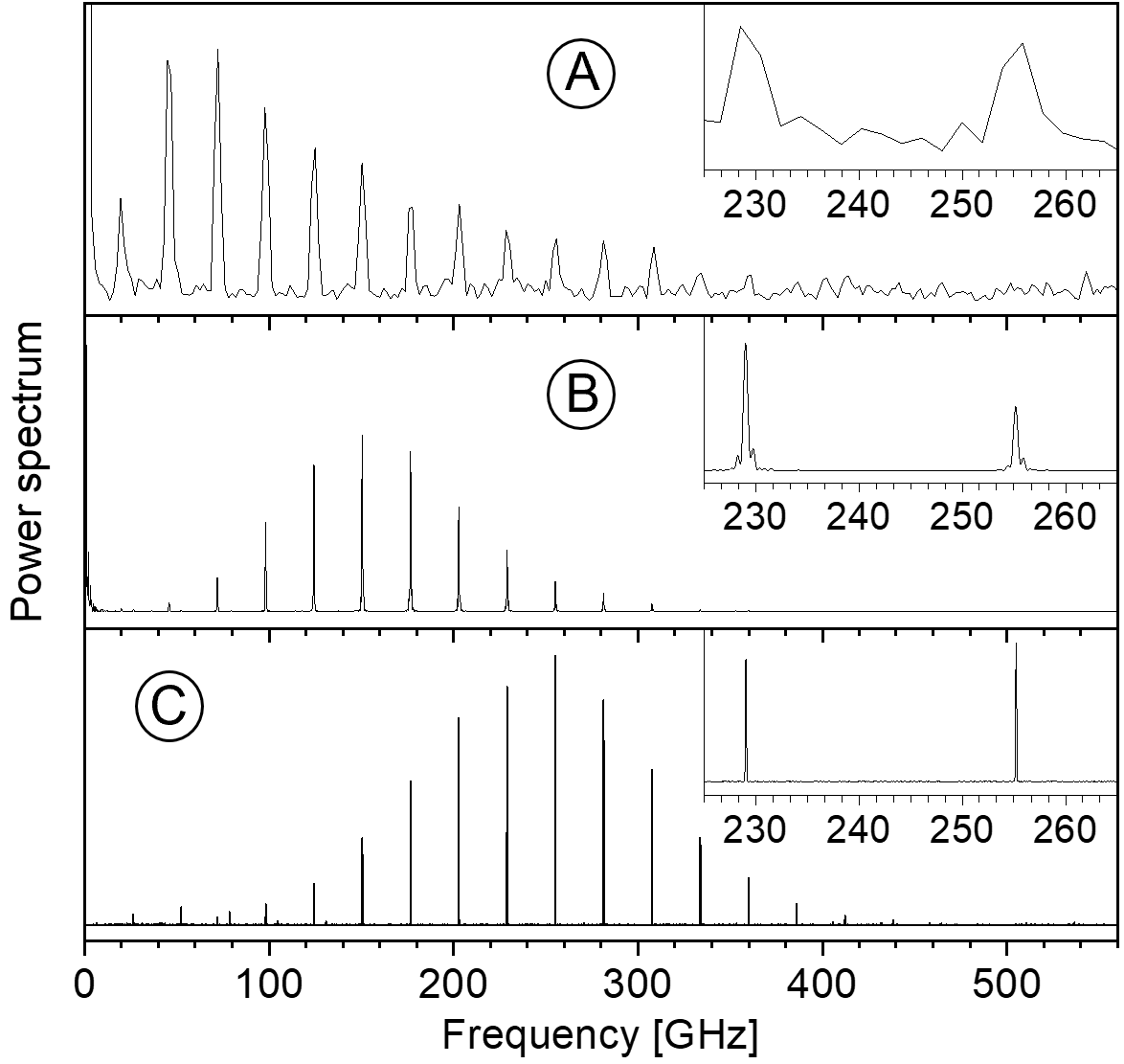}
  \caption[Fig23]{\circleds{A} First CRASY rotational Raman spectrum, measured for \ce{CS2} in the year 2009. The scanned opto-mechanical delay range was 15\,cm ($t_{range} = 0.5$\,ns), resulting in a resolution of 1.2\,GHz. \circleds{B} The spectrum measured with a longer stage and 60\,cm scan range ($t_{range} = 2$\,ns) gave a spectral resolution of 305\,MHz. \circleds{C} The spectrum measured with a 16x folded delay stage and 4.8\,m scan range ($t_{range} = 16$\,ns) gave a nominal resolution of 38\,MHz.}
  \label{Fig23_Interferometer_Folded_beam_Data}
\end{figure}

Fig.\ \ref{Fig23_Interferometer_Folded_beam_Data} shows the resulting improvement in spectroscopic resolution. Measured rotational spectra reached a resolution $\le$\,60\,MHz \cite{Schroter2013,Schroter2015,Schultz2015}, outperforming the resolution of the best previously reported rotational Raman spectra\cite{Frey2011} by a significant margin and approaching the performance of high-resolution FTIR experiments. The achieved resolution remained somewhat below the non-apodized resolution limit of 38\,MHz FWHM because the flatness of the delay stage was insufficient to exploit the full scan range.

A second increase in resolution, this time by several orders-of-magnitude, was obtained by combining opto-mechanical and electronic delays.\cite{Schroter2018,Schultz2023} Electronically generated delays do not deliver the picosecond timing accuracy that is required to observed rotational frequencies in the GHz regime. E.g., the Stanford Research Systems DG535, a common precision delay generator, is specified with a typical timing accuracy of 500\,ps. Electronic delays were therefore only used to control the selection of specific pulses from a femtosecond Ti:Sa laser oscillator, which served as alignment and probe pulses. The oscillator frequency stability therefore determined the timing accuracy of electronically delayed laser pulses.

\begin{figure}[htb]
\centering
  \includegraphics[width=8.3cm]{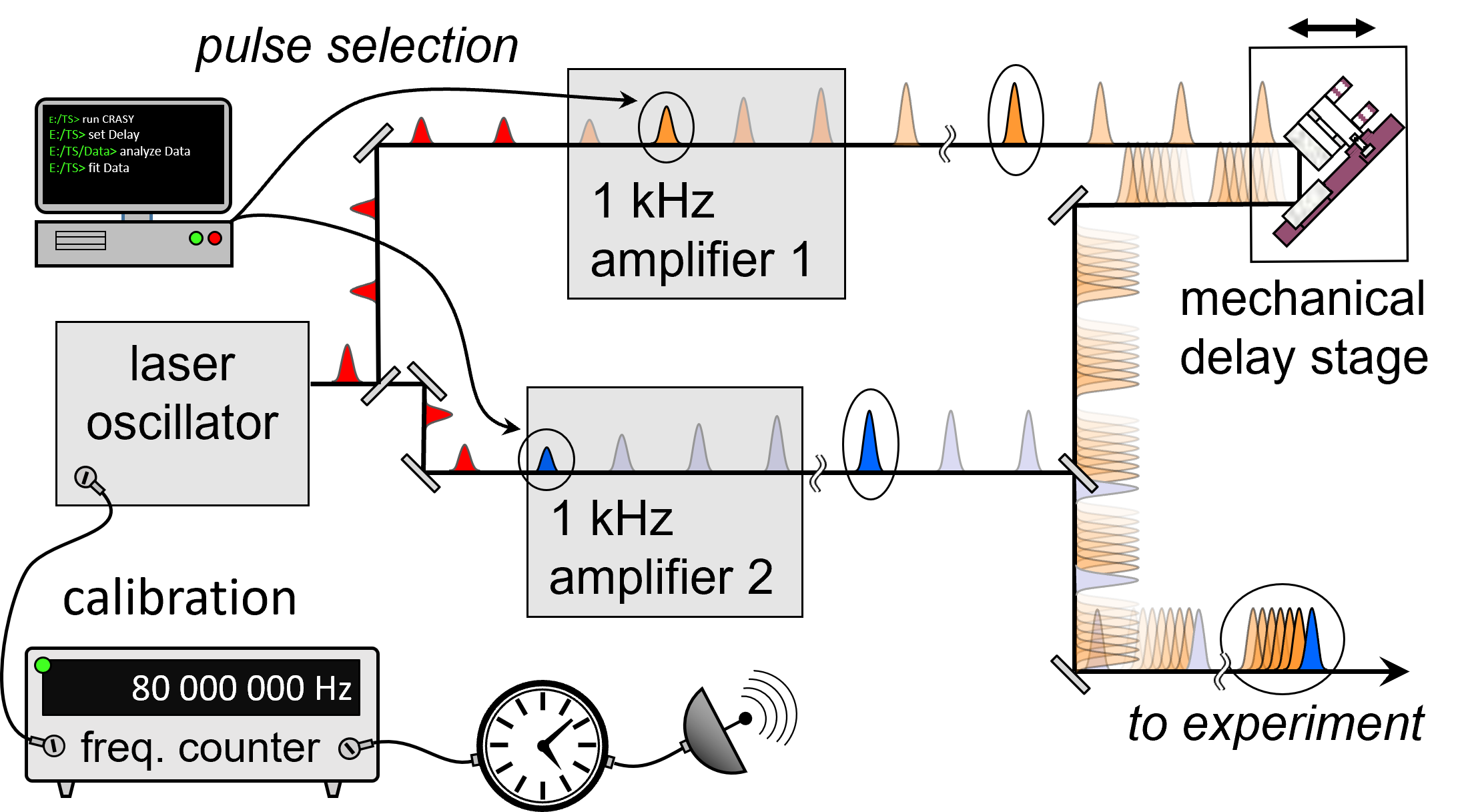}
  \caption{ Combined opto-mechanical and electronic delays for CRASY. An opto-mechanical delay stage controls delays within a range of 16\,ns by moving a set of mirrors. Computer-controlled electronic pulse-selection adds discrete delays in multiples of 12.5\,ns by selecting laser oscillator pulses for amplifier 1 (alignment pulse) and amplifier 2 (probe pulse). A frequency counter monitors the oscillator repetition rate against a GPS-stabilized clock. [Adapted from Ref.\ \citenum{Schroter2018}.]}
  \label{Fig24_Long_interferometric_delay}
\end{figure}

The experimental scheme for combined mechanical and electronic delays is shown in Fig.\ \ref{Fig24_Long_interferometric_delay}. A laser oscillator (Coherent, Vitara-T) emitted a continuous pulse train with 80\,MHz repetition rate, which corresponds to a 12.5\,ns pulse-to-pulse delay. Electronically controlled delays were used to select specific oscillator pulses, which were then amplified in two separate regenerative amplifiers (Libra-USP-1K-HE-200). After frequency conversion and attenuation, the pulses were recombined to excite and probe molecular rotation. Electronic pulse selection therefore added delays in discrete 12.5\,ns steps.

\subsection{Calibration and Accuracy}
\label{sec:Calibration and Accuracy}
Extended delay scans, for high-resolution CRASY experiments, were based on the repeated mechanical scanning of 12.5\,ns time delays with femtosecond or picosecond step-size and the incremental addition of discrete 12.5\,ns pulse-selection delays. This experimental design allowed to scan near-arbitrary delays with femtosecond step size and accuracy.

\begin{figure}[htb]
\centering
  \includegraphics[width=8.3cm]{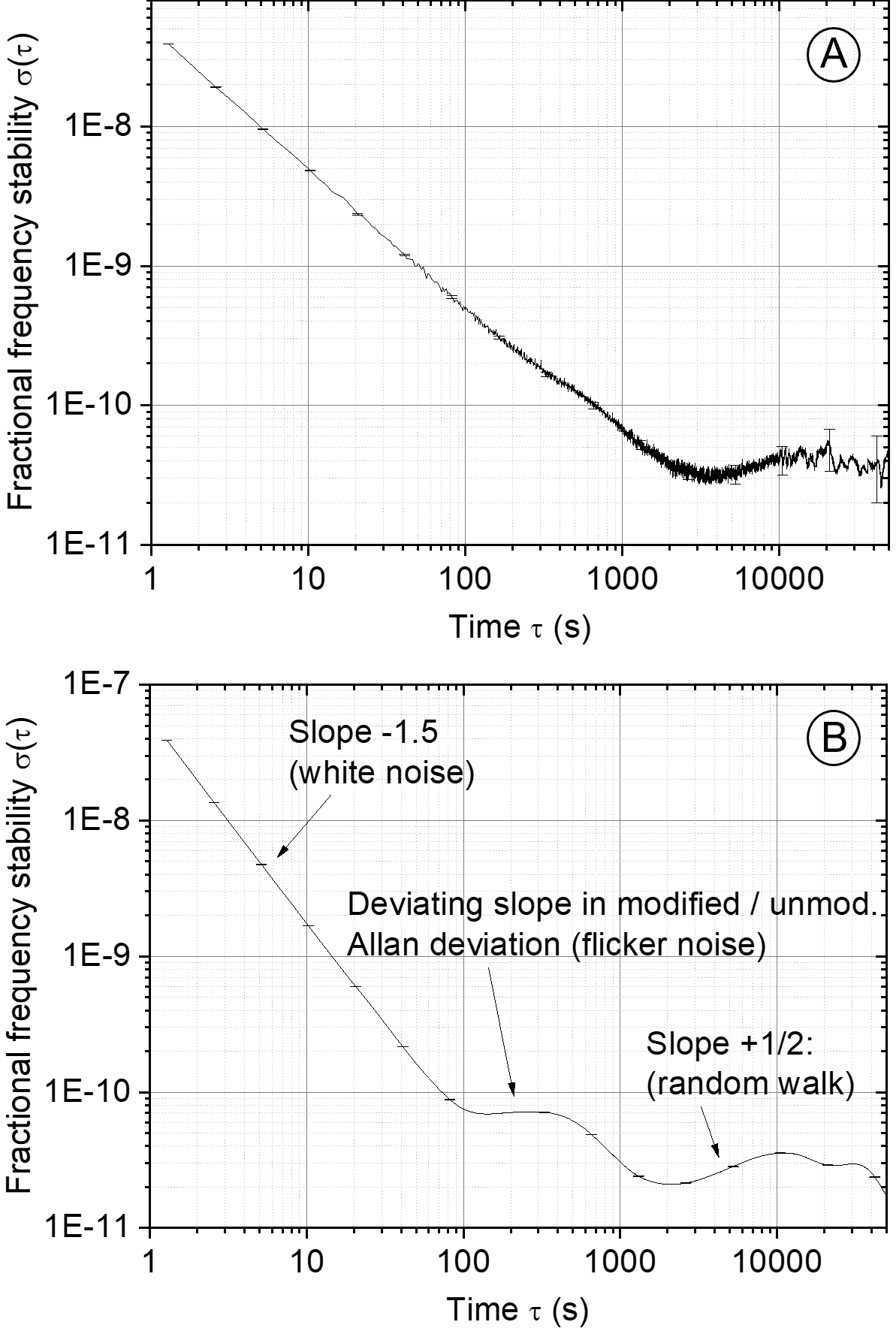}
  \caption[Fig_25]{ Allan deviation \circleds{A} and modified Allan deviation \circleds{B} of the Coherent Vitara-T laser oscillator frequency, measured against a GPS-stabilized clock. The modified Allan deviation allows to diagnose the type of noise, e.g., the characteristic white (or quantum) noise caused by the discrete pulse counting error up to 100\,s measurement time. Additional flicker noise at longer measurement times may reflect the noise floor of the Leo Bodnar clock or drifts in the oscillator frequency. [Adapted from Ref.\ \citenum{Schultz2023}.]}
  \label{Fig25_Allan_Deviation}
\end{figure}

The accuracy of pulse-selection delays is determined by the frequency stability of the laser oscillator.\cite{Schroter2018,Schultz2023} This stability was characterized against a GPS-calibrated clock and Fig.\ \ref{Fig25_Allan_Deviation} shows a representative Allan deviation,\cite{Riley2008} measured over the course of one day. Frequency deviations between oscillator and clock reached $\Delta \nu / \nu < 10^{-10}$, with the latter representing the expected noise floor for the inexpensive reference clock (Leo Bodnar GPSDO). Note that larger frequency deviations at short observation times merely represent the discrete counting noise of the frequency counter and do not reflect actual frequency deviations.

A continuous monitoring of the oscillator repetition rate during CRASY measurements allowed to calibrate the discrete 12.5\,ns delays with extraordinary accuracy, limited ultimately by the quality of the reference clock. Other uncertainties may arise due to the limited accuracy of the position encoder in the opto-mechanical delay stage, changes in the air refractive index, air pressure, and air humidity. The latter uncertainties only play a role across the opto-mechanical scan range and become insignificant if a larger number of pulse-selection delays are sampled. Small Doppler shifts may occur due to the molecular beam velocity if the laser beams are not aligned perpendicular to the molecular beam. High-resolution CRASY data determined the \ce{CS2} rotational constant with a relative uncertainty of $\Delta \nu / \nu \approx 2 \cdot 10^{-7}$, corresponding to a 700\,Hz uncertainty for a 3.17\,GHz rotational constant.\cite{Schroter2018}

The direct calibration of rotational Raman spectra against a reference clock created a time-domain equivalent to frequency comb spectroscopy.\cite{Schroter2018} The continuous 80\,MHz pulse-train created by the laser oscillator corresponds to a frequency comb, with the time and frequency properties connected by their Fourier relation. In frequency comb spectroscopy,\cite{Hansch2006} the length and thereby the frequency properties of the oscillator are locked to a reference clock via an active feedback loop.\footnote{The wavelength $\Lambda$ of each comb line is an exact integer fraction of the oscillator cavity length.} The stable comb line frequencies then allow to perform spectroscopy with an extraordinarily long interaction time $\Delta t$ and correspondingly low frequency uncertainty $\Delta \nu$. In CRASY measurements, the laser oscillator is allowed to drift freely, but is characterized against a reference clock. Any frequency drift in the oscillator is then corrected in a feed-forward scheme, adjusting the opto-mechanical delays accordingly. The long interaction time is then achieved via an extended time-domain scan range, with an extraordinary timing accuracy for each delay value.

\subsection{High-Resolution Rotational Raman Spectra}
\label{sec:High-Resolution Rotational Raman Spectra}
\begin{figure}[htb]
\centering
  \includegraphics[width=8.3cm]{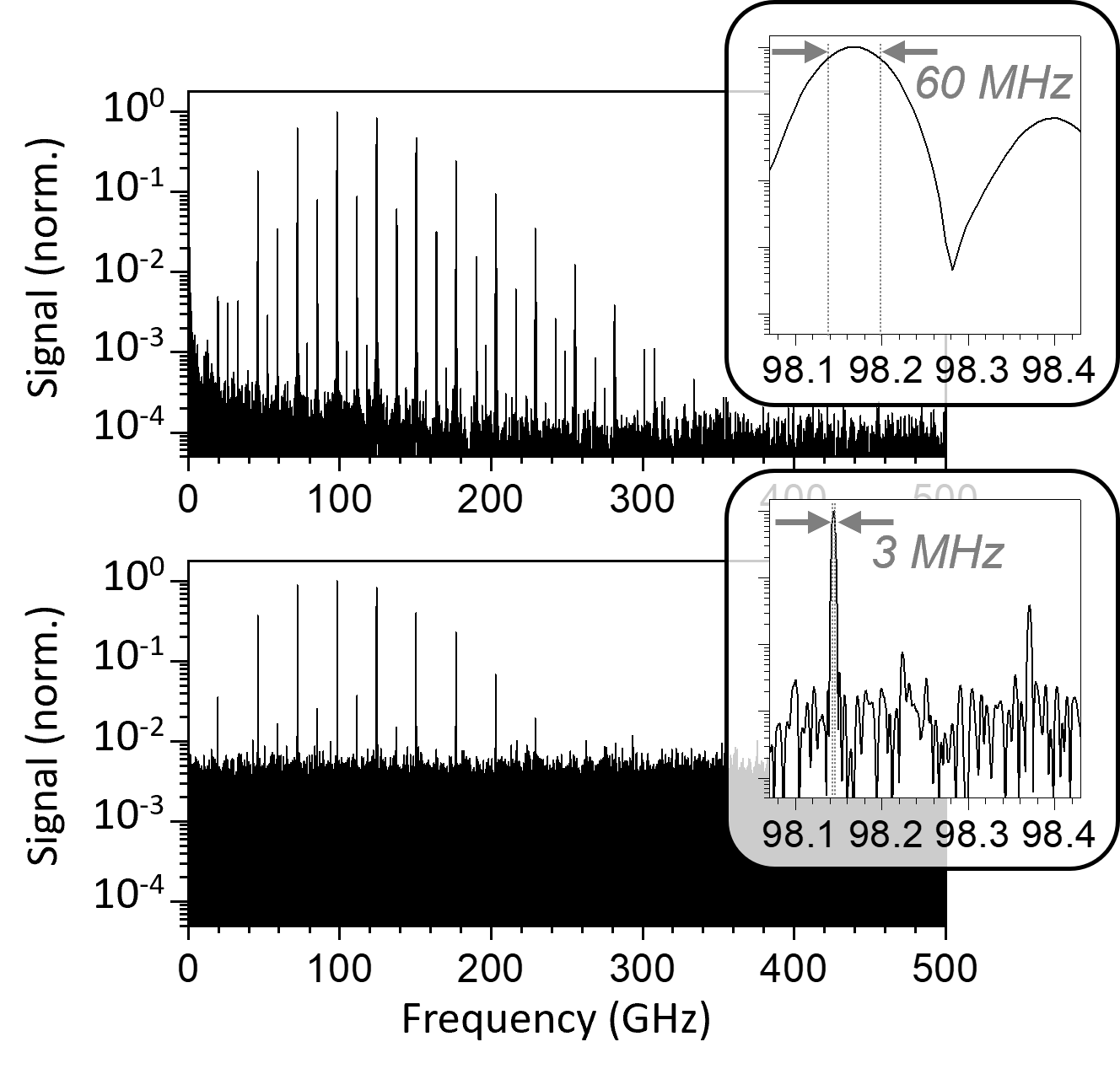}
  \caption{ High resolution rotational Raman spectra for \ce{CS2}, obtained with continuous sampling of a 15.3\,ns delay range (top) or random sparse sampling of a 312.8\,ns delay range (bottom). Insets with 2000-fold enlarged abscissa reveal the resolution increase. The enhanced resolution is coupled to a loss in signal contrast, as evident from the elevated noise in the bottom trace. Note the logarithmic ordinate. [Adapted from Ref.\ \citenum{Schroter2018}.] }
  \label{Fig26_High_resolution_Spectra_CS2}
\end{figure}

Fig.\ \ref{Fig26_High_resolution_Spectra_CS2} compares spectra obtained by scanning a purely opto-mechanical delay range of 15.3\,ns versus scanning a combined opto-mechanical and electronic delay range of more than 300\,ns. The increased scan range resulted in a corresponding resolution increase, reducing the observed spectral line-widths from 60\,MHz to 3\,MHz.\cite{Schroter2018} Note that the increased resolution came at the cost of a reduced signal-to-noise ratio (SNR). Both measurements sampled a similar number of mass spectra, with a data acquisition time of about 20\,h. In the low-resolution measurement, this corresponded to a complete sampling of the delay axis with a 1\,ps step size. For the high-resolution measurement, a corresponding sampling of the 300\,ns delay axis would be excessively costly and generate impractical amounts of data.\footnote{A typical mass spectrum for CRASY contains 120\,000 points along the mass axis and is stored with a data depth of 2 bytes per mass channel. This corresponds to 240\,kB of data per mass spectrum and 3.8\,GB of data in a scan containing 16\,000 mass spectra.} Random sparse sampling reduced the number of measured mass spectra, i.e., only 15\,000 mass spectra were measured at random delays. Random sampling adds noise across the full spectral range, but does not otherwise affect observed signals.

\begin{figure}[htb]
\centering
  \includegraphics[width=8.3cm]{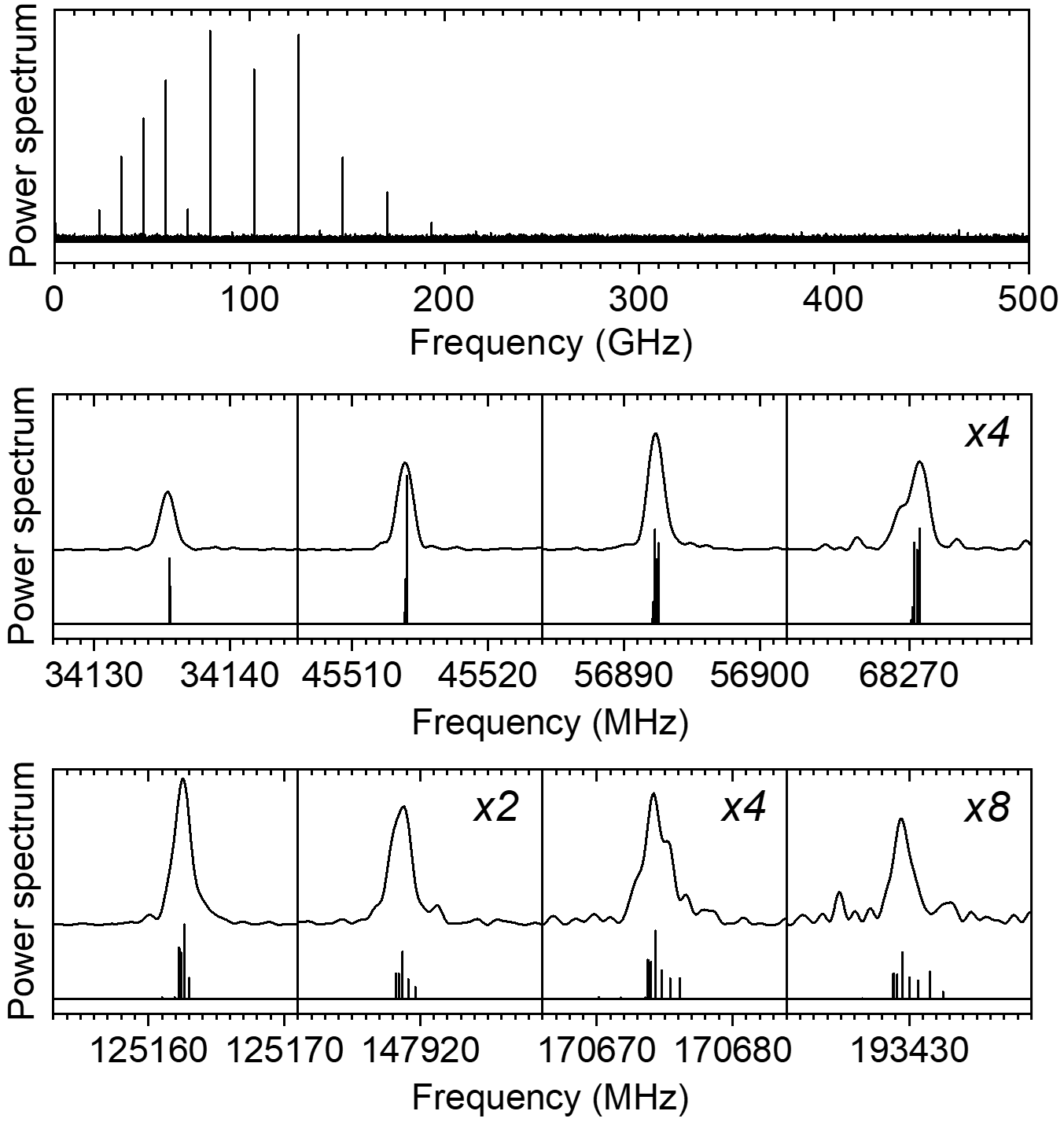}
  \caption{ (Top) High-resolution rotational Raman spectrum for benzene (mass channel 78\,u), obtained from a CRASY data set with 1\,\mus\ scan range. (Middle, bottom) Sections of the spectrum with 30\,000-fold enlarged abscissa are compared to fitted line positions and show the presence of K-splitting. }
  \label{Fig27_High_resolution_Spectrum_Benzene}
\end{figure}

For a mixed sample of benzene and carbon disulfide, mass CRASY data was obtained with a 1\,ps effective step size across a 1\,\mus\ scan range (\SI{2}{\percent} sparse sampling). The expected unapodized rotational resolution corresponds to 610\,kHz FWHM, achieved over a spectral range of 500\,GHz. The observed effective line-width for \ce{CS2} transitions reached 1\,MHz FWHM. As shown in Fig.\ \ref{Fig27_High_resolution_Spectrum_Benzene}, benzene lines were broadened and showed structure due to K-splitting: the slightly different distortion constants along the otherwise degenerate rotational axes caused a splitting of K$_a$ and K$_c$ states. K-splitting had never been observed in a symmetric top molecule of this size and the only reported K-splitting we found in the literature was for the case of ammonia,\cite{Dowling1968,Cloppenburg1979} where the rotational frequencies and K-splitting constant are many orders-of-magnitude larger.

Benzene is the prototypical aromatic molecule and there exists an abundance of preceding high-resolution measurements and literature values for the benzene rotational constants. Rotational constants were analyzed based on FTIR,\cite{Kauppinen1980} laser-based rovibrational and rovibronic spectroscopy,\cite{Pliva1982CanPhy,Pliva1982JMol,Junttila1991,Domenech1991,Okruss1999,Doi2004b} and rotational coherence spectroscopy.\cite{Riehn2001,Jarzeba2002,Matylitsky2002} But even the most recent literature values disagreed far beyond their stated uncertainty limits. The observation of K-splitting in CRASY data\cite{Lee2019} offers an explanation for this inconsistency: Unresolved K splitting creates a temperature-dependent shift of the observed line positions and this shift cannot be neglected in the analysis.

CRASY data obtained with the high-resolution set-up routinely reached a resolution in the single-MHz regime \cite{Schroter2018,Ozer2020,Lee2021,Lee2019,Schultz2023}. The longest scanned delay reached 10\,\mus, which corresponds to an effective interferometer range of 3\,km.\cite{Schultz2023} This represents a two order-of-magnitude improvement on previously reported interferometer scans, achieved with large FTIR interferometers.\cite{Amyay2010,Albert2011,Albert2015,Albert2018}

\begin{figure}[htb]
\centering
  \includegraphics[width=8.3cm]{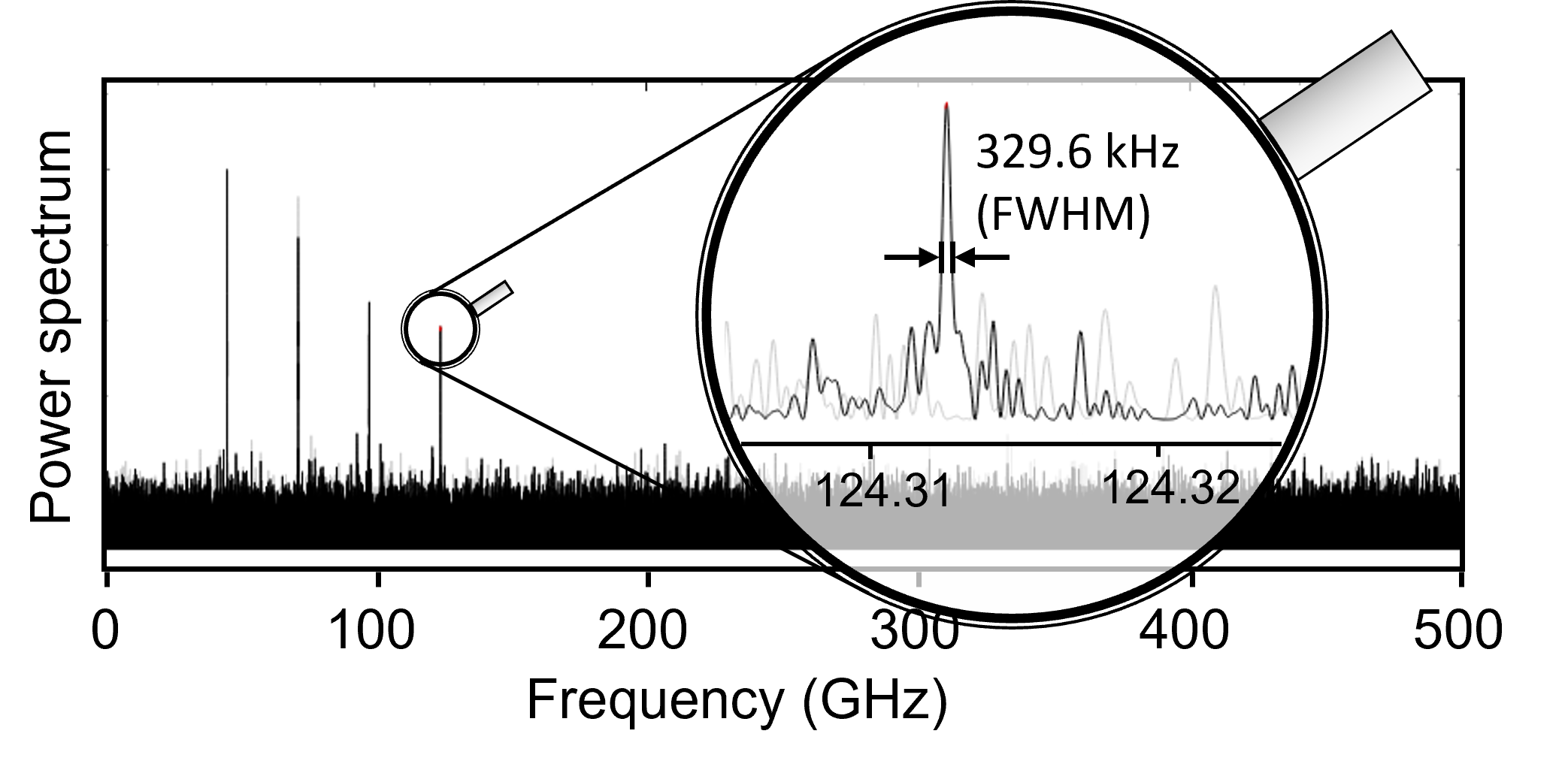}
  \caption{ High resolution rotational Raman spectrum for \ce{CS2}. The enlarged inset shows a 330\,kHz FWHM effective resolution for the J = 6--8 transition. [Adapted from Ref.\ \citenum{Schultz2023}.]}
  \label{Fig28_300kHz_Spectrum}
\end{figure}

Fig.\ \ref{Fig28_300kHz_Spectrum} shows the highest resolution rotational Raman spectrum, obtained with a 10\,\mus\ scan of a sample containing benzene and residual \ce{CS2}. The achieved effective resolution of 330\,kHz FWHM remained significantly below the non-apodized resolution limit of 61\,kHz. The resolution loss was due to an insufficient tracking of the molecular beam. Molecules in the seeded helium beam traveled with a speed of $\approx$\,1100\,m/s and were tracked by moving the pump beam position inside the spectrometer. Tracking was only achieved over a distance of few millimeters, reducing the effective scan range to <\,3\,\mus. This caused a loss of signal at long delays and a corresponding loss of spectroscopic resolution. The signal loss degraded the SNR to the point where only four lines for \ce{CS2} were resolved with good signal contrast.

Table \ref{tab:Spectroscopic_resolution} compares the resolution limit of competing spectroscopic techniques for the characterization of rotational spectra. Note that CRASY currently represents the highest-resolution method for non-dipolar molecules. FTMW shows significantly better resolution but can only be performed for dipolar species and across a much smaller spectroscopic range. Modern FTMW experiments can reach a spectral range of tens of GHz\cite{Shipman2011}, about one order-of-magnitude below the spectral coverage obtained in CRASY experiments. FTIR measurements can cover a significantly larger spectral range but the achieved effective resolution typically remains well below the theoretical resolution limit due to Doppler broadening. E.g., a low-temperature measurement of fluoroform, measured with the with the Bruker 125 HR Zurich Prototype spectrometer, reached a Doppler-limited resolution of 0.00172\cm\ (52\,MHz), significantly below the nominal resolution of 0.001\cm (30\,MHz).

\newcommand{\rowstyle}[1]{\gdef\currentrowstyle{#1}#1\ignorespaces}%
\begin{table}[htb]
 \small
 {\centering
 \caption{ Resolution of common methods for linear and Raman spectroscopy. [Adapted from Ref.\ \citenum{Schultz2023}.]}
 \begin{tabular}{lrl}
   \toprule
{Spectroscopic method}           &   \mc{2}{c}{Resolution limit}\\
   \midrule
\rowstyle{\itshape}
  Raman, laser--FT\A             &     300&MHz\cite{Weber2011}\\
\rowstyle{\itshape}
  Raman, RCS\A                   &     150&MHz\cite{Frey2011,Weber2011}\\
\rowstyle{\itshape}
  Coherent anti-Stokes Raman\A   &      30&MHz\cite{Weber2011}\\
                FTIR\A           &      16&MHz\cite{Albert2011}\\
\rowstyle{\itshape}
  Raman, high resolution CRASY\B &     330&kHz\cite{Schultz2023}\\
                FTMW\B           &    few &kHz\cite{Grabow2011,Shipman2011}\\
   \bottomrule
 \end{tabular}\\
     \footnotesize{ \A Theoretical resolution limit. \B Effective resolution. }
 \label{tab:Spectroscopic_resolution}}
\end{table}%

Note that Table \ref{tab:Spectroscopic_resolution} omitted frequency comb measurements\cite{Hansch2006,Diddams2020} and Ramsey spectroscopy. The latter is not suitable for the broad-band characterization of spectra and we do not expect that Ramsey measurements will play a role for the characterization of larger molecules. In the field of frequency comb spectroscopy, dual-comb or direct comb spectroscopy (DCS, see Refs.\ \citenum{Foltynowicz2011,Gambetta2016,Muraviev2020} and references therein) allow the rapid, broad-band, and high-resolution characterization of molecular spectra. DCS experiments are performed with the unamplified output of laser oscillators, i.e., with low pulse energies. DCS therefore requires extended interaction times with significant molecular sample densities. To our knowledge, the resolution of all reported DCS spectra is therefore subject to significant Doppler broadening and the effective spectral resolution is not competitive with high-resolution gas-phase experiments. DCS spectroscopy uses the full oscillator bandwidth and is therefore a direct frequency-domain equivalent to the time-domain techniques of RCS and CRASY.

The ability to use sparse sampling and trade signal resolution versus signal contrast or spectral range is, in principle, available for all Fourier-transform measurements. Indeed, sparse sampling was previously discussed in the field of multi-dimensional NMR.\cite{Pelczer1991,Hoch2014} But in most experimental schemes, spectroscopic resolution is fundamentally limited by decoherence of the observed molecular states. This imposes a hard limit to the achievable resolution and reduces the utility of sparse sampling.

In mass-CRASY experiments, a skimmed molecular beam creates near-perfect collision-free conditions and rotational wave packets remain coherent until the molecules collide with the spectrometer wall. Doppler broadening is negligible if the molecular beam is well-collimated. The achievable resolution is therefore only limited by the experimental challenges of tracking the molecular beam for an extended observation time and of acquiring the required large quantities of data within a reasonable measurement period.

\begin{figure}[htb]
\centering
  \includegraphics[width=8.3cm]{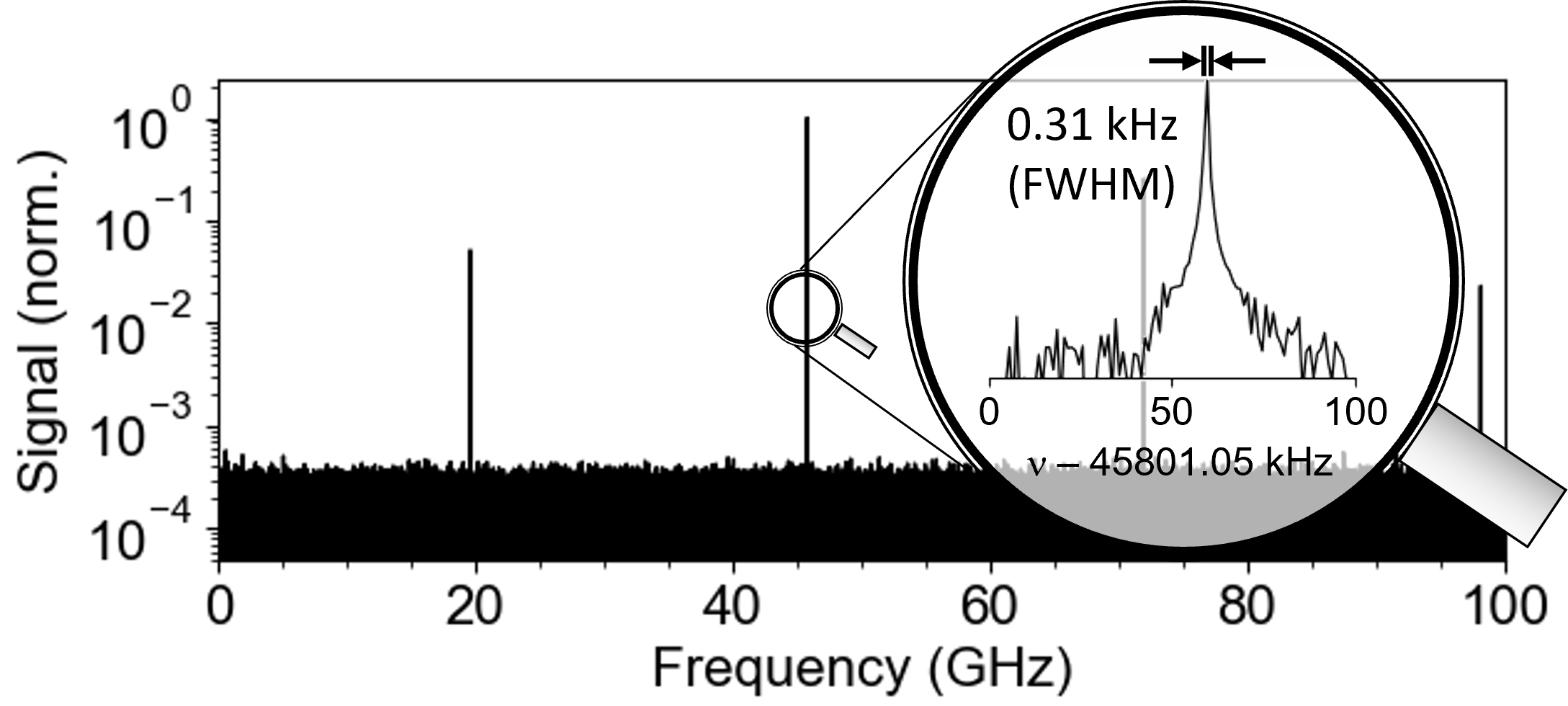}
  \caption{ Simulated high resolution rotational Raman spectrum for \ce{CS2}, based on random sampling of 100\,000 data points across a 2\,ms scan range. Note the logarithmic ordinate.}
  \label{Fig29_60kHz_Simulation}
\end{figure}

Fig.\ \ref{Fig29_60kHz_Simulation} shows a simulated rotational Raman spectrum for \ce{CS2}, as expected for the sparse sampling of 100\,000 mass spectra across a 2\,ms delay range with a nominal 5\,ps step size. The simulation assumed a typical ion count rate (1 count per shot, 1000 shots per mass spectrum) and modulation contrast (\SI{10}{\percent}) of past CRASY measurements and shows that excellent signal contrast can be achieved with an extremely low sampling rate of \SI{0.025}{\percent}. At such low sampling, the sampling noise exceeds the inherent experimental noise and the latter plays a negligible role. With current data acquisition speeds, such a measurement would require approximately 2.5 days of measurement time.

To achieve the simulated sub-KHz resolution, a well-collimated molecular beam must be tracked over an extended distance, e.g., a molecular beam seeded in krypton carrier gas should be tracked across a distance of $\approx$\,0.5\,m to achieve a 2\,ms tracking time. This represents no particular experimental challenge, but requires a specialized vacuum chamber for molecular beam tracking.

\begin{figure}[htb]
\centering
  \includegraphics[width=8.3cm]{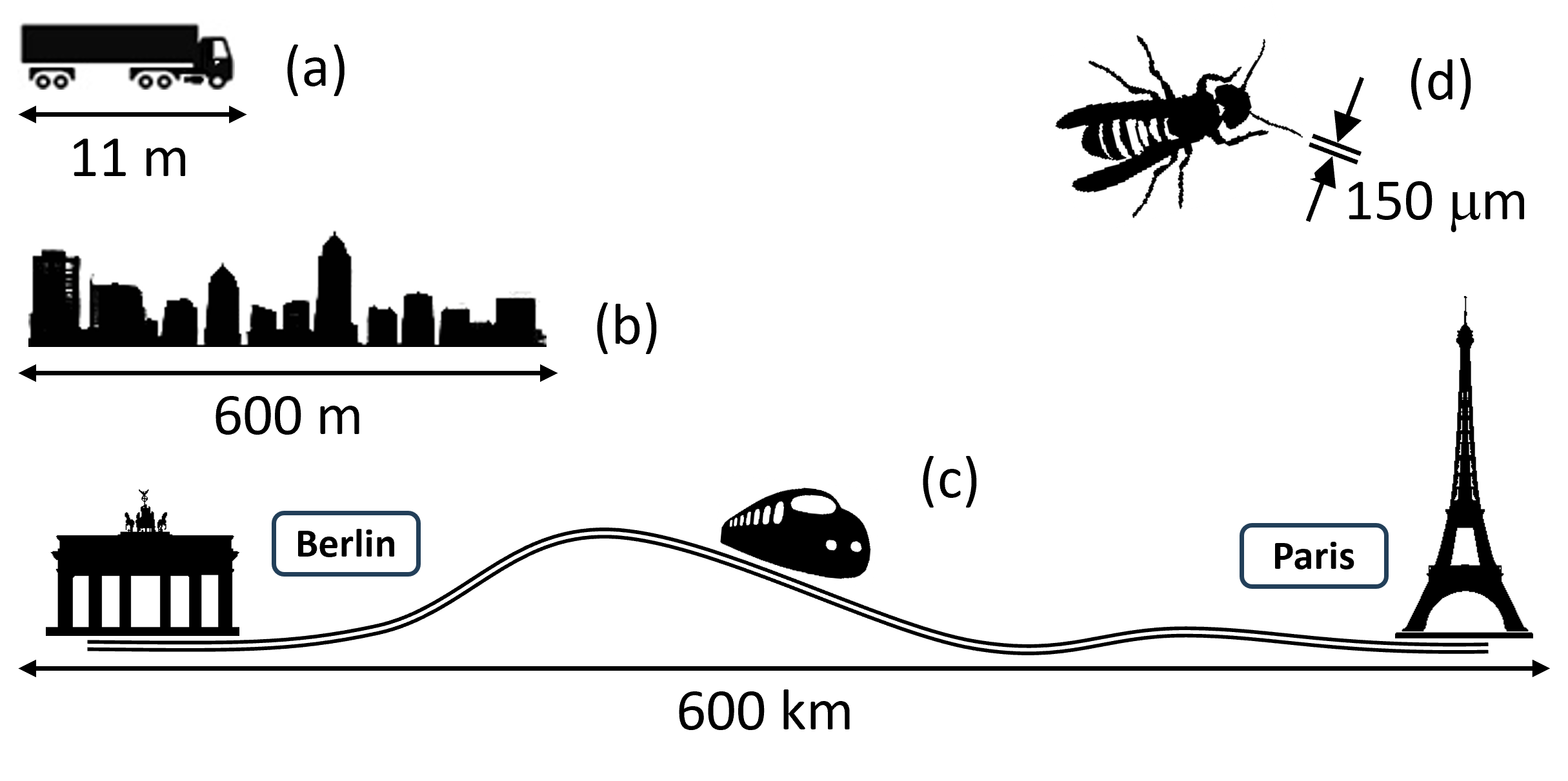}
  \caption{ Illustration of the length scales for high-resolution scanned interferometric spectroscopy. (a) The longest FTIR scans covered 11.7\,m scan range (16\,MHz resolution).\cite{Albert2018} (b) The longest successful CRASY scan covered >\,600\,m scan range (0.3\,kHz resolution).\cite{Schultz2023} (c) Required 600\,km scan range for a 0.3\,kHz resolution. (d) The required 150\,\mum\ interferometer step size for a 500\,GHz spectral range.}
  \label{Fig30_Distance_Scales_for_HR-CRASY}
\end{figure}

Fig.\ \ref{Fig30_Distance_Scales_for_HR-CRASY} illustrates the interferometer step size and scan distance required for high-resolution measurements with broad spectral range. Highest resolution FTIR measurements scanned $\approx$\,10\,m interferometer ranges with sub-\mum\ step size, to achieve a nominal resolution near 0.001\cm (30\,MHz) across a spectral range of several hundred wavenumbers.\cite{Albert2018} The highest resolution CRASY data,\cite{Schultz2023} shown in Fig.\ \ref{Fig28_300kHz_Spectrum}, was based on a 600\,m scan with 150\,\mum\ step size. A 2\,ms scan range, as required for the simulated spectrum in Fig.\ \ref{Fig29_60kHz_Simulation}, corresponds to a 600\,km scan range, almost comparable to the 880\,km distance between Berlin and Paris.

\section{Outlook: The Future of CRASY Experiments}
\label{sec:Outlook: Future CRASY Experiments}

On a fundamental level, the CRASY method represents a \emph{weak measurement} of rotational coherence, observed via the \emph{strong measurement} of a second spectroscopic observable. The excitation of rotational coherence via impulsive Raman process and its probing via resonant photoionization and mass detection, as described above, merely represents one convenient implementation of CRASY. The value of correlating other spectroscopic observables should be readily apparent and we hope that a broader field of CRASY experiments will develop in due course. The following paragraphs will discuss the feasibility of alternative excitation and probing schemes.

The excitation of rotational coherence can be achieved with any ultrafast excitation scheme, but must reach a sufficient excitation density to create a detectable modulation of the probed signal. Impulsive vibrational excitation, i.e., the Raman excitation of low-lying rovibrational states, is easily achieved with femtosecond laser pulses. A Gaussian 50\,fs laser pulse has a minimal spectral bandwidth of $\approx$\,300\,cm$^{-1}$,\cite{Diels2006} sufficient to excite low-lying vibronic states. Indeed, we already observed probable aliasing artifacts from \ce{CS2} rovibrational states when using sub-50 fs pulses for wavepacket excitation.

The linear rotational excitation of dipolar molecules may become feasible with the development of increasingly intense few-cycle THz pulses,\cite{Salen2019} or if radio wave pulses, as used for chirped pulse FTMW,\cite{Park2016} reach sufficient intensity. A fundamentally different, but very general approach would be the use of hole burning for wavepacket excitation. The excitation of long-lived electronic or vibrational states would create detectable rotational wavepacket interference in the ground and excited states and the resulting superposition of spectra might be difficult to resolve. But excitation of short-lived or dissociative electronic states would effectively remove the burned-out population from the observed wavepacket dynamics and create a pure ground state rotational wavepacket. Excitation via hole-burning might be particularly attractive to perform action spectroscopy on cold, trapped molecular ions: Electronic excitation could burn out an aligned sub-population and the remaining population could be probed by photofragmentation or photoionization. Electronic transitions of a selected chromophore, e.g., an aromatic amino acid in a peptide, might be used for the excitation and probing of rotational coherence.

The probe process can be based on any interaction between the coherently excited molecular ensemble and an ultrafast probe pulse. The probe pulse duration must be shorter than highest excited transition frequency, e.g., in the picosecond regime to observe $<$\,THz transition frequencies for small molecules in a cold molecular beam. Signals might readily be generated via IR, visible, UV, or XUV excitation to correlate rotational Raman spectra with vibrational or electronic spectra. A large number of spectra must sample the time delay axis to obtain a reasonable spectroscopic resolution and bandwidth. This is easier with high-contrast and background-free signals, e.g., collecting photons, electrons, or ions.

\begin{figure}[htb]
\centering
  \includegraphics[width=7.4cm]{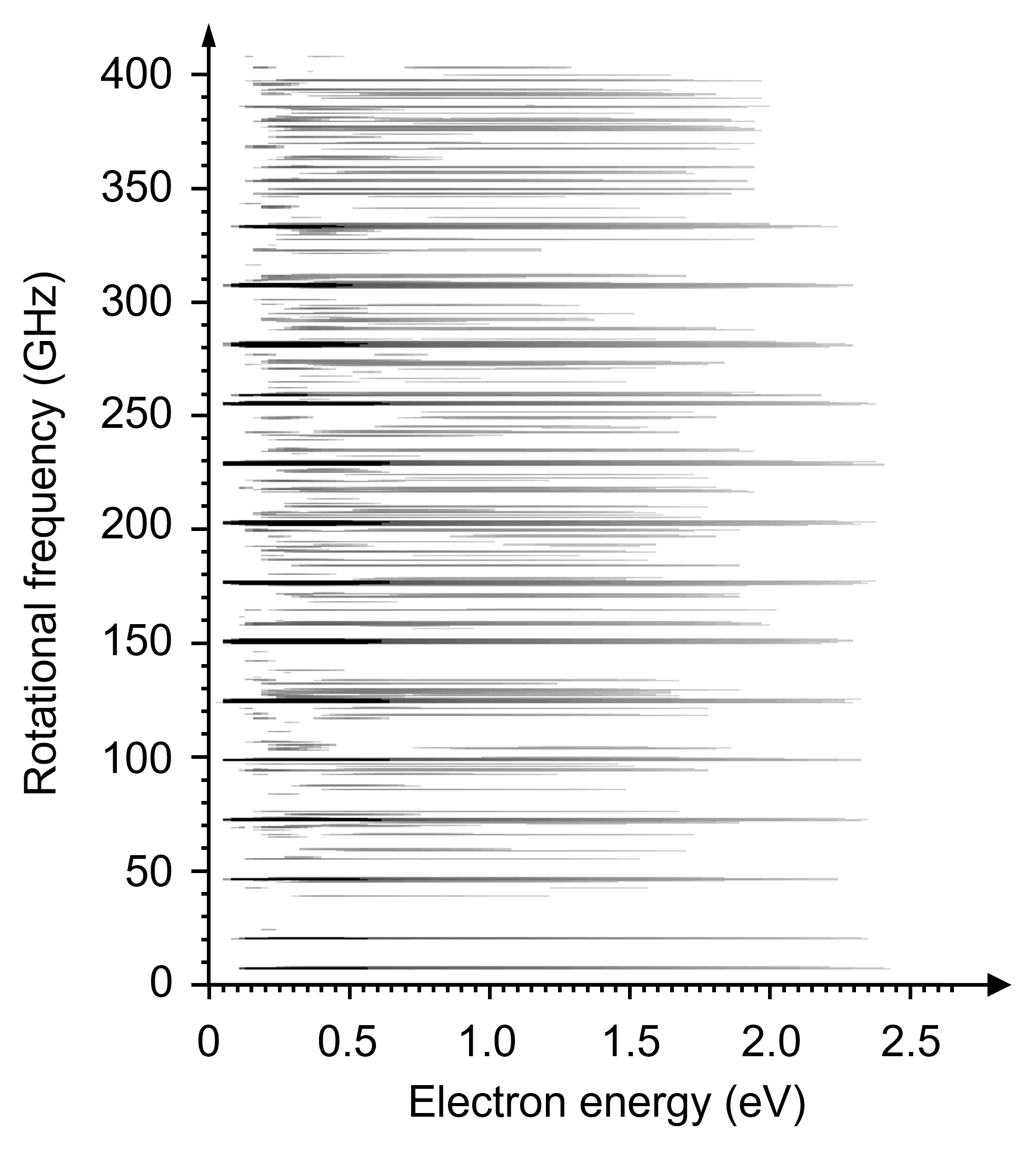}
  \caption{ 2-Dimensional representation of electron-CRASY data for carbon disulfide isotopologues. To resolve rare isotopologue signals, signal amplitudes are plotted on a non-linear scale. The dense frequency spectrum arises from the sum of all isotopologue contributions.}
  \label{Fig31_PES-CRASY_for_CS2}
\end{figure}

\begin{figure}[htb]
\centering
  \includegraphics[width=8.3cm]{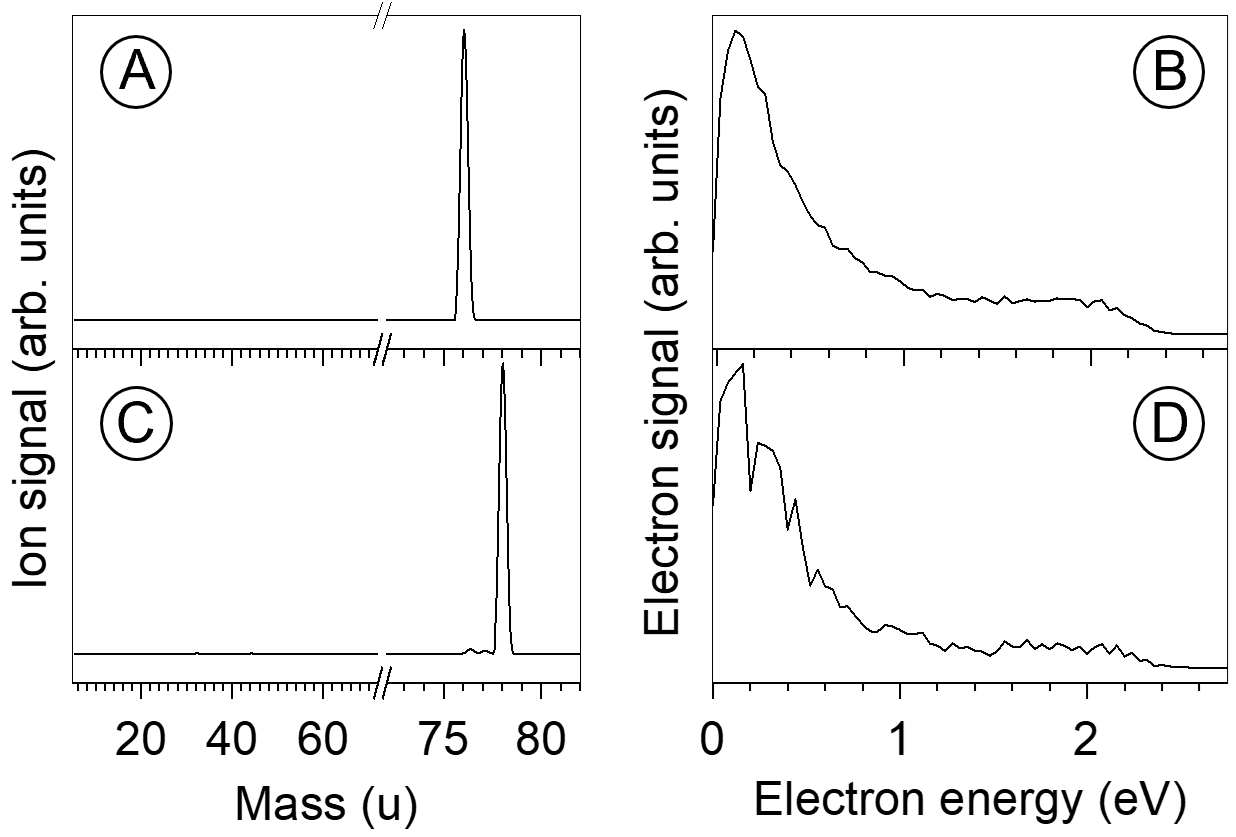}
  \caption{ (A,B) Mass and electron spectra correlated to the rotational Raman frequencies of the 5 strongest transition lines for the \ce{^{12}C^{32}S2} isotopologue. (C,D) Mass and electron spectra correlated to the rotational Raman frequencies of the 5 strongest transition lines for the \ce{^{34}S^{12}C^{32}S} isotopologue. }
  \label{Fig32_PES_for_CS2_isotopologues}
\end{figure}

Fig.\ \ref{Fig31_PES-CRASY_for_CS2} shows experimental results for the correlation of rotational Raman spectra with photoelectron spectra (electron-CRASY). The two dimensional data representation is equivalent to that shown for mass-CRASY data in Fig.\ \ref{Fig1_Correlated_Spectroscopy} and was obtained by Fourier-transformation of delay-dependent electron signal modulation. The data was measured in a repurposed electron-ion coincidence spectrometer,\cite{Stert1999,Gador2007,Samoylova2009,Smith2010a} with simultaneous detection of photoelectrons and ions. Fig.\ \ref{Fig32_PES_for_CS2_isotopologues} shows frequency-filtered mass- and electron spectra for the strongest transition lines of the \ce{^{12}C^{32}S2} and \ce{^{34}S ^{12}C^{32}S} isotopologues. In this data, electron and ions signals are correlated indirectly through their correlated rotational transition frequencies. This indirect correlation offers significant advantages over direct correlation in coincidence measurements:\cite{Gador2007,Samoylova2009,Smith2010a} Coincidence measurements require very low count-rates and therefore have inherently long signal acquisition times, unsuited to the characterization of complex samples, whereas CRASY data can be collected with large signal count rates.

The further development of electron-CRASY experiments might be particularly interesting for the time-resolved characterization of photochemical processes in molecular isomers or clusters. Time-resolved electron spectroscopy is a very powerful tool for the analysis of excited state reaction mechanisms\cite{Stolow2004}, but attempts to resolve processes in biomolecular clusters proved problematic due to the abundance of cluster isomers and fragmentation.\cite{Gador2007,Samoylova2009,Smith2010} Correlation of time-resolved electron spectra with rotational spectra (dynamics-CRASY) should allow the assignment of structure-specific and size-specific electronic properties and reactions.

\begin{figure*}[htb]
\centering
  \includegraphics[width=15.cm]{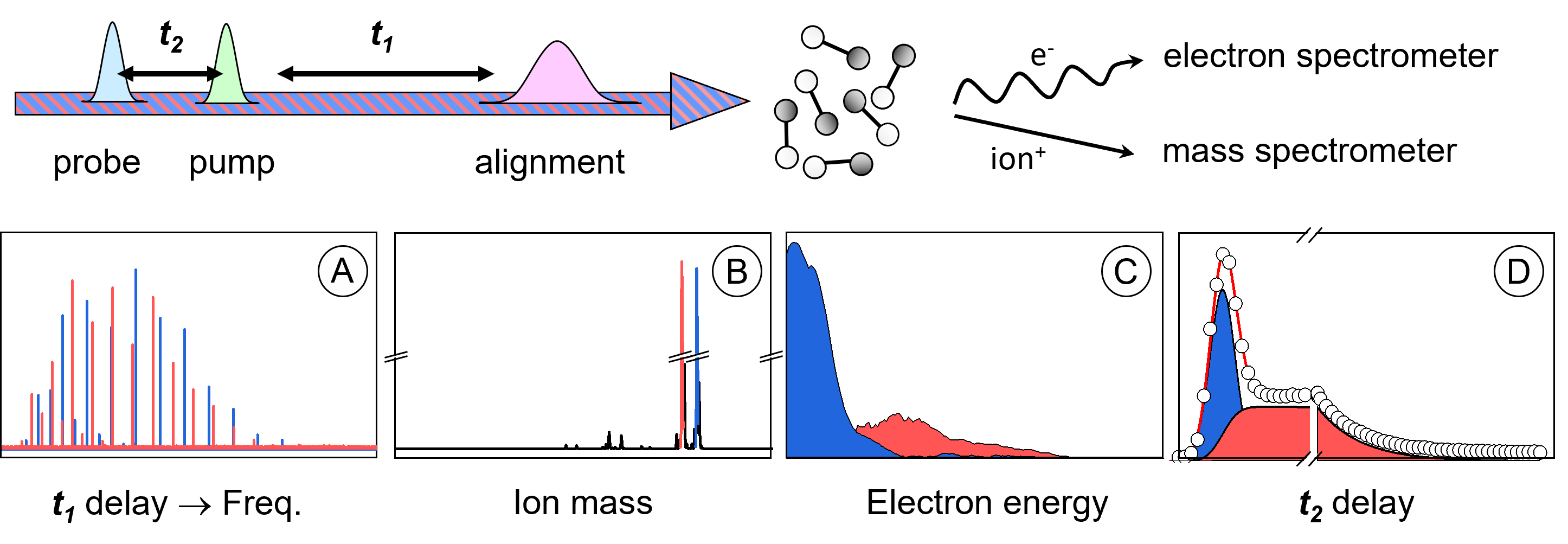}
  \caption[Fig33]{Dynamics-CRASY experiment for the study of excited state photochemical reactions. (Top) Rotational alignment is followed by electronic excitation with a pump pulse, ionization with a probe pulse, and electron and ion detection. The scan of $t_1$ delays and the FT of signal modulations resolves rotational spectra \circleds{A} and correlates detected ion masses \circleds{B} and electron energies \circleds{C} with the structure of reaction educts. The scan of $t_2$ delays \circleds{D} correlates observed ion masses and electron energies with excited state lifetimes. }
  \label{Fig33_CRASY_Reaction_Microscope.png}
\end{figure*}

Fig.\ \ref{Fig33_CRASY_Reaction_Microscope.png} illustrates the three-pulse experiment required for a dynamics-CRASY measurement and the expected types of correlated data. By using separate laser pulses for electronic excitation (pump) and photoionization (probe), an additional pump-probe time delay $t_2$ can be sampled to resolve excited state lifetimes. $t_2$-Dependent changes in electron signals resolve electronic state populations and reveal electronic transitions. Corresponding changes in ion signals reveal associated fragmentation processes. As in all CRASY experiments, the scan of a delay $t_1$ and subsequent FT of signal modulations in electron and ion signals correlates all observables with the rotational spectra of the neutral precursor molecules.

The probing of rotational alignment with absorption measurements in the IR, visible, or UV regime might be more challenging. To yield high-resolution data, CRASY measurements require the Doppler-free environment of a molecular beam with a low density of target molecules. But molecular beam sample densities are too low to meaningfully resolve absorption signals. Measurements might be performed in gas cells with long absorption path length, akin to high-resolution FTIR experiments\cite{Albert2018}. But in this case, Doppler broadening seriously limits the achievable rotational resolution, especially for larger molecules that require high evaporation temperatures.

Fluorescence detection schemes were previously discussed in the context of RCS spectroscopy\cite{Hartland1991} and are based on a time-resolved, stimulated Raman-induced fluorescence depletion scheme to correlate fluorescence with ground-state rotational properties. Fluorescence excitation is tied to a well-defined electronic transition moment and therefore shows the same alignment sensitivity as the mass- or electron-CRASY experiments presented above. Fluorescence excitation is a linear process, as opposed to the two-photon ionization used for mass-or electron-CRASY and fluorescence-CRASY could therefore be performed with much larger signal collection rates than mass-CRASY experiments.

As discussed in Section \ref{sec:High-Resolution, High-Accuracy Rotational Raman Spectra}, sparse sampling allows to trade spectroscopic resolution and spectral range versus resolution. Sub-MHz resolution was achieved with an ion count range near one ion per laser shot. Detecting fluorescence signals with orders-of-magnitude increased signal collection rates could improve the SNR and be leveraged to achieve corresponding improvements in rotational Raman resolution. We expect that the achievable resolution in such experiments will be limited only by the experimental ability to track a skimmed molecular beam.

Correlation with diffraction experiments may be possible in the future, based on modern femtosecond or picosecond free-electron lasers\cite{Schultz2014} or ultrafast electron sources.\cite{Gliserin2015} We can already find a growing literature describing diffraction data for non-adiabatically aligned molecules, a related experiment performed at a fixed delay between alignment and probe pulses. For example, ultrafast electron diffraction experiments on aligned carbon disulfide resolved asymmetric diffraction patterns and found evidence for structural deformation, at high alignment laser intensities.\cite{Yang2015,Yang2015b} Ultrafast X-ray pulses from free electron lasers were used to observe diffraction signals from aligned diiodo-benzonitrile molecules\cite{Kupper2014}. The proper sampling of a pump-probe delay range, as required for diffraction-CRASY, is therefore merely constrained by the data acquisition speed and will become realistic with sufficiently intense X-ray photon pulses.

It is difficult to achieve very high degrees of nonadiabatic molecular alignment without inadvertent excitation of electronic states through higher-order transitions.\cite{Yang2015b} Diffraction experiments on aligned molecules therefore have limited contrast . As shown in Fig.\ \ref{Fig5_Wavepacket}, CRASY data allows to exhaustively characterize the amplitudes and phases for all wavepacket constituents. The strong spectroscopic observable, e.g., a diffraction signal, can then be calculated for any chosen set of rotational amplitudes and phases to predict the signal for an 'ideal' aligned and anti-aligned state. The quality of the projected signal is only limited by the width of the observed wavepacket and  can greatly exceed the level of instantaneous alignment that might be observed at any particular time delay. In the best case, this could allow a full 3-dimensional molecular structure analysis based on diffraction measurements in a cold molecular beam.

The excitation of rotational coherence could also be achieved with a variety of techniques. A very general approach would be the use of hole burning, e.g., via short-lived or dissociative electronic states. The excitation of long-lived electronic or vibrational states would also lead to detectable wavepacket interference, but the excited states must be expected to contribute to the detected signals and the resulting superposition of ground and excited state spectra my be difficult to resolve. Excitation via hole-burning might be particularly attractive to perform action spectroscopy on cold, trapped molecular ions: Electronic excitation could burn out an aligned sub-population\footnote{Rapid internal conversion after electronic excitation would lead to a very dense continuum of states that would not show any resolvable coherence. Hence, the coherence in the unexcited population could be observed with delay dependent probing.} and the remaining population could be probed by time-delayed fragmentation or ionization.

The work of Richard Ernst on multidimensional NMR inspired the development of CRASY. The German word ernst (serious, stern, austere) is an antonym of verr\"uckt (crazy), yet I want to close with the sentence Ernst used to conclude his Nobel lecture:\cite{Ernst1992} "I am not aware of any other field of science outside of magnetic resonance that offers so much freedom and opportunities for a creative mind to invent and explore new experimental schemes that can be fruitfully applied in a variety of disciplines." I feel the same way about CRASY and hope that the presented work can offer similar inspiration to future scientists.



\begin{acknowledgement}
The author acknowledges funding support from the National Research Foundation of Korea under Grants No. NRF-2018R1D1A1A02042720 and Samsung Science and Technology Foundation under Grants No. SSTF-BA2001-08.
\end{acknowledgement}
\small
\bibliography{Review_CRASY_g}

\providecommand{\latin}[1]{#1}
\makeatletter
\providecommand{\doi}
  {\begingroup\let\do\@makeother\dospecials
  \catcode`\{=1 \catcode`\}=2 \doi@aux}
\providecommand{\doi@aux}[1]{\endgroup\texttt{#1}}
\makeatother
\providecommand*\mcitethebibliography{\thebibliography}
\csname @ifundefined\endcsname{endmcitethebibliography}
  {\let\endmcitethebibliography\endthebibliography}{}
\begin{mcitethebibliography}{166}
\providecommand*\natexlab[1]{#1}
\providecommand*\mciteSetBstSublistMode[1]{}
\providecommand*\mciteSetBstMaxWidthForm[2]{}
\providecommand*\mciteBstWouldAddEndPuncttrue
  {\def\EndOfBibitem{\unskip.}}
\providecommand*\mciteBstWouldAddEndPunctfalse
  {\let\EndOfBibitem\relax}
\providecommand*\mciteSetBstMidEndSepPunct[3]{}
\providecommand*\mciteSetBstSublistLabelBeginEnd[3]{}
\providecommand*\EndOfBibitem{}
\mciteSetBstSublistMode{f}
\mciteSetBstMaxWidthForm{subitem}{(\alph{mcitesubitemcount})}
\mciteSetBstSublistLabelBeginEnd
  {\mcitemaxwidthsubitemform\space}
  {\relax}
  {\relax}

\bibitem[Ernst(1992)]{Ernst1992}
Ernst,~R.~R. Nuclear-Magnetic-Resonance Fourier-Transform Spectroscopy (Nobel
  Lecture). \emph{Angewandte Chemie-International Edition in English}
  \textbf{1992}, \emph{31}, 805--823\relax
\mciteBstWouldAddEndPuncttrue
\mciteSetBstMidEndSepPunct{\mcitedefaultmidpunct}
{\mcitedefaultendpunct}{\mcitedefaultseppunct}\relax
\EndOfBibitem
\bibitem[W\"uthrich(2003)]{Wuthrich2003}
W\"uthrich,~K. NMR studies of structure and function of biological
  macromolecules (Nobel Lecture). \emph{Journal of Biomolecular Nmr}
  \textbf{2003}, \emph{27}, 13--39\relax
\mciteBstWouldAddEndPuncttrue
\mciteSetBstMidEndSepPunct{\mcitedefaultmidpunct}
{\mcitedefaultendpunct}{\mcitedefaultseppunct}\relax
\EndOfBibitem
\bibitem[Schultz(2004)]{SFB450}
Schultz,~T. SFB-450, Projekt A4: Analyse und Steuerung ultraschneller
  photoinduzierter Elementarprozesse in Molek\"ulclustern.
  \url{http://users.physik.fu-berlin.de/~abt/sfb450/english.html}, 2004\relax
\mciteBstWouldAddEndPuncttrue
\mciteSetBstMidEndSepPunct{\mcitedefaultmidpunct}
{\mcitedefaultendpunct}{\mcitedefaultseppunct}\relax
\EndOfBibitem
\bibitem[Zewail(2000)]{Zewail2000}
Zewail,~A.~H. Femtochemistry: Atomic-Scale Dynamics of the Chemical Bond.
  \emph{The Journal of Physical Chemistry A} \textbf{2000}, \emph{104},
  5660--5694\relax
\mciteBstWouldAddEndPuncttrue
\mciteSetBstMidEndSepPunct{\mcitedefaultmidpunct}
{\mcitedefaultendpunct}{\mcitedefaultseppunct}\relax
\EndOfBibitem
\bibitem[Zewail(2000)]{Zewail2000a}
Zewail,~A.~H. Femtochemistry: Atomic-Scale Dynamics of the Chemical Bond Using
  Ultrafast Lasers (Nobel Lecture). \emph{Angewandte Chemie International
  Edition} \textbf{2000}, \emph{39}, 2586--2631\relax
\mciteBstWouldAddEndPuncttrue
\mciteSetBstMidEndSepPunct{\mcitedefaultmidpunct}
{\mcitedefaultendpunct}{\mcitedefaultseppunct}\relax
\EndOfBibitem
\bibitem[Reimann \latin{et~al.}(2021)Reimann, Woerner, and
  Elsaesser]{Reimann2021}
Reimann,~K.; Woerner,~M.; Elsaesser,~T. Two-dimensional terahertz spectroscopy
  of condensed-phase molecular systems. \emph{The Journal of Chemical Physics}
  \textbf{2021}, \emph{154}, 120901\relax
\mciteBstWouldAddEndPuncttrue
\mciteSetBstMidEndSepPunct{\mcitedefaultmidpunct}
{\mcitedefaultendpunct}{\mcitedefaultseppunct}\relax
\EndOfBibitem
\bibitem[Hamm and Zanni(2011)Hamm, and Zanni]{Hamm2011}
Hamm,~P.; Zanni,~M. \emph{Concepts and Methods of 2D Infrared Spectroscopy};
  Cambridge University Press: Cambridge DB - Cambridge Core Cambridge
  University Press, 2011\relax
\mciteBstWouldAddEndPuncttrue
\mciteSetBstMidEndSepPunct{\mcitedefaultmidpunct}
{\mcitedefaultendpunct}{\mcitedefaultseppunct}\relax
\EndOfBibitem
\bibitem[Jonas(2003)]{Jonas2003a}
Jonas,~D.~M. Two-dimensional femtosecond spectroscopy. \emph{Annual Review of
  Physical Chemistry} \textbf{2003}, \emph{54}, 425--63\relax
\mciteBstWouldAddEndPuncttrue
\mciteSetBstMidEndSepPunct{\mcitedefaultmidpunct}
{\mcitedefaultendpunct}{\mcitedefaultseppunct}\relax
\EndOfBibitem
\bibitem[Schultz \latin{et~al.}(2015)Schultz, Schr\"oter, and Lee]{Schultz2015}
Schultz,~T.; Schr\"oter,~C.; Lee,~J.~C. Crazy for CRASY: Combining
  High-resolution and Ultrafast Measurements. \emph{Journal of the Korean
  Physical Society} \textbf{2015}, \emph{24}, 25--29\relax
\mciteBstWouldAddEndPuncttrue
\mciteSetBstMidEndSepPunct{\mcitedefaultmidpunct}
{\mcitedefaultendpunct}{\mcitedefaultseppunct}\relax
\EndOfBibitem
\bibitem[Flygare and Schmalz(1976)Flygare, and Schmalz]{Flygare1976}
Flygare,~W.~H.; Schmalz,~T.~G. Transient experiments and relaxation processes
  involving rotational states. \emph{Accounts of Chemical Research}
  \textbf{1976}, \emph{9}, 385--392\relax
\mciteBstWouldAddEndPuncttrue
\mciteSetBstMidEndSepPunct{\mcitedefaultmidpunct}
{\mcitedefaultendpunct}{\mcitedefaultseppunct}\relax
\EndOfBibitem
\bibitem[Grabow(2011)]{Grabow2011}
Grabow,~J. In \emph{Handbook of High-Resolution Spectroscopy}; Quack,~M.,
  Merkt,~F., Eds.; John Wiley \& Sons, Ltd: Chichester, UK, 2011; Vol.~2; pp
  723--799\relax
\mciteBstWouldAddEndPuncttrue
\mciteSetBstMidEndSepPunct{\mcitedefaultmidpunct}
{\mcitedefaultendpunct}{\mcitedefaultseppunct}\relax
\EndOfBibitem
\bibitem[Shipman and Pate(2011)Shipman, and Pate]{Shipman2011}
Shipman,~S.~T.; Pate,~B.~H. \emph{Handbook of High-Resolution Spectroscopy};
  John Wiley \& Sons, Ltd: Chichester, UK, 2011; Vol.~2; pp 801--828\relax
\mciteBstWouldAddEndPuncttrue
\mciteSetBstMidEndSepPunct{\mcitedefaultmidpunct}
{\mcitedefaultendpunct}{\mcitedefaultseppunct}\relax
\EndOfBibitem
\bibitem[Park and Field(2016)Park, and Field]{Park2016}
Park,~G.~B.; Field,~R.~W. Perspective: The first ten years of broadband chirped
  pulse Fourier transform microwave spectroscopy. \emph{The Journal of Chemical
  Physics J. Chem. Phys.} \textbf{2016}, \emph{144}, 200901\relax
\mciteBstWouldAddEndPuncttrue
\mciteSetBstMidEndSepPunct{\mcitedefaultmidpunct}
{\mcitedefaultendpunct}{\mcitedefaultseppunct}\relax
\EndOfBibitem
\bibitem[Heritage \latin{et~al.}(1975)Heritage, Gustafson, and
  Lin]{Heritage1975}
Heritage,~J.~P.; Gustafson,~T.~K.; Lin,~C.~H. Observation of Coherent Transient
  Birefringence in Cs2 Vapor. \emph{Physical Review Letters} \textbf{1975},
  \emph{34}, 1299--1302\relax
\mciteBstWouldAddEndPuncttrue
\mciteSetBstMidEndSepPunct{\mcitedefaultmidpunct}
{\mcitedefaultendpunct}{\mcitedefaultseppunct}\relax
\EndOfBibitem
\bibitem[Baskin \latin{et~al.}(1986)Baskin, Felker, and Zewail]{Baskin1986}
Baskin,~J.~S.; Felker,~P.~M.; Zewail,~A.~H. Doppler-free time-resolved
  polarization spectroscopy of large molecules: Measurement of excited state
  rotational constants. \emph{J. Chem. Phys.} \textbf{1986}, \emph{84},
  4708--4710\relax
\mciteBstWouldAddEndPuncttrue
\mciteSetBstMidEndSepPunct{\mcitedefaultmidpunct}
{\mcitedefaultendpunct}{\mcitedefaultseppunct}\relax
\EndOfBibitem
\bibitem[Felker \latin{et~al.}(1986)Felker, Baskin, and Zewail]{Felker1986}
Felker,~P.~M.; Baskin,~J.~S.; Zewail,~A.~H. Rephasing of Collisionless
  Molecular Rotational Coherence in Large Molecules. \emph{Journal of Physical
  Chemistry} \textbf{1986}, \emph{90}, 724--728\relax
\mciteBstWouldAddEndPuncttrue
\mciteSetBstMidEndSepPunct{\mcitedefaultmidpunct}
{\mcitedefaultendpunct}{\mcitedefaultseppunct}\relax
\EndOfBibitem
\bibitem[Baskin \latin{et~al.}(1987)Baskin, Felker, and Zewail]{Baskin1987}
Baskin,~J.~S.; Felker,~P.~M.; Zewail,~A.~H. Purely Rotational Coherence Effect
  and Time-Resolved Sub-Doppler Spectroscopy of Large Molecules .2.
  Experimental. \emph{Journal of Chemical Physics} \textbf{1987}, \emph{86},
  2483--2499\relax
\mciteBstWouldAddEndPuncttrue
\mciteSetBstMidEndSepPunct{\mcitedefaultmidpunct}
{\mcitedefaultendpunct}{\mcitedefaultseppunct}\relax
\EndOfBibitem
\bibitem[Felker and Zewail(1987)Felker, and Zewail]{Felker1987}
Felker,~P.~M.; Zewail,~A.~H. Purely Rotational Coherence Effect and
  Time-Resolved Sub-Doppler Spectroscopy of Large Molecules .1. Theoretical.
  \emph{Journal of Chemical Physics} \textbf{1987}, \emph{86}, 2460--2482\relax
\mciteBstWouldAddEndPuncttrue
\mciteSetBstMidEndSepPunct{\mcitedefaultmidpunct}
{\mcitedefaultendpunct}{\mcitedefaultseppunct}\relax
\EndOfBibitem
\bibitem[Felker(1992)]{Felker1992}
Felker,~P.~M. Rotational Coherence Spectroscopy - Studies of the Geometries of
  Large Gas-Phase Species by Picosecond Time-Domain Methods. \emph{Journal of
  Physical Chemistry} \textbf{1992}, \emph{96}, 7844--7857\relax
\mciteBstWouldAddEndPuncttrue
\mciteSetBstMidEndSepPunct{\mcitedefaultmidpunct}
{\mcitedefaultendpunct}{\mcitedefaultseppunct}\relax
\EndOfBibitem
\bibitem[Frey \latin{et~al.}(2011)Frey, Kummli, Lobsiger, and
  Leutwyler]{Frey2011}
Frey,~H.-M.; Kummli,~D.; Lobsiger,~S.; Leutwyler,~S. In \emph{Handbook of
  High-Resolution Spectroscopy}; Quack,~M., Merkt,~F., Eds.; John Wiley \&
  Sons, Ltd: Chichester, UK, 2011; Vol.~2; pp 1237--1265\relax
\mciteBstWouldAddEndPuncttrue
\mciteSetBstMidEndSepPunct{\mcitedefaultmidpunct}
{\mcitedefaultendpunct}{\mcitedefaultseppunct}\relax
\EndOfBibitem
\bibitem[Riehn \latin{et~al.}(2000)Riehn, Degen, Weichert, Bolte, Egert,
  Brutschy, Tarakeshwar, and Kim]{Riehn2000}
Riehn,~C.; Degen,~A.; Weichert,~A.; Bolte,~M.; Egert,~E.; Brutschy,~B.;
  Tarakeshwar,~P.; Kim,~K.~S. Molecular structure of p-cyclohexylaniline.
  Comparison of results obtained by X-ray diffraction with gas phase laser
  experiments and ab initio calculations. \emph{Journal of Physical Chemistry
  A} \textbf{2000}, \emph{104}, 11593--11600\relax
\mciteBstWouldAddEndPuncttrue
\mciteSetBstMidEndSepPunct{\mcitedefaultmidpunct}
{\mcitedefaultendpunct}{\mcitedefaultseppunct}\relax
\EndOfBibitem
\bibitem[Riehn(2002)]{Riehn2002}
Riehn,~C. High-resolution pump-probe rotational coherencespectroscopy -
  rotational constants and structure of ground and electronically excited
  states of large molecular systems. \emph{Chemical Physics} \textbf{2002},
  \emph{283}, 297--329\relax
\mciteBstWouldAddEndPuncttrue
\mciteSetBstMidEndSepPunct{\mcitedefaultmidpunct}
{\mcitedefaultendpunct}{\mcitedefaultseppunct}\relax
\EndOfBibitem
\bibitem[Frey \latin{et~al.}(2004)Frey, Kummli, Keller, Leist, and
  Leutwyler]{Frey2004}
Frey,~H.~M.; Kummli,~D.; Keller,~M.; Leist,~R.; Leutwyler,~S.
  \emph{Femtochemistry and Femtobiology: Ultrafast Events in Molecular
  Science}; 2004; pp 261--264\relax
\mciteBstWouldAddEndPuncttrue
\mciteSetBstMidEndSepPunct{\mcitedefaultmidpunct}
{\mcitedefaultendpunct}{\mcitedefaultseppunct}\relax
\EndOfBibitem
\bibitem[Rouz\'{e}e \latin{et~al.}(2006)Rouz\'{e}e, Guerin, Boudon, Lavorel,
  and Faucher]{Rouzee2006}
Rouz\'{e}e,~A.; Guerin,~S.; Boudon,~V.; Lavorel,~B.; Faucher,~O. Field-free
  one-dimensional alignment of ethylene molecule. \emph{Physical Review A}
  \textbf{2006}, \emph{73}, 033418\relax
\mciteBstWouldAddEndPuncttrue
\mciteSetBstMidEndSepPunct{\mcitedefaultmidpunct}
{\mcitedefaultendpunct}{\mcitedefaultseppunct}\relax
\EndOfBibitem
\bibitem[Stapelfeldt and Seideman(2003)Stapelfeldt, and
  Seideman]{Stapelfeldt2003}
Stapelfeldt,~H.; Seideman,~T. Colloquium: Aligning molecules with strong laser
  pulses. \emph{Reviews of Modern Physics} \textbf{2003}, \emph{75}, 543 --
  557\relax
\mciteBstWouldAddEndPuncttrue
\mciteSetBstMidEndSepPunct{\mcitedefaultmidpunct}
{\mcitedefaultendpunct}{\mcitedefaultseppunct}\relax
\EndOfBibitem
\bibitem[Peronne \latin{et~al.}(2003)Peronne, Poulsen, Bisgaard, Stapelfeldt,
  and Seideman]{Peronne2003}
Peronne,~E.; Poulsen,~M.~D.; Bisgaard,~C.~Z.; Stapelfeldt,~H.; Seideman,~T.
  Nonadiabatic alignment of asymmetric top molecules: Field-free alignment of
  iodobenzene. \emph{Physical Review Letters} \textbf{2003}, \emph{91}\relax
\mciteBstWouldAddEndPuncttrue
\mciteSetBstMidEndSepPunct{\mcitedefaultmidpunct}
{\mcitedefaultendpunct}{\mcitedefaultseppunct}\relax
\EndOfBibitem
\bibitem[Peronne \latin{et~al.}(2004)Peronne, Poulsen, Stapelfeldt, Bisgaard,
  Hamilton, and Seideman]{Peronne2004}
Peronne,~E.; Poulsen,~M.~D.; Stapelfeldt,~H.; Bisgaard,~C.~Z.; Hamilton,~E.;
  Seideman,~T. Nonadiabatic laser-induced alignment of iodobenzene molecules.
  \emph{Physical Review A} \textbf{2004}, \emph{70}\relax
\mciteBstWouldAddEndPuncttrue
\mciteSetBstMidEndSepPunct{\mcitedefaultmidpunct}
{\mcitedefaultendpunct}{\mcitedefaultseppunct}\relax
\EndOfBibitem
\bibitem[Poulsen \latin{et~al.}(2004)Poulsen, Peronne, Stapelfeldt, Bisgaard,
  Viftrup, Hamilton, and Seideman]{Poulsen2004}
Poulsen,~M.~D.; Peronne,~E.; Stapelfeldt,~H.; Bisgaard,~C.~Z.; Viftrup,~S.~S.;
  Hamilton,~E.; Seideman,~T. Nonadiabatic alignment of asymmetric top
  molecules: Rotational revivals. \emph{Journal of Chemical Physics}
  \textbf{2004}, \emph{121}, 783--791\relax
\mciteBstWouldAddEndPuncttrue
\mciteSetBstMidEndSepPunct{\mcitedefaultmidpunct}
{\mcitedefaultendpunct}{\mcitedefaultseppunct}\relax
\EndOfBibitem
\bibitem[Hamilton \latin{et~al.}(2005)Hamilton, Seideman, Ejdrup, Poulsen,
  Bisgaard, Viftrup, and Stapelfeldt]{Hamilton2005}
Hamilton,~E.; Seideman,~T.; Ejdrup,~T.; Poulsen,~M.~D.; Bisgaard,~C.~Z.;
  Viftrup,~S.~S.; Stapelfeldt,~H. Alignment of symmetric top molecules by short
  laser pulses. \emph{Physical Review A} \textbf{2005}, \emph{72}\relax
\mciteBstWouldAddEndPuncttrue
\mciteSetBstMidEndSepPunct{\mcitedefaultmidpunct}
{\mcitedefaultendpunct}{\mcitedefaultseppunct}\relax
\EndOfBibitem
\bibitem[Holmegaard \latin{et~al.}(2007)Holmegaard, Viftrup, Kumarappan,
  Bisgaard, Stapelfeldt, Hamilton, and Seideman]{Holmegaard2007}
Holmegaard,~L.; Viftrup,~S.~S.; Kumarappan,~V.; Bisgaard,~C.~Z.;
  Stapelfeldt,~H.; Hamilton,~E.; Seideman,~T. Control of rotational wave-packet
  dynamics in asymmetric top molecules. \emph{Physical Review A} \textbf{2007},
  \emph{75}, 051403\relax
\mciteBstWouldAddEndPuncttrue
\mciteSetBstMidEndSepPunct{\mcitedefaultmidpunct}
{\mcitedefaultendpunct}{\mcitedefaultseppunct}\relax
\EndOfBibitem
\bibitem[Viftrup \latin{et~al.}(2007)Viftrup, Kumarappan, Trippel, Stapelfeldt,
  Hamilton, and Seideman]{Viftrup2007}
Viftrup,~S.~S.; Kumarappan,~V.; Trippel,~S.; Stapelfeldt,~H.; Hamilton,~E.;
  Seideman,~T. Holding and spinning molecules in space. \emph{Physical Review
  Letters} \textbf{2007}, \emph{99}\relax
\mciteBstWouldAddEndPuncttrue
\mciteSetBstMidEndSepPunct{\mcitedefaultmidpunct}
{\mcitedefaultendpunct}{\mcitedefaultseppunct}\relax
\EndOfBibitem
\bibitem[Wu \latin{et~al.}(2011)Wu, Vredenborg, Ulrich, Schmidt, Meckel, Voss,
  Sann, Kim, Jahnke, and D\"orner]{Wu2011}
Wu,~J.; Vredenborg,~A.; Ulrich,~B.; Schmidt,~L. P.~H.; Meckel,~M.; Voss,~S.;
  Sann,~H.; Kim,~H.; Jahnke,~T.; D\"orner,~R. Nonadiabatic alignment of van der
  Waals--force-bound argon dimers by femtosecond laser pulses. \emph{Physical
  Review A} \textbf{2011}, \emph{83}, 061403\relax
\mciteBstWouldAddEndPuncttrue
\mciteSetBstMidEndSepPunct{\mcitedefaultmidpunct}
{\mcitedefaultendpunct}{\mcitedefaultseppunct}\relax
\EndOfBibitem
\bibitem[Veltheim \latin{et~al.}(2014)Veltheim, Borchers, Steinmeyer, and
  Rottke]{Veltheim2014}
Veltheim,~A.~v.; Borchers,~B.; Steinmeyer,~G.; Rottke,~H. Imaging the impulsive
  alignment of noble-gas dimers via Coulomb explosion. \emph{Physical Review A}
  \textbf{2014}, \emph{89}, 023432\relax
\mciteBstWouldAddEndPuncttrue
\mciteSetBstMidEndSepPunct{\mcitedefaultmidpunct}
{\mcitedefaultendpunct}{\mcitedefaultseppunct}\relax
\EndOfBibitem
\bibitem[Mizuse \latin{et~al.}(2022)Mizuse, Sato, Tobata, and
  Ohshima]{Mizuse2022}
Mizuse,~K.; Sato,~U.; Tobata,~Y.; Ohshima,~Y. Rotational spectroscopy of the
  argon dimer by time-resolved Coulomb explosion imaging of rotational wave
  packets. \emph{Phys. Chem. Chem. Phys.} \textbf{2022}, \emph{24},
  11014--11022\relax
\mciteBstWouldAddEndPuncttrue
\mciteSetBstMidEndSepPunct{\mcitedefaultmidpunct}
{\mcitedefaultendpunct}{\mcitedefaultseppunct}\relax
\EndOfBibitem
\bibitem[Galinis \latin{et~al.}(2014)Galinis, Cacho, Chapman, Ellis, Lewerenz,
  Mendoza~Luna, Minns, Mladenović, Rouzée, Springate, Turcu, Watkins, and von
  Haeften]{Galinis2014}
Galinis,~G.; Cacho,~C.; Chapman,~R.~T.; Ellis,~A.~M.; Lewerenz,~M.;
  Mendoza~Luna,~L.~G.; Minns,~R.~S.; Mladenović,~M.; Rouzée,~A.;
  Springate,~E.; Turcu,~I.~E.; Watkins,~M.~J.; von Haeften,~K. Probing the
  Structure and Dynamics of Molecular Clusters Using Rotational Wave Packets.
  \emph{Physical Review Letters} \textbf{2014}, \emph{113}, 043004\relax
\mciteBstWouldAddEndPuncttrue
\mciteSetBstMidEndSepPunct{\mcitedefaultmidpunct}
{\mcitedefaultendpunct}{\mcitedefaultseppunct}\relax
\EndOfBibitem
\bibitem[Chatterley \latin{et~al.}(2020)Chatterley, Baatrup, Schouder, and
  Stapelfeldt]{Chatterley2020}
Chatterley,~A.~S.; Baatrup,~M.~O.; Schouder,~C.~A.; Stapelfeldt,~H.
  Laser-induced alignment dynamics of gas phase CS2 dimers. \emph{Phys. Chem.
  Chem. Phys.} \textbf{2020}, \emph{22}, 3245--3253\relax
\mciteBstWouldAddEndPuncttrue
\mciteSetBstMidEndSepPunct{\mcitedefaultmidpunct}
{\mcitedefaultendpunct}{\mcitedefaultseppunct}\relax
\EndOfBibitem
\bibitem[Chatterley \latin{et~al.}(2019)Chatterley, Schouder, Christiansen,
  Shepperson, Rasmussen, and Stapelfeldt]{Chatterley2019}
Chatterley,~A.~S.; Schouder,~C.; Christiansen,~L.; Shepperson,~B.;
  Rasmussen,~M.~H.; Stapelfeldt,~H. Long-lasting field-free alignment of large
  molecules inside helium nanodroplets. \emph{Nature Communications}
  \textbf{2019}, \emph{10}, 133\relax
\mciteBstWouldAddEndPuncttrue
\mciteSetBstMidEndSepPunct{\mcitedefaultmidpunct}
{\mcitedefaultendpunct}{\mcitedefaultseppunct}\relax
\EndOfBibitem
\bibitem[Chatterley \latin{et~al.}(2020)Chatterley, Christiansen, Schouder,
  Jørgensen, Shepperson, Cherepanov, Bighin, Zillich, Lemeshko, and
  Stapelfeldt]{Chatterley2020a}
Chatterley,~A.~S.; Christiansen,~L.; Schouder,~C.~A.; Jørgensen,~A.~V.;
  Shepperson,~B.; Cherepanov,~I.~N.; Bighin,~G.; Zillich,~R.~E.; Lemeshko,~M.;
  Stapelfeldt,~H. Rotational Coherence Spectroscopy of Molecules in Helium
  Nanodroplets: Reconciling the Time and the Frequency Domains. \emph{Physical
  Review Letters} \textbf{2020}, \emph{125}, 013001\relax
\mciteBstWouldAddEndPuncttrue
\mciteSetBstMidEndSepPunct{\mcitedefaultmidpunct}
{\mcitedefaultendpunct}{\mcitedefaultseppunct}\relax
\EndOfBibitem
\bibitem[Schouder \latin{et~al.}(2021)Schouder, Chatterley, Johny,
  H\"ubschmann, Al-Refaie, Calvo, K\"upper, and Stapelfeldt]{Schouder2021}
Schouder,~C.; Chatterley,~A.~S.; Johny,~M.; H\"ubschmann,~F.; Al-Refaie,~A.~F.;
  Calvo,~F.; K\"upper,~J.; Stapelfeldt,~H. Laser-induced Coulomb explosion
  imaging of (C$_6$H$_5$Br)$_2$ and C$_6$H$_5$Br-I$_2$ dimers in helium
  nanodroplets using a Tpx3Cam. \emph{Journal of Physics B: Atomic, Molecular
  and Optical Physics} \textbf{2021}, \emph{54}, 184001\relax
\mciteBstWouldAddEndPuncttrue
\mciteSetBstMidEndSepPunct{\mcitedefaultmidpunct}
{\mcitedefaultendpunct}{\mcitedefaultseppunct}\relax
\EndOfBibitem
\bibitem[Schouder \latin{et~al.}(2022)Schouder, Chatterley, Pickering, and
  Stapelfeldt]{Schouder2022}
Schouder,~C.~A.; Chatterley,~A.~S.; Pickering,~J.~D.; Stapelfeldt,~H.
  Laser-Induced Coulomb Explosion Imaging of Aligned Molecules and Molecular
  Dimers. \emph{Annu. Rev. Phys. Chem.} \textbf{2022}, \emph{73},
  323--347\relax
\mciteBstWouldAddEndPuncttrue
\mciteSetBstMidEndSepPunct{\mcitedefaultmidpunct}
{\mcitedefaultendpunct}{\mcitedefaultseppunct}\relax
\EndOfBibitem
\bibitem[Cho(2008)]{Cho2008}
Cho,~M. Coherent Two-Dimensional Optical Spectroscopy. \emph{Chem. Rev.}
  \textbf{2008}, \emph{108}, 1331--1418\relax
\mciteBstWouldAddEndPuncttrue
\mciteSetBstMidEndSepPunct{\mcitedefaultmidpunct}
{\mcitedefaultendpunct}{\mcitedefaultseppunct}\relax
\EndOfBibitem
\bibitem[Fritzsch \latin{et~al.}(2020)Fritzsch, Hume, Minnes, Baker, Burley,
  and Hunt]{Fritzsch2020}
Fritzsch,~R.; Hume,~S.; Minnes,~L.; Baker,~M.~J.; Burley,~G.~A.; Hunt,~N.~T.
  Two-dimensional infrared spectroscopy: an emerging analytical tool?
  \emph{Analyst} \textbf{2020}, \emph{145}, 2014--2024\relax
\mciteBstWouldAddEndPuncttrue
\mciteSetBstMidEndSepPunct{\mcitedefaultmidpunct}
{\mcitedefaultendpunct}{\mcitedefaultseppunct}\relax
\EndOfBibitem
\bibitem[Tollerud and Davis(2017)Tollerud, and Davis]{Tollerud2017}
Tollerud,~J.~O.; Davis,~J.~A. Coherent multi-dimensional spectroscopy:
  Experimental considerations, direct comparisons and new capabilities.
  \emph{Progress in Quantum Electronics} \textbf{2017}, \emph{55}, 1--34\relax
\mciteBstWouldAddEndPuncttrue
\mciteSetBstMidEndSepPunct{\mcitedefaultmidpunct}
{\mcitedefaultendpunct}{\mcitedefaultseppunct}\relax
\EndOfBibitem
\bibitem[de~Vries and Hobza(2007)de~Vries, and Hobza]{deVries2007}
de~Vries,~M.; Hobza,~P. Gas-Phase Spectroscopy of Biomolecular Building Blocks.
  \emph{Annual review of physical chemistry} \textbf{2007}, \emph{58},
  585--612\relax
\mciteBstWouldAddEndPuncttrue
\mciteSetBstMidEndSepPunct{\mcitedefaultmidpunct}
{\mcitedefaultendpunct}{\mcitedefaultseppunct}\relax
\EndOfBibitem
\bibitem[Fang \latin{et~al.}(2015)Fang, Ivanisevic, Benton, Johnson, Patti,
  Hoang, Uritboonthai, Kurczy, and Siuzdak]{Fang2015_GCMS}
Fang,~M.; Ivanisevic,~J.; Benton,~H.~P.; Johnson,~C.~H.; Patti,~G.~J.;
  Hoang,~L.~T.; Uritboonthai,~W.; Kurczy,~M.~E.; Siuzdak,~G. Thermal
  Degradation of Small Molecules: A Global Metabolomic Investigation.
  \emph{Anal. Chem.} \textbf{2015}, \emph{87}, 10935--10941\relax
\mciteBstWouldAddEndPuncttrue
\mciteSetBstMidEndSepPunct{\mcitedefaultmidpunct}
{\mcitedefaultendpunct}{\mcitedefaultseppunct}\relax
\EndOfBibitem
\bibitem[Glish and Burinsky(2008)Glish, and Burinsky]{Glish2008}
Glish,~G.~L.; Burinsky,~D.~J. Hybrid mass spectrometers for tandem mass
  spectrometry. \emph{Journal of the American Society for Mass Spectrometry}
  \textbf{2008}, \emph{19}, 161--172\relax
\mciteBstWouldAddEndPuncttrue
\mciteSetBstMidEndSepPunct{\mcitedefaultmidpunct}
{\mcitedefaultendpunct}{\mcitedefaultseppunct}\relax
\EndOfBibitem
\bibitem[van Agthoven \latin{et~al.}(2019)van Agthoven, Lam, O'Connor, Rolando,
  and Delsuc]{Agthoven2019}
van Agthoven,~M.~A.; Lam,~Y. P.~Y.; O'Connor,~P.~B.; Rolando,~C.; Delsuc,~M.-A.
  Two-dimensional mass spectrometry: new perspectives for tandem mass
  spectrometry. \emph{European Biophysics Journal} \textbf{2019}, \emph{48},
  213--229\relax
\mciteBstWouldAddEndPuncttrue
\mciteSetBstMidEndSepPunct{\mcitedefaultmidpunct}
{\mcitedefaultendpunct}{\mcitedefaultseppunct}\relax
\EndOfBibitem
\bibitem[Schultz and Vrakking(2014)Schultz, and Vrakking]{Schultz2014}
Schultz,~T.; Vrakking,~M.~E. \emph{Attosecond and XUV Physics}; Wiley-VCH,
  2014\relax
\mciteBstWouldAddEndPuncttrue
\mciteSetBstMidEndSepPunct{\mcitedefaultmidpunct}
{\mcitedefaultendpunct}{\mcitedefaultseppunct}\relax
\EndOfBibitem
\bibitem[Schroter \latin{et~al.}(2011)Schroter, Kosma, and
  Schultz]{Schroter2011}
Schroter,~C.; Kosma,~K.; Schultz,~T. CRASY: Mass- or Electron-Correlated
  Rotational Alignment Spectroscopy. \emph{Science} \textbf{2011}, \emph{333},
  1011--1015\relax
\mciteBstWouldAddEndPuncttrue
\mciteSetBstMidEndSepPunct{\mcitedefaultmidpunct}
{\mcitedefaultendpunct}{\mcitedefaultseppunct}\relax
\EndOfBibitem
\bibitem[Bartholdi and Ernst(1973)Bartholdi, and Ernst]{Bartholdi1973}
Bartholdi,~E.; Ernst,~R. Fourier spectroscopy and the causality principle.
  \emph{Journal of Magnetic Resonance (1969)} \textbf{1973}, \emph{11},
  9--19\relax
\mciteBstWouldAddEndPuncttrue
\mciteSetBstMidEndSepPunct{\mcitedefaultmidpunct}
{\mcitedefaultendpunct}{\mcitedefaultseppunct}\relax
\EndOfBibitem
\bibitem[Gordy and Cook(1984)Gordy, and Cook]{Gordy1984}
Gordy,~W.; Cook,~R. \emph{Microwave Molecular Spectra}; JOHN WILEY \& SONS: New
  York, 1984\relax
\mciteBstWouldAddEndPuncttrue
\mciteSetBstMidEndSepPunct{\mcitedefaultmidpunct}
{\mcitedefaultendpunct}{\mcitedefaultseppunct}\relax
\EndOfBibitem
\bibitem[Western(2010)]{Western2010}
Western,~C. PGOPHER - a Program for Simulating Rotational Structure. 2010\relax
\mciteBstWouldAddEndPuncttrue
\mciteSetBstMidEndSepPunct{\mcitedefaultmidpunct}
{\mcitedefaultendpunct}{\mcitedefaultseppunct}\relax
\EndOfBibitem
\bibitem[Western and Billinghurst(2017)Western, and Billinghurst]{Western2017}
Western,~C.~M.; Billinghurst,~B.~E. Automatic assignment and fitting of spectra
  with PGOPHER. \emph{Physical Chemistry Chemical Physics} \textbf{2017},
  \emph{19}, 10222--10226\relax
\mciteBstWouldAddEndPuncttrue
\mciteSetBstMidEndSepPunct{\mcitedefaultmidpunct}
{\mcitedefaultendpunct}{\mcitedefaultseppunct}\relax
\EndOfBibitem
\bibitem[Press \latin{et~al.}(1992)Press, Teukolsky, Vetterling, and
  Flannery]{NumericalRecipes}
Press,~W.~H.; Teukolsky,~S.~A.; Vetterling,~W.~T.; Flannery,~B.~P.
  \emph{Numerical Recipes in C}, 2nd ed.; Cambridge University Press:
  Cambridge, USA, 1992\relax
\mciteBstWouldAddEndPuncttrue
\mciteSetBstMidEndSepPunct{\mcitedefaultmidpunct}
{\mcitedefaultendpunct}{\mcitedefaultseppunct}\relax
\EndOfBibitem
\bibitem[Gailly and Adler()Gailly, and Adler]{ZLIB}
Gailly,~J.; Adler,~M. ZLIB - A Massively Spiffy Yet Delicately Unobtrusive
  Compression Library. \url{http://zlib.net/}, Accessed: 2022-06-13\relax
\mciteBstWouldAddEndPuncttrue
\mciteSetBstMidEndSepPunct{\mcitedefaultmidpunct}
{\mcitedefaultendpunct}{\mcitedefaultseppunct}\relax
\EndOfBibitem
\bibitem[Schultz(2018)]{Figshare_CrasyDataAnalysis}
Schultz,~T. Crasy Data and Analysis.
  \url{https://doi.org/10.6084/m9.figshare.5886406.v1}, 2018\relax
\mciteBstWouldAddEndPuncttrue
\mciteSetBstMidEndSepPunct{\mcitedefaultmidpunct}
{\mcitedefaultendpunct}{\mcitedefaultseppunct}\relax
\EndOfBibitem
\bibitem[Michael and E.(2011)Michael, and E.]{Michael2011}
Michael,~B.; E.,~W.~M. Isotopic compositions of the elements 2009 (IUPAC
  Technical Report). \emph{Pure and Applied Chemistry} \textbf{2011},
  \emph{83}, 397\relax
\mciteBstWouldAddEndPuncttrue
\mciteSetBstMidEndSepPunct{\mcitedefaultmidpunct}
{\mcitedefaultendpunct}{\mcitedefaultseppunct}\relax
\EndOfBibitem
\bibitem[Fleming \latin{et~al.}(2014)Fleming, Manz, Sato, and
  Takayanagi]{Fleming2014}
Fleming,~D.~G.; Manz,~J.; Sato,~K.; Takayanagi,~T. Fundamental Change in the
  Nature of Chemical Bonding by Isotopic Substitution. \emph{Angewandte Chemie
  International Edition} \textbf{2014}, \emph{53}, 13706--13709\relax
\mciteBstWouldAddEndPuncttrue
\mciteSetBstMidEndSepPunct{\mcitedefaultmidpunct}
{\mcitedefaultendpunct}{\mcitedefaultseppunct}\relax
\EndOfBibitem
\bibitem[Wolfsberg(1969)]{Wolfsberg1969}
Wolfsberg,~M. Isotope effects. \emph{Annual Review of Physical Chemistry}
  \textbf{1969}, \emph{20}, 449--478\relax
\mciteBstWouldAddEndPuncttrue
\mciteSetBstMidEndSepPunct{\mcitedefaultmidpunct}
{\mcitedefaultendpunct}{\mcitedefaultseppunct}\relax
\EndOfBibitem
\bibitem[Cheng \latin{et~al.}(1996)Cheng, Hardwick, and Dyke]{Cheng1996}
Cheng,~C.-L.~C.; Hardwick,~J.~L.; Dyke,~T.~R. High-Resolution
  Vibration-Rotation Spectroscopy of $^{12}C^{34}S_2$ and $^{13}C^{34}S_2$ at
  400 cm$^{-1}$. \emph{Journal of Molecular Spectroscopy} \textbf{1996},
  \emph{179}, 205--218\relax
\mciteBstWouldAddEndPuncttrue
\mciteSetBstMidEndSepPunct{\mcitedefaultmidpunct}
{\mcitedefaultendpunct}{\mcitedefaultseppunct}\relax
\EndOfBibitem
\bibitem[Ahonen \latin{et~al.}(1997)Ahonen, Alanko, Horneman, Koivusaari, Paso,
  Tolonen, and Anttila]{Ahonen1997}
Ahonen,~T.; Alanko,~S.; Horneman,~V.-M.; Koivusaari,~M.; Paso,~R.;
  Tolonen,~A.-M.; Anttila,~R. A Long Path Cell for the Fourier Spectrometer
  Bruker IFS 120 HR: Application to the Weak $\nu_1 + \nu_2$ and $3\nu_2$ Bands
  of Carbon Disulfide. \emph{Journal of Molecular Spectroscopy} \textbf{1997},
  \emph{181}, 279--286\relax
\mciteBstWouldAddEndPuncttrue
\mciteSetBstMidEndSepPunct{\mcitedefaultmidpunct}
{\mcitedefaultendpunct}{\mcitedefaultseppunct}\relax
\EndOfBibitem
\bibitem[Winther \latin{et~al.}(1988)Winther, Heyne, and
  Guarnieri]{Winther1988}
Winther,~F.; Heyne,~U.; Guarnieri,~A. The Infrared Spectrum of CS$_2$ in the
  $\nu_3$ Band Region. \emph{Z. Naturforsch} \textbf{1988}, \emph{43a},
  215--218\relax
\mciteBstWouldAddEndPuncttrue
\mciteSetBstMidEndSepPunct{\mcitedefaultmidpunct}
{\mcitedefaultendpunct}{\mcitedefaultseppunct}\relax
\EndOfBibitem
\bibitem[Horneman \latin{et~al.}(2005)Horneman, Anttila, Alanko, and
  Pietila]{Horneman2005}
Horneman,~V.~M.; Anttila,~R.; Alanko,~S.; Pietila,~J. Transferring calibration
  from CO2 laser lines to far infrared water lines with the aid of the v(2)
  band of OCS and the v(2), v(1)-v(2), and v(1)+v(2) bands of (CS2)-C-13:
  Molecular constants of (CS2)-C-13. \emph{Journal of Molecular Spectroscopy}
  \textbf{2005}, \emph{234}, 238--254\relax
\mciteBstWouldAddEndPuncttrue
\mciteSetBstMidEndSepPunct{\mcitedefaultmidpunct}
{\mcitedefaultendpunct}{\mcitedefaultseppunct}\relax
\EndOfBibitem
\bibitem[Schr\"oter \latin{et~al.}(2018)Schr\"oter, Lee, and
  Schultz]{Schroter2018}
Schr\"oter,~C.; Lee,~J.~C.; Schultz,~T. Mass-correlated rotational Raman
  spectra with high resolution, broad bandwidth, and absolute frequency
  accuracy. \emph{Proceedings of the National Academy of Sciences}
  \textbf{2018}, \emph{115}, 5072 -- 5076\relax
\mciteBstWouldAddEndPuncttrue
\mciteSetBstMidEndSepPunct{\mcitedefaultmidpunct}
{\mcitedefaultendpunct}{\mcitedefaultseppunct}\relax
\EndOfBibitem
\bibitem[Schroter \latin{et~al.}(2015)Schroter, Choi, and
  Schultz]{Schroter2015}
Schroter,~C.; Choi,~C.~M.; Schultz,~T. CRASY: Correlated Rotational Alignment
  Spectroscopy Reveals Atomic Scrambling in Ionic States of Butadiene.
  \emph{Journal of Physical Chemistry A} \textbf{2015}, \emph{119},
  1309--1314\relax
\mciteBstWouldAddEndPuncttrue
\mciteSetBstMidEndSepPunct{\mcitedefaultmidpunct}
{\mcitedefaultendpunct}{\mcitedefaultseppunct}\relax
\EndOfBibitem
\bibitem[\"Ozer \latin{et~al.}(2020)\"Ozer, Heo, Lee, Schroter, and
  Schultz]{Ozer2020}
\"Ozer,~B.~R.; Heo,~I.; Lee,~J.~C.; Schroter,~C.; Schultz,~T. De novo structure
  determination of butadiene by isotope-resolved rotational Raman spectroscopy.
  \emph{Phys. Chem. Chem. Phys.} \textbf{2020}, \relax
\mciteBstWouldAddEndPunctfalse
\mciteSetBstMidEndSepPunct{\mcitedefaultmidpunct}
{}{\mcitedefaultseppunct}\relax
\EndOfBibitem
\bibitem[Heo \latin{et~al.}(2022)Heo, Lee, \"Ozer, and Schultz]{Heo2022a}
Heo,~I.; Lee,~J.~C.; \"Ozer,~B.~R.; Schultz,~T. Structure of benzene from
  mass-correlated rotational Raman spectroscopy. \emph{RSC Adv.} \textbf{2022},
  \emph{12}, 21406--21416\relax
\mciteBstWouldAddEndPuncttrue
\mciteSetBstMidEndSepPunct{\mcitedefaultmidpunct}
{\mcitedefaultendpunct}{\mcitedefaultseppunct}\relax
\EndOfBibitem
\bibitem[Heo \latin{et~al.}(2022)Heo, Lee, \"Ozer, and Schultz]{Heo2022b}
Heo,~I.; Lee,~J.~C.; \"Ozer,~B.~R.; Schultz,~T. Mass-Correlated High-Resolution
  Spectra and the Structure of Benzene. \emph{J. Phys. Chem. Lett.}
  \textbf{2022}, \emph{13}, 8278--8283\relax
\mciteBstWouldAddEndPuncttrue
\mciteSetBstMidEndSepPunct{\mcitedefaultmidpunct}
{\mcitedefaultendpunct}{\mcitedefaultseppunct}\relax
\EndOfBibitem
\bibitem[Rukiye~\"Ozer \latin{et~al.}(2024)Rukiye~\"Ozer, Chan~Lee, Heo, and
  Schultz]{Ozer2024}
Rukiye~\"Ozer,~B.; Chan~Lee,~J.; Heo,~I.; Schultz,~T. Mass-correlated
  rotational Raman spectra and the structure of furan. \emph{Journal of Raman
  Spectroscopy} \textbf{2024}, \emph{in print}\relax
\mciteBstWouldAddEndPuncttrue
\mciteSetBstMidEndSepPunct{\mcitedefaultmidpunct}
{\mcitedefaultendpunct}{\mcitedefaultseppunct}\relax
\EndOfBibitem
\bibitem[Zinn \latin{et~al.}(2015)Zinn, Betz, Medcraft, and Schnell]{Zinn2015}
Zinn,~S.; Betz,~T.; Medcraft,~C.; Schnell,~M. Structure determination of
  trans-cinnamaldehyde by broadband microwave spectroscopy. \emph{Phys. Chem.
  Chem. Phys.} \textbf{2015}, \emph{17}, 16080--16085\relax
\mciteBstWouldAddEndPuncttrue
\mciteSetBstMidEndSepPunct{\mcitedefaultmidpunct}
{\mcitedefaultendpunct}{\mcitedefaultseppunct}\relax
\EndOfBibitem
\bibitem[Gougoula \latin{et~al.}(2019)Gougoula, Medcraft, Alkorta, Walker, and
  Legon]{Gougoula2019}
Gougoula,~E.; Medcraft,~C.; Alkorta,~I.; Walker,~N.~R.; Legon,~A.~C. A
  chalcogen-bonded complex H$_3$N$\cdot\cdot\cdot$S=C=S formed by ammonia and
  carbon disulfide characterised by chirped-pulse, broadband microwave
  spectroscopy. \emph{J. Chem. Phys.} \textbf{2019}, \emph{150}, 084307\relax
\mciteBstWouldAddEndPuncttrue
\mciteSetBstMidEndSepPunct{\mcitedefaultmidpunct}
{\mcitedefaultendpunct}{\mcitedefaultseppunct}\relax
\EndOfBibitem
\bibitem[Fatima \latin{et~al.}(2020)Fatima, Perez, Arenas, Schnell, and
  Steber]{Perez2020}
Fatima,~M.; Perez,~C.; Arenas,~B.~E.; Schnell,~M.; Steber,~A.~L. Benchmarking a
  new segmented K-band chirped-pulse microwave spectrometer and its application
  to the conformationally rich amino alcohol isoleucinol. \emph{Phys. Chem.
  Chem. Phys.} \textbf{2020}, \emph{22}, 17042--17051\relax
\mciteBstWouldAddEndPuncttrue
\mciteSetBstMidEndSepPunct{\mcitedefaultmidpunct}
{\mcitedefaultendpunct}{\mcitedefaultseppunct}\relax
\EndOfBibitem
\bibitem[Lide and Paul(1974)Lide, and Paul]{Lide1974}
Lide,~D.; Paul,~M.~E.
  \url{https://nap.nationalacademies.org/read/20111/chapter/1}, 1974; Accessed:
  2022-06-13\relax
\mciteBstWouldAddEndPuncttrue
\mciteSetBstMidEndSepPunct{\mcitedefaultmidpunct}
{\mcitedefaultendpunct}{\mcitedefaultseppunct}\relax
\EndOfBibitem
\bibitem[W\"uthrich(1990)]{Wuthrich1990}
W\"uthrich,~K. Protein structure determination in solution by NMR spectroscopy.
  \emph{Journal of Biological Chemistry} \textbf{1990}, \emph{265},
  22059--22062\relax
\mciteBstWouldAddEndPuncttrue
\mciteSetBstMidEndSepPunct{\mcitedefaultmidpunct}
{\mcitedefaultendpunct}{\mcitedefaultseppunct}\relax
\EndOfBibitem
\bibitem[Gruene and Mugnaioli(2021)Gruene, and Mugnaioli]{Gruene2021}
Gruene,~T.; Mugnaioli,~E. 3D Electron Diffraction for Chemical Analysis:
  Instrumentation Developments and Innovative Applications. \emph{Chemical
  Reviews} \textbf{2021}, \emph{121}, 11823--11834, PMID: 34533919\relax
\mciteBstWouldAddEndPuncttrue
\mciteSetBstMidEndSepPunct{\mcitedefaultmidpunct}
{\mcitedefaultendpunct}{\mcitedefaultseppunct}\relax
\EndOfBibitem
\bibitem[Watson \latin{et~al.}(1999)Watson, Roytburg, and Ulrich]{Watson1999}
Watson,~J.~K.; Roytburg,~A.; Ulrich,~W. Least-Squares Mass-Dependence Molecular
  Structures. \emph{Journal of Molecular Spectroscopy} \textbf{1999},
  \emph{196}, 102--119\relax
\mciteBstWouldAddEndPuncttrue
\mciteSetBstMidEndSepPunct{\mcitedefaultmidpunct}
{\mcitedefaultendpunct}{\mcitedefaultseppunct}\relax
\EndOfBibitem
\bibitem[Piccardo \latin{et~al.}(2015)Piccardo, Penocchio, Puzzarini, Biczysko,
  and Barone]{Piccardo2015}
Piccardo,~M.; Penocchio,~E.; Puzzarini,~C.; Biczysko,~M.; Barone,~V.
  Semi-Experimental Equilibrium Structure Determinations by Employing
  B3LYP/SNSD Anharmonic Force Fields: Validation and Application to Semirigid
  Organic Molecules. \emph{J. Phys. Chem. A} \textbf{2015}, \emph{119},
  2058--2082\relax
\mciteBstWouldAddEndPuncttrue
\mciteSetBstMidEndSepPunct{\mcitedefaultmidpunct}
{\mcitedefaultendpunct}{\mcitedefaultseppunct}\relax
\EndOfBibitem
\bibitem[Kraitchman(1953)]{Kraitchman1953}
Kraitchman,~J. Determination of Molecular Structure from Microwave
  Spectroscopic Data. \emph{American Journal of Physics} \textbf{1953},
  \emph{21}, 17--24\relax
\mciteBstWouldAddEndPuncttrue
\mciteSetBstMidEndSepPunct{\mcitedefaultmidpunct}
{\mcitedefaultendpunct}{\mcitedefaultseppunct}\relax
\EndOfBibitem
\bibitem[Costain(1958)]{Costain1958}
Costain,~C. Determination of Molecular Structures from Ground State Rotational
  Constants. \emph{Journal of Chemical Physics} \textbf{1958}, \emph{29},
  864\relax
\mciteBstWouldAddEndPuncttrue
\mciteSetBstMidEndSepPunct{\mcitedefaultmidpunct}
{\mcitedefaultendpunct}{\mcitedefaultseppunct}\relax
\EndOfBibitem
\bibitem[van Eijck(1982)]{Eijck1982}
van Eijck,~B. Influence of molecular vibrations on substitution coordinates.
  \emph{Journal of Molecular Spectroscopy} \textbf{1982}, \emph{91},
  348--362\relax
\mciteBstWouldAddEndPuncttrue
\mciteSetBstMidEndSepPunct{\mcitedefaultmidpunct}
{\mcitedefaultendpunct}{\mcitedefaultseppunct}\relax
\EndOfBibitem
\bibitem[Demaison and Rudolph(2002)Demaison, and Rudolph]{Demaison2002}
Demaison,~J.; Rudolph,~H. When Is the Substitution Structure Not Reliable?
  \emph{Journal of Molecular Spectroscopy} \textbf{2002}, \emph{215},
  78--84\relax
\mciteBstWouldAddEndPuncttrue
\mciteSetBstMidEndSepPunct{\mcitedefaultmidpunct}
{\mcitedefaultendpunct}{\mcitedefaultseppunct}\relax
\EndOfBibitem
\bibitem[Laurie and Herschbach(1962)Laurie, and Herschbach]{Laurie1962}
Laurie,~V.~W.; Herschbach,~D.~R. Influence of Vibrations on Molecular Structure
  Determinations. II. Average Structures Derived from Spectroscopic Data.
  \emph{J. Chem. Phys.} \textbf{1962}, \emph{37}, 1687--1693\relax
\mciteBstWouldAddEndPuncttrue
\mciteSetBstMidEndSepPunct{\mcitedefaultmidpunct}
{\mcitedefaultendpunct}{\mcitedefaultseppunct}\relax
\EndOfBibitem
\bibitem[Kisiel()]{STRFIT}
Kisiel,~Z. STRFIT - GENERAL STRUCTURE FITTING PROGRAM, URL
  http://www.ifpan.edu.pl/~kisiel/struct/struct.htm.
  \url{http://www.ifpan.edu.pl/~kisiel/struct/struct.htm}, Accessed:
  2021-10-30\relax
\mciteBstWouldAddEndPuncttrue
\mciteSetBstMidEndSepPunct{\mcitedefaultmidpunct}
{\mcitedefaultendpunct}{\mcitedefaultseppunct}\relax
\EndOfBibitem
\bibitem[Kisiel(2003)]{Kisiel2003}
Kisiel,~Z. Least-squares mass-dependence molecular structures for selected
  weakly bound intermolecular clusters. \emph{Journal of Molecular
  Spectroscopy} \textbf{2003}, \emph{218}, 58--67\relax
\mciteBstWouldAddEndPuncttrue
\mciteSetBstMidEndSepPunct{\mcitedefaultmidpunct}
{\mcitedefaultendpunct}{\mcitedefaultseppunct}\relax
\EndOfBibitem
\bibitem[Schultz(2022)]{Figshare_FitMOI}
Schultz,~T. Molecular Structure Fit from Rotational Constants.
  \url{https://doi.org/10.6084/m9.figshare.19336628}, 2022\relax
\mciteBstWouldAddEndPuncttrue
\mciteSetBstMidEndSepPunct{\mcitedefaultmidpunct}
{\mcitedefaultendpunct}{\mcitedefaultseppunct}\relax
\EndOfBibitem
\bibitem[Demaison \latin{et~al.}(2011)Demaison, Csaszar, Margules, and
  Rudolph]{Demaison2011}
Demaison,~J.; Csaszar,~A.~G.; Margules,~L.~D.; Rudolph,~H.~D. Equilibrium
  Structures of Heterocyclic Molecules with Large Principal Axis Rotations upon
  Isotopic Substitution. \emph{J. Phys. Chem. A} \textbf{2011}, \emph{115},
  14078--14091\relax
\mciteBstWouldAddEndPuncttrue
\mciteSetBstMidEndSepPunct{\mcitedefaultmidpunct}
{\mcitedefaultendpunct}{\mcitedefaultseppunct}\relax
\EndOfBibitem
\bibitem[Craig \latin{et~al.}(2004)Craig, Davis, Hanson, Moore, Weidenbaum, and
  Lock]{Craig2004}
Craig,~N.~C.; Davis,~J.~L.; Hanson,~K.~A.; Moore,~M.~C.; Weidenbaum,~K.~J.;
  Lock,~M. Analysis of the rotational structure in bands in the high-resolution
  infrared spectra of butadiene and butadiene-2,3-d$_2$: refinement in
  assignments of fundamentals. \emph{Journal of Molecular Structure}
  \textbf{2004}, \emph{695-696}, 59--69\relax
\mciteBstWouldAddEndPuncttrue
\mciteSetBstMidEndSepPunct{\mcitedefaultmidpunct}
{\mcitedefaultendpunct}{\mcitedefaultseppunct}\relax
\EndOfBibitem
\bibitem[Craig \latin{et~al.}(2004)Craig, Hanson, Pierce, Saylor, and
  Sams]{Craig2004a}
Craig,~N.; Hanson,~K.; Pierce,~W.; Saylor,~S.; Sams,~R. Rotational analysis of
  bands in the high-resolution infrared spectra of the three species of
  butadiene-1,4-d$_2$; refinement of the assignments of the vibrational
  fundamentals. \emph{Journal of Molecular Spectroscopy} \textbf{2004},
  \emph{228}, 401--413\relax
\mciteBstWouldAddEndPuncttrue
\mciteSetBstMidEndSepPunct{\mcitedefaultmidpunct}
{\mcitedefaultendpunct}{\mcitedefaultseppunct}\relax
\EndOfBibitem
\bibitem[Craig \latin{et~al.}(2006)Craig, Groner, and McKean]{Craig2006}
Craig,~N.~C.; Groner,~P.; McKean,~D.~C. Equilibrium Structures for Butadiene
  and Ethylene: Compelling Evidence for pi-Electron Delocalization in
  Butadiene. \emph{J. Phys. Chem. A} \textbf{2006}, \emph{110},
  7461--7469\relax
\mciteBstWouldAddEndPuncttrue
\mciteSetBstMidEndSepPunct{\mcitedefaultmidpunct}
{\mcitedefaultendpunct}{\mcitedefaultseppunct}\relax
\EndOfBibitem
\bibitem[Craig \latin{et~al.}(2006)Craig, Moore, Patchen, and Sams]{Craig2006a}
Craig,~N.~C.; Moore,~M.~C.; Patchen,~A.~K.; Sams,~R.~L. Analysis of rotational
  structure in the high-resolution infrared spectrum and assignment of
  vibrational fundamentals of butadiene-2,3-$^{13}$C$_2$. \emph{Journal of
  Molecular Spectroscopy} \textbf{2006}, \emph{235}, 181--189\relax
\mciteBstWouldAddEndPuncttrue
\mciteSetBstMidEndSepPunct{\mcitedefaultmidpunct}
{\mcitedefaultendpunct}{\mcitedefaultseppunct}\relax
\EndOfBibitem
\bibitem[Kveseth \latin{et~al.}(1980)Kveseth, Seip, and A.~Kohl]{Kveseth1980}
Kveseth,~K.; Seip,~R.; A.~Kohl,~D. \emph{Conformational Analysis. The Structure
  and Torsional Potential of 1,3-Butadiene as Studied by Gas Electron
  Diffraction}, 34th ed.; Acta Chemica Scandinavica A; 1980; Vol.~34\relax
\mciteBstWouldAddEndPuncttrue
\mciteSetBstMidEndSepPunct{\mcitedefaultmidpunct}
{\mcitedefaultendpunct}{\mcitedefaultseppunct}\relax
\EndOfBibitem
\bibitem[Pliva \latin{et~al.}(1990)Pliva, Johns, and Goodman]{Pliva1990}
Pliva,~J.; Johns,~J.; Goodman,~L. Infrared bands of isotopic benzenes:
  $\nu_{13}$ of $^{12}$C$_{6}$D$_{6}$ and $\nu_{12}$ of $^{13}$C$_{6}$H$_{6}$.
  \emph{Journal of Molecular Spectroscopy} \textbf{1990}, \emph{140},
  214--225\relax
\mciteBstWouldAddEndPuncttrue
\mciteSetBstMidEndSepPunct{\mcitedefaultmidpunct}
{\mcitedefaultendpunct}{\mcitedefaultseppunct}\relax
\EndOfBibitem
\bibitem[Kunishige \latin{et~al.}(2015)Kunishige, Katori, Baba, Nakajima, and
  Endo]{Kunishige2015}
Kunishige,~S.; Katori,~T.; Baba,~M.; Nakajima,~M.; Endo,~Y. Spectroscopic study
  on deuterated benzenes. I. Microwave spectra and molecular structure in the
  ground state. \emph{J. Chem. Phys.} \textbf{2015}, \emph{143}, 244302\relax
\mciteBstWouldAddEndPuncttrue
\mciteSetBstMidEndSepPunct{\mcitedefaultmidpunct}
{\mcitedefaultendpunct}{\mcitedefaultseppunct}\relax
\EndOfBibitem
\bibitem[CCC(2020)]{CCCBDB}
Russell,~D.~J., Ed. \emph{NIST Computational Chemistry Comparison and Benchmark
  Database}; NIST, 2020\relax
\mciteBstWouldAddEndPuncttrue
\mciteSetBstMidEndSepPunct{\mcitedefaultmidpunct}
{\mcitedefaultendpunct}{\mcitedefaultseppunct}\relax
\EndOfBibitem
\bibitem[Ceselin \latin{et~al.}(2021)Ceselin, Barone, and
  Tasinato]{Ceselin2021}
Ceselin,~G.; Barone,~V.; Tasinato,~N. Supplementary Table S2 in: Accurate
  Biomolecular Structures by the Nano-LEGO Approach: Pick the Bricks and Build
  Your Geometry. \emph{J. Chem. Theory Comput.} \textbf{2021}, \emph{17},
  7290--7311\relax
\mciteBstWouldAddEndPuncttrue
\mciteSetBstMidEndSepPunct{\mcitedefaultmidpunct}
{\mcitedefaultendpunct}{\mcitedefaultseppunct}\relax
\EndOfBibitem
\bibitem[Pliva and Pine(1982)Pliva, and Pine]{Pliva1982JMol}
Pliva,~J.; Pine,~A. The spectrum of benzene in the 3-$\mu$m region: The
  $\nu_{12}$ fundamental band. \emph{Journal of Molecular Spectroscopy}
  \textbf{1982}, \emph{93}, 209--236\relax
\mciteBstWouldAddEndPuncttrue
\mciteSetBstMidEndSepPunct{\mcitedefaultmidpunct}
{\mcitedefaultendpunct}{\mcitedefaultseppunct}\relax
\EndOfBibitem
\bibitem[Pliva and Pine(1987)Pliva, and Pine]{Pliva1987}
Pliva,~J.; Pine,~A.~S. Analysis of the 3-$\mu$m bands of benzene. \emph{Journal
  of Molecular Spectroscopy} \textbf{1987}, \emph{126}, 82--98\relax
\mciteBstWouldAddEndPuncttrue
\mciteSetBstMidEndSepPunct{\mcitedefaultmidpunct}
{\mcitedefaultendpunct}{\mcitedefaultseppunct}\relax
\EndOfBibitem
\bibitem[Pliva \latin{et~al.}(1989)Pliva, Johns, and Goodman]{Pliva1989a}
Pliva,~J.; Johns,~J.; Goodman,~L. Infrared bands of $^{13}$C$_6$H$_6$:
  $\nu_{13}$ and $\nu_{14}$. \emph{Journal of Molecular Spectroscopy}
  \textbf{1989}, \emph{134}, 227--233\relax
\mciteBstWouldAddEndPuncttrue
\mciteSetBstMidEndSepPunct{\mcitedefaultmidpunct}
{\mcitedefaultendpunct}{\mcitedefaultseppunct}\relax
\EndOfBibitem
\bibitem[Pliva \latin{et~al.}(1991)Pliva, Johns, and Goodman]{Pliva1991}
Pliva,~J.; Johns,~J.; Goodman,~L. Infrared bands of isotopic benzenes:
  $\nu_{13}$ and $\nu_{14}$ of $^{13}$C$_6$D$_6$. \emph{Journal of Molecular
  Spectroscopy} \textbf{1991}, \emph{148}, 427--435\relax
\mciteBstWouldAddEndPuncttrue
\mciteSetBstMidEndSepPunct{\mcitedefaultmidpunct}
{\mcitedefaultendpunct}{\mcitedefaultseppunct}\relax
\EndOfBibitem
\bibitem[Baba \latin{et~al.}(2011)Baba, Kowaka, Nagashima, Ishimoto, Goto, and
  Nakayama]{Baba2011}
Baba,~M.; Kowaka,~Y.; Nagashima,~U.; Ishimoto,~T.; Goto,~H.; Nakayama,~N.
  Geometrical structure of benzene and naphthalene: Ultrahigh-resolution laser
  spectroscopy and ab initio calculation. \emph{The Journal of Chemical
  Physics} \textbf{2011}, \emph{135}, 054305\relax
\mciteBstWouldAddEndPuncttrue
\mciteSetBstMidEndSepPunct{\mcitedefaultmidpunct}
{\mcitedefaultendpunct}{\mcitedefaultseppunct}\relax
\EndOfBibitem
\bibitem[Hirano \latin{et~al.}(2021)Hirano, Nagashima, and Baba]{Hirano2021}
Hirano,~T.; Nagashima,~U.; Baba,~M. Ro-vibrationally averaged molecular
  structure of benzene: Why almost the same bond lengths are observed for the
  CH and CD bonds? \emph{Journal of Molecular Structure} \textbf{2021},
  \emph{1243}, 130537\relax
\mciteBstWouldAddEndPuncttrue
\mciteSetBstMidEndSepPunct{\mcitedefaultmidpunct}
{\mcitedefaultendpunct}{\mcitedefaultseppunct}\relax
\EndOfBibitem
\bibitem[Udagawa \latin{et~al.}(2023)Udagawa, Tanaka, Hirano, Kuwahata,
  Tachikawa, Baba, and Nagashima]{Udagawa2023}
Udagawa,~T.; Tanaka,~H.; Hirano,~T.; Kuwahata,~K.; Tachikawa,~M.; Baba,~M.;
  Nagashima,~U. Direct Elucidation of the Vibrationally Averaged Structure of
  Benzene: A Path Integral Molecular Dynamics Study. \emph{J. Phys. Chem. A}
  \textbf{2023}, \emph{127}, 894--901\relax
\mciteBstWouldAddEndPuncttrue
\mciteSetBstMidEndSepPunct{\mcitedefaultmidpunct}
{\mcitedefaultendpunct}{\mcitedefaultseppunct}\relax
\EndOfBibitem
\bibitem[Esselman \latin{et~al.}(2023)Esselman, Zdanovskaia, Owen, Stanton,
  Woods, and McMahon]{Esselman2023}
Esselman,~B.~J.; Zdanovskaia,~M.~A.; Owen,~A.~N.; Stanton,~J.~F.; Woods,~R.~C.;
  McMahon,~R.~J. Precise Equilibrium Structure of Benzene. \emph{Journal of the
  American Chemical Society J. Am. Chem. Soc.} \textbf{2023}, \emph{145},
  21785--21797\relax
\mciteBstWouldAddEndPuncttrue
\mciteSetBstMidEndSepPunct{\mcitedefaultmidpunct}
{\mcitedefaultendpunct}{\mcitedefaultseppunct}\relax
\EndOfBibitem
\bibitem[Bak \latin{et~al.}(1962)Bak, Christensen, B.Dixon, Hansen-Nygaard,
  Andersen, and Schottl\"ander]{Bak1962}
Bak,~B.; Christensen,~D.; B.Dixon,~W.; Hansen-Nygaard,~L.; Andersen,~J.~R.;
  Schottl\"ander,~M. The complete structure of furan. \emph{J. Mol. Spectrosc.}
  \textbf{1962}, \emph{9}, 124--129\relax
\mciteBstWouldAddEndPuncttrue
\mciteSetBstMidEndSepPunct{\mcitedefaultmidpunct}
{\mcitedefaultendpunct}{\mcitedefaultseppunct}\relax
\EndOfBibitem
\bibitem[Barnum \latin{et~al.}(2021)Barnum, Lee, and McGuire]{Barnum2021}
Barnum,~T.~J.; Lee,~K. L.~K.; McGuire,~B.~A. Chirped-Pulse Fourier Transform
  Millimeter-Wave Spectroscopy of Furan, Isotopologues, and Vibrational Excited
  States. \emph{ACS Earth Space Chem.} \textbf{2021}, \emph{5},
  2986--2994\relax
\mciteBstWouldAddEndPuncttrue
\mciteSetBstMidEndSepPunct{\mcitedefaultmidpunct}
{\mcitedefaultendpunct}{\mcitedefaultseppunct}\relax
\EndOfBibitem
\bibitem[Pretsch \latin{et~al.}(2000)Pretsch, B\"uhlmann, and
  Affolter]{Pretsch2000}
Pretsch,~E.; B\"uhlmann,~P.; Affolter,~C. \emph{Structure Determination of
  Organic Compounds}; Springer Verlag, 2000\relax
\mciteBstWouldAddEndPuncttrue
\mciteSetBstMidEndSepPunct{\mcitedefaultmidpunct}
{\mcitedefaultendpunct}{\mcitedefaultseppunct}\relax
\EndOfBibitem
\bibitem[But(2010)]{ButadieneProduct}
\emph{Butadiene Product Stewardship Guidance Manual}; American Chemistry
  Council, 2010; pp II--2\relax
\mciteBstWouldAddEndPuncttrue
\mciteSetBstMidEndSepPunct{\mcitedefaultmidpunct}
{\mcitedefaultendpunct}{\mcitedefaultseppunct}\relax
\EndOfBibitem
\bibitem[Lee \latin{et~al.}(2019)Lee, Lee, and Schultz]{Lee2019}
Lee,~J.~C.; Lee,~D.~E.; Schultz,~T. High-resolution rotational Raman
  spectroscopy of benzene. \emph{Phys. Chem. Chem. Phys.} \textbf{2019},
  \emph{21}, 2857--2860\relax
\mciteBstWouldAddEndPuncttrue
\mciteSetBstMidEndSepPunct{\mcitedefaultmidpunct}
{\mcitedefaultendpunct}{\mcitedefaultseppunct}\relax
\EndOfBibitem
\bibitem[Fang \latin{et~al.}(2011)Fang, Gong, Zhang, Shan, Liu, Wang, and
  Sheng]{Fang2011}
Fang,~W.; Gong,~L.; Zhang,~Q.; Shan,~X.; Liu,~F.; Wang,~Z.; Sheng,~L.
  Dissociative photoionization of 1,3-butadiene: Experimental and theoretical
  insights. \emph{The Journal of Chemical Physics} \textbf{2011}, \emph{134},
  174306\relax
\mciteBstWouldAddEndPuncttrue
\mciteSetBstMidEndSepPunct{\mcitedefaultmidpunct}
{\mcitedefaultendpunct}{\mcitedefaultseppunct}\relax
\EndOfBibitem
\bibitem[Senent \latin{et~al.}(2016)Senent, Hochlaf, and Carvajal]{Senent2016}
Senent,~M.~L.; Hochlaf,~M.; Carvajal,~M. Spectroscopy and Dynamics of
  Medium-Sized Molecules and Clusters: Theory, Experiment, and Applications.
  \emph{The Journal of Physical Chemistry A} \textbf{2016}, \emph{120},
  475--476\relax
\mciteBstWouldAddEndPuncttrue
\mciteSetBstMidEndSepPunct{\mcitedefaultmidpunct}
{\mcitedefaultendpunct}{\mcitedefaultseppunct}\relax
\EndOfBibitem
\bibitem[Zeng and Johnson(2021)Zeng, and Johnson]{Zeng2021}
Zeng,~H.~J.; Johnson,~M.~A. Demystifying the Diffuse Vibrational Spectrum of
  Aqueous Protons Through Cold Cluster Spectroscopy. \emph{Annu. Rev. Phys.
  Chem.} \textbf{2021}, \emph{72}, 667--691\relax
\mciteBstWouldAddEndPuncttrue
\mciteSetBstMidEndSepPunct{\mcitedefaultmidpunct}
{\mcitedefaultendpunct}{\mcitedefaultseppunct}\relax
\EndOfBibitem
\bibitem[Gregoire \latin{et~al.}(2001)Gregoire, Dedonder-Lardeux, Jouvet,
  Martrenchard, and Solgadi]{Gregoire2001}
Gregoire,~G.; Dedonder-Lardeux,~C.; Jouvet,~C.; Martrenchard,~S.; Solgadi,~D.
  Has the Excited State Proton Transfer Ever Been Observed in
  Phenol-(NH$_3$)$_n$ Molecular Clusters? \emph{J. Phys. Chem. A}
  \textbf{2001}, \emph{105}, 5971--5976\relax
\mciteBstWouldAddEndPuncttrue
\mciteSetBstMidEndSepPunct{\mcitedefaultmidpunct}
{\mcitedefaultendpunct}{\mcitedefaultseppunct}\relax
\EndOfBibitem
\bibitem[David \latin{et~al.}(2002)David, Dedonder-Lardeux, and
  Jouvet]{David2002}
David,~O.; Dedonder-Lardeux,~C.; Jouvet,~C. Is there an excited state proton
  transfer in phenol (or 1-naphthol)-ammonia clusters? Hydrogen detachment and
  transfer to solvent: a key for non-radiative processes in clusters.
  \emph{International Reviews in Physical Chemistry} \textbf{2002}, \emph{21},
  499--523\relax
\mciteBstWouldAddEndPuncttrue
\mciteSetBstMidEndSepPunct{\mcitedefaultmidpunct}
{\mcitedefaultendpunct}{\mcitedefaultseppunct}\relax
\EndOfBibitem
\bibitem[Schultz \latin{et~al.}(2004)Schultz, Samoylova, Radloff, Hertel,
  Sobolewski, and Domcke]{Schultz2004}
Schultz,~T.; Samoylova,~E.; Radloff,~W.; Hertel,~I.~V.; Sobolewski,~A.~L.;
  Domcke,~W. Efficient deactivation of a model base pair via excited-state
  hydrogen transfer. \emph{Science} \textbf{2004}, \emph{306}, 1765--1768\relax
\mciteBstWouldAddEndPuncttrue
\mciteSetBstMidEndSepPunct{\mcitedefaultmidpunct}
{\mcitedefaultendpunct}{\mcitedefaultseppunct}\relax
\EndOfBibitem
\bibitem[Ritze \latin{et~al.}(2005)Ritze, Lippert, Samoylova, Smith, Hertel,
  Radloff, and Schultz]{Ritze2005}
Ritze,~H.~H.; Lippert,~H.; Samoylova,~E.; Smith,~V.~R.; Hertel,~I.~V.;
  Radloff,~W.; Schultz,~T. Relevance of pi sigma(*) states in the photoinduced
  processes of adenine, adenine dimer, and adenine-water complexes.
  \emph{Journal of Chemical Physics} \textbf{2005}, \emph{122}\relax
\mciteBstWouldAddEndPuncttrue
\mciteSetBstMidEndSepPunct{\mcitedefaultmidpunct}
{\mcitedefaultendpunct}{\mcitedefaultseppunct}\relax
\EndOfBibitem
\bibitem[Samoylova \latin{et~al.}(2005)Samoylova, Lippert, Ullrich, Hertel,
  Radloff, and Schultz]{Samoylova2005}
Samoylova,~E.; Lippert,~H.; Ullrich,~S.; Hertel,~I.~V.; Radloff,~W.;
  Schultz,~T. Dynamics of photoinduced processes in adenine and thymine base
  pairs. \emph{Journal of the American Chemical Society} \textbf{2005},
  \emph{127}, 1782--1786\relax
\mciteBstWouldAddEndPuncttrue
\mciteSetBstMidEndSepPunct{\mcitedefaultmidpunct}
{\mcitedefaultendpunct}{\mcitedefaultseppunct}\relax
\EndOfBibitem
\bibitem[Gador \latin{et~al.}(2007)Gador, Samoylova, Smith, Stolow, Rayner,
  Radloff, Hertel, and Schultz]{Gador2007}
Gador,~N.; Samoylova,~E.; Smith,~V.~R.; Stolow,~A.; Rayner,~D.~M.;
  Radloff,~W.~G.; Hertel,~I.~V.; Schultz,~T. Electronic structure of adenine
  and thymine base pairs studied by femtosecond electron-ion coincidence
  spectroscopy. \emph{Journal of Physical Chemistry A} \textbf{2007},
  \emph{111}, 11743--11749\relax
\mciteBstWouldAddEndPuncttrue
\mciteSetBstMidEndSepPunct{\mcitedefaultmidpunct}
{\mcitedefaultendpunct}{\mcitedefaultseppunct}\relax
\EndOfBibitem
\bibitem[Samoylova \latin{et~al.}(2008)Samoylova, Schultz, Hertel, and
  Radloff]{Samoylova2008}
Samoylova,~E.; Schultz,~T.; Hertel,~I.~V.; Radloff,~W. Analysis of ultrafast
  relaxation in photoexcited DNA base pairs of adenine and thymine.
  \emph{Chemical Physics} \textbf{2008}, \emph{347}, 376--382\relax
\mciteBstWouldAddEndPuncttrue
\mciteSetBstMidEndSepPunct{\mcitedefaultmidpunct}
{\mcitedefaultendpunct}{\mcitedefaultseppunct}\relax
\EndOfBibitem
\bibitem[Smith \latin{et~al.}(2010)Smith, Samoylova, Ritze, Radloff, and
  Schultz]{Smith2010a}
Smith,~V.~R.; Samoylova,~E.; Ritze,~H.~H.; Radloff,~W.; Schultz,~T. Excimer
  states in microhydrated adenine clusters. \emph{Physical Chemistry Chemical
  Physics} \textbf{2010}, \emph{12}, 9632--9636\relax
\mciteBstWouldAddEndPuncttrue
\mciteSetBstMidEndSepPunct{\mcitedefaultmidpunct}
{\mcitedefaultendpunct}{\mcitedefaultseppunct}\relax
\EndOfBibitem
\bibitem[Lee \latin{et~al.}(2021)Lee, \"Ozer, and Schultz]{Lee2021}
Lee,~J.~C.; \"Ozer,~B.~R.; Schultz,~T. CRASY: correlated rotational alignment
  spectroscopy of pyridine. The rotational Raman spectrum of pyridine and
  asymmetric fragmentation of pyridine dimer cations. \emph{Phys. Chem. Chem.
  Phys.} \textbf{2021}, \emph{23}, 10621--10628\relax
\mciteBstWouldAddEndPuncttrue
\mciteSetBstMidEndSepPunct{\mcitedefaultmidpunct}
{\mcitedefaultendpunct}{\mcitedefaultseppunct}\relax
\EndOfBibitem
\bibitem[Piacenza and Grimme(2005)Piacenza, and Grimme]{Piacenza2005}
Piacenza,~M.; Grimme,~S. Van der Waals interactions in aromatic systems:
  Structure and energetics of dimers and trimers of pyridine.
  \emph{Chemphyschem} \textbf{2005}, \emph{6}, 1554--1558\relax
\mciteBstWouldAddEndPuncttrue
\mciteSetBstMidEndSepPunct{\mcitedefaultmidpunct}
{\mcitedefaultendpunct}{\mcitedefaultseppunct}\relax
\EndOfBibitem
\bibitem[Hohenstein and Sherrill(2009)Hohenstein, and Sherrill]{Hohenstein2009}
Hohenstein,~E.~G.; Sherrill,~C.~D. Effects of Heteroatoms on Aromatic pi-pi
  Interactions: Benzene-Pyridine and Pyridine Dimer. \emph{Journal of Physical
  Chemistry A} \textbf{2009}, \emph{113}, 878--886\relax
\mciteBstWouldAddEndPuncttrue
\mciteSetBstMidEndSepPunct{\mcitedefaultmidpunct}
{\mcitedefaultendpunct}{\mcitedefaultseppunct}\relax
\EndOfBibitem
\bibitem[Zhang \latin{et~al.}(2014)Zhang, Gao, Yao, Li, and Tao]{Zhang2014}
Zhang,~J.; Gao,~Y.; Yao,~W.; Li,~S.; Tao,~F.-M. Ab initio study of the pi-pi
  interaction in the pyridine dimer: Effects of bond functions.
  \emph{Computational and Theoretical Chemistry} \textbf{2014}, \emph{1049},
  82-- 89\relax
\mciteBstWouldAddEndPuncttrue
\mciteSetBstMidEndSepPunct{\mcitedefaultmidpunct}
{\mcitedefaultendpunct}{\mcitedefaultseppunct}\relax
\EndOfBibitem
\bibitem[Sieranski(2017)]{Sieranski2017}
Sieranski,~T. Discovering the stacking landscape of a pyridine-pyridine system.
  \emph{Journal of Molecular Modeling} \textbf{2017}, \emph{23}, 338\relax
\mciteBstWouldAddEndPuncttrue
\mciteSetBstMidEndSepPunct{\mcitedefaultmidpunct}
{\mcitedefaultendpunct}{\mcitedefaultseppunct}\relax
\EndOfBibitem
\bibitem[Sinnokrot and Sherrill(2004)Sinnokrot, and Sherrill]{Sinnokrot2004}
Sinnokrot,~M.~O.; Sherrill,~C.~D. Highly Accurate Coupled Cluster Potential
  Energy Curves for the Benzene Dimer: Sandwich, T-Shaped, and
  Parallel-Displaced Configurations. \emph{J. Phys. Chem. A} \textbf{2004},
  \emph{108}, 10200--10207\relax
\mciteBstWouldAddEndPuncttrue
\mciteSetBstMidEndSepPunct{\mcitedefaultmidpunct}
{\mcitedefaultendpunct}{\mcitedefaultseppunct}\relax
\EndOfBibitem
\bibitem[Arunan and Gutowsky(1993)Arunan, and Gutowsky]{Arunan1993}
Arunan,~E.; Gutowsky,~H.~S. The rotational spectrum, structure and dynamics of
  a benzene dimer. \emph{The Journal of Chemical Physics} \textbf{1993},
  \emph{98}, 4294--4296\relax
\mciteBstWouldAddEndPuncttrue
\mciteSetBstMidEndSepPunct{\mcitedefaultmidpunct}
{\mcitedefaultendpunct}{\mcitedefaultseppunct}\relax
\EndOfBibitem
\bibitem[Schnell \latin{et~al.}(2013)Schnell, Erlekam, Bunker, Helden, Grabow,
  Meijer, and Avoird]{Schnell2013}
Schnell,~M.; Erlekam,~U.; Bunker,~P.~R.; Helden,~G.; Grabow,~J.-U.; Meijer,~G.;
  Avoird,~A. Unraveling the internal dynamics of the benzene dimer: a combined
  theoretical and microwave spectroscopy study. \emph{Phys. Chem. Chem. Phys.}
  \textbf{2013}, \emph{15}, 10207--10223\relax
\mciteBstWouldAddEndPuncttrue
\mciteSetBstMidEndSepPunct{\mcitedefaultmidpunct}
{\mcitedefaultendpunct}{\mcitedefaultseppunct}\relax
\EndOfBibitem
\bibitem[Schnell \latin{et~al.}(2013)Schnell, Erlekam, Bunker, Helden, Grabow,
  Meijer, and Avoird]{Schnell2013b}
Schnell,~M.; Erlekam,~U.; Bunker,~P.~R.; Helden,~G.; Grabow,~J.-U.; Meijer,~G.;
  Avoird,~A. Structure of the Benzene Dimer - Governed by Dynamics.
  \emph{Angewandte Chemie International Edition} \textbf{2013}, \emph{52},
  5180--5183\relax
\mciteBstWouldAddEndPuncttrue
\mciteSetBstMidEndSepPunct{\mcitedefaultmidpunct}
{\mcitedefaultendpunct}{\mcitedefaultseppunct}\relax
\EndOfBibitem
\bibitem[Albert \latin{et~al.}(2011)Albert, Albert, and Quack]{Albert2011}
Albert,~S.; Albert,~K.; Quack,~M. In \emph{Handbook of High-Resolution
  Spectroscopy}; Quack,~M., Merkt,~F., Eds.; John Wiley \& Sons, Ltd:
  Chichester, UK, 2011; Vol.~2; pp 965--1019\relax
\mciteBstWouldAddEndPuncttrue
\mciteSetBstMidEndSepPunct{\mcitedefaultmidpunct}
{\mcitedefaultendpunct}{\mcitedefaultseppunct}\relax
\EndOfBibitem
\bibitem[Albert \latin{et~al.}(2018)Albert, Bauerecker, Bekhtereva, Bolotova,
  Hollenstein, Quack, and Ulenikov]{Albert2018}
Albert,~S.; Bauerecker,~S.; Bekhtereva,~E.~S.; Bolotova,~I.~B.;
  Hollenstein,~H.; Quack,~M.; Ulenikov,~O.~N. High resolution FTIR spectroscopy
  of fluoroform \ce{^{12}CHF3} and critical analysis of the infrared spectrum
  from 25 to 1500 cm$^{-1}$. \emph{Molecular Physics} \textbf{2018},
  \emph{116}, 1091--1107\relax
\mciteBstWouldAddEndPuncttrue
\mciteSetBstMidEndSepPunct{\mcitedefaultmidpunct}
{\mcitedefaultendpunct}{\mcitedefaultseppunct}\relax
\EndOfBibitem
\bibitem[Schroter \latin{et~al.}(2013)Schroter, Kosma, and
  Schultz]{Schroter2013}
Schroter,~C.; Kosma,~K.; Schultz,~T. Correlated Rotational Alignment
  Spectroscopy of Isolated Molecules and Molecular Mixtures. \emph{EPJ Web of
  Conferences} \textbf{2013}, \emph{41}\relax
\mciteBstWouldAddEndPuncttrue
\mciteSetBstMidEndSepPunct{\mcitedefaultmidpunct}
{\mcitedefaultendpunct}{\mcitedefaultseppunct}\relax
\EndOfBibitem
\bibitem[Schultz \latin{et~al.}(2023)Schultz, Heo, Lee, and
  \"Ozer]{Schultz2023}
Schultz,~T.; Heo,~I.; Lee,~J.~C.; \"Ozer,~B.~R. Molecular-beam spectroscopy
  with an infinite interferometer: spectroscopic resolution and accuracy.
  \emph{Journal of the Korean Physical Society} \textbf{2023}, \emph{82},
  919--927\relax
\mciteBstWouldAddEndPuncttrue
\mciteSetBstMidEndSepPunct{\mcitedefaultmidpunct}
{\mcitedefaultendpunct}{\mcitedefaultseppunct}\relax
\EndOfBibitem
\bibitem[Riley(2008)]{Riley2008}
Riley,~W. \emph{Handbook of Frequency Stability Analysis}; National Institute
  of Standards and Technology Special Publication 1065; U. S. GOVERNMENT
  PRINTING OFFICE: Washington, 2008\relax
\mciteBstWouldAddEndPuncttrue
\mciteSetBstMidEndSepPunct{\mcitedefaultmidpunct}
{\mcitedefaultendpunct}{\mcitedefaultseppunct}\relax
\EndOfBibitem
\bibitem[Hansch(2006)]{Hansch2006}
Hansch,~T.~W. Nobel Lecture: Passion for precision. \emph{Reviews of Modern
  Physics} \textbf{2006}, \emph{78}, 1297--1309\relax
\mciteBstWouldAddEndPuncttrue
\mciteSetBstMidEndSepPunct{\mcitedefaultmidpunct}
{\mcitedefaultendpunct}{\mcitedefaultseppunct}\relax
\EndOfBibitem
\bibitem[Dowling(1968)]{Dowling1968}
Dowling,~J.~M. The rotation-inversion spectrum of ammonia. \emph{Journal of
  Molecular Spectroscopy} \textbf{1968}, \emph{27}, 527--538\relax
\mciteBstWouldAddEndPuncttrue
\mciteSetBstMidEndSepPunct{\mcitedefaultmidpunct}
{\mcitedefaultendpunct}{\mcitedefaultseppunct}\relax
\EndOfBibitem
\bibitem[Cloppenburg \latin{et~al.}(1979)Cloppenburg, Manczak, Prockl,
  Schr\"oter, and Strey]{Cloppenburg1979}
Cloppenburg,~B.; Manczak,~K.; Prockl,~H.; Schr\"oter,~H.~W.; Strey,~G.
  Resolution of K-Splitting in the Rotation-Inversion Raman Spectrum of Ammonia
  NH$_3$. \emph{Zeitschrift f\"ur Naturforschung A} \textbf{1979}, \emph{34},
  1160--1163\relax
\mciteBstWouldAddEndPuncttrue
\mciteSetBstMidEndSepPunct{\mcitedefaultmidpunct}
{\mcitedefaultendpunct}{\mcitedefaultseppunct}\relax
\EndOfBibitem
\bibitem[Kauppinen \latin{et~al.}(1980)Kauppinen, Jensen, and
  Brodersen]{Kauppinen1980}
Kauppinen,~J.; Jensen,~P.; Brodersen,~S. Determination of the $B_0$ constant of
  C$_6$H$_6$. \emph{Journal of Molecular Spectroscopy} \textbf{1980},
  \emph{83}, 161--174\relax
\mciteBstWouldAddEndPuncttrue
\mciteSetBstMidEndSepPunct{\mcitedefaultmidpunct}
{\mcitedefaultendpunct}{\mcitedefaultseppunct}\relax
\EndOfBibitem
\bibitem[Pliva and Johns(1983)Pliva, and Johns]{Pliva1982CanPhy}
Pliva,~J.; Johns,~J. W.~C. The $\nu_{13}$ fundamental band of benzene.
  \emph{Canadian Journal of Physics} \textbf{1983}, \emph{61}, 269--277\relax
\mciteBstWouldAddEndPuncttrue
\mciteSetBstMidEndSepPunct{\mcitedefaultmidpunct}
{\mcitedefaultendpunct}{\mcitedefaultseppunct}\relax
\EndOfBibitem
\bibitem[Junttila \latin{et~al.}(1991)Junttila, Domenech, Fraser, and
  Pine]{Junttila1991}
Junttila,~M.; Domenech,~J.; Fraser,~G.~T.; Pine,~A. Molecular-beam optothermal
  spectroscopy of the 9.6-$\mu$m $\nu_{14}$ band of benzene. \emph{Journal of
  Molecular Spectroscopy} \textbf{1991}, \emph{147}, 513--520\relax
\mciteBstWouldAddEndPuncttrue
\mciteSetBstMidEndSepPunct{\mcitedefaultmidpunct}
{\mcitedefaultendpunct}{\mcitedefaultseppunct}\relax
\EndOfBibitem
\bibitem[Domenech \latin{et~al.}(1991)Domenech, Junttila, and
  Pine]{Domenech1991}
Domenech,~J.; Junttila,~M.; Pine,~A. Molecular-beam spectrum of the 3.3-$\mu$m
  $\nu_{12}$ band of benzene. \emph{Journal of Molecular Spectroscopy}
  \textbf{1991}, \emph{149}, 391--398\relax
\mciteBstWouldAddEndPuncttrue
\mciteSetBstMidEndSepPunct{\mcitedefaultmidpunct}
{\mcitedefaultendpunct}{\mcitedefaultseppunct}\relax
\EndOfBibitem
\bibitem[Okruss \latin{et~al.}(1999)Okruss, M{\"u}ller, and Hese]{Okruss1999}
Okruss,~M.; M{\"u}ller,~R.; Hese,~A. High-Resolution UV Laser Spectroscopy of
  Jet-Cooled Benzene Molecules: Complete Rotational Analysis of the $S_1
  \leftarrow S_0\ 6^1_0$ (l=$\pm$1) Band. \emph{Journal of Molecular
  Spectroscopy} \textbf{1999}, \emph{193}, 293--305\relax
\mciteBstWouldAddEndPuncttrue
\mciteSetBstMidEndSepPunct{\mcitedefaultmidpunct}
{\mcitedefaultendpunct}{\mcitedefaultseppunct}\relax
\EndOfBibitem
\bibitem[Doi \latin{et~al.}(2004)Doi, Baba, Kasahara, and Kat\^o]{Doi2004b}
Doi,~A.; Baba,~M.; Kasahara,~S.; Kat\^o,~H. Sub-Doppler rotationally resolved
  spectroscopy of the $S_1 ^1B_{2u}(\nu_6$=$1) \leftarrow S_0
  ^1A_{1g}(\nu$=$0)$ bands of benzene and benzene-$d_6$. \emph{Journal of
  Molecular Spectroscopy} \textbf{2004}, \emph{227}, 180--186\relax
\mciteBstWouldAddEndPuncttrue
\mciteSetBstMidEndSepPunct{\mcitedefaultmidpunct}
{\mcitedefaultendpunct}{\mcitedefaultseppunct}\relax
\EndOfBibitem
\bibitem[Riehn \latin{et~al.}(2001)Riehn, Weichert, and Brutschy]{Riehn2001}
Riehn,~C.; Weichert,~A.; Brutschy,~B. Probing benzene in a new way:
  High-resolution time-resolved rotational spectroscopy. \emph{Journal of
  Physical Chemistry A} \textbf{2001}, \emph{105}, 5618--5621\relax
\mciteBstWouldAddEndPuncttrue
\mciteSetBstMidEndSepPunct{\mcitedefaultmidpunct}
{\mcitedefaultendpunct}{\mcitedefaultseppunct}\relax
\EndOfBibitem
\bibitem[Jarzeba \latin{et~al.}(2002)Jarzeba, Matylitsky, Weichert, and
  Riehn]{Jarzeba2002}
Jarzeba,~W.; Matylitsky,~V.~V.; Weichert,~A.; Riehn,~C. Rotational coherence
  spectroscopy of benzene by femtosecond degenerate four-wave mixing.
  \emph{Physical Chemistry Chemical Physics} \textbf{2002}, \emph{4},
  451--454\relax
\mciteBstWouldAddEndPuncttrue
\mciteSetBstMidEndSepPunct{\mcitedefaultmidpunct}
{\mcitedefaultendpunct}{\mcitedefaultseppunct}\relax
\EndOfBibitem
\bibitem[Matylitsky \latin{et~al.}(2002)Matylitsky, Jarzeba, Riehn, and
  Brutschy]{Matylitsky2002}
Matylitsky,~V.~V.; Jarzeba,~W.; Riehn,~C.; Brutschy,~B. Femtosecond degenerate
  four-wave mixing study of benzene in the gas phase. \emph{Journal of Raman
  Spectroscopy} \textbf{2002}, \emph{33}, 877--883\relax
\mciteBstWouldAddEndPuncttrue
\mciteSetBstMidEndSepPunct{\mcitedefaultmidpunct}
{\mcitedefaultendpunct}{\mcitedefaultseppunct}\relax
\EndOfBibitem
\bibitem[Amyay \latin{et~al.}(2010)Amyay, Herman, Fayt, Fusina, and
  Predoi-Cross]{Amyay2010}
Amyay,~B.; Herman,~M.; Fayt,~A.; Fusina,~L.; Predoi-Cross,~A. High resolution
  FTIR investigation of $^{12}$C$_2$H$_2$ in the FIR spectral range using
  synchrotron radiation. \emph{Chemical Physics Letters} \textbf{2010},
  \emph{491}, 17--19\relax
\mciteBstWouldAddEndPuncttrue
\mciteSetBstMidEndSepPunct{\mcitedefaultmidpunct}
{\mcitedefaultendpunct}{\mcitedefaultseppunct}\relax
\EndOfBibitem
\bibitem[Albert \latin{et~al.}(2015)Albert, Keppler, Lerch, Quack, and
  Wokaun]{Albert2015}
Albert,~S.; Keppler,~K.; Lerch,~P.; Quack,~M.; Wokaun,~A. Synchrotron-based
  highest resolution FTIR spectroscopy of chlorobenzene. \emph{Journal of
  Molecular Spectroscopy} \textbf{2015}, \emph{315}, 92--101, Spectroscopy with
  Synchrotron Radiation\relax
\mciteBstWouldAddEndPuncttrue
\mciteSetBstMidEndSepPunct{\mcitedefaultmidpunct}
{\mcitedefaultendpunct}{\mcitedefaultseppunct}\relax
\EndOfBibitem
\bibitem[Weber(2011)]{Weber2011}
Weber,~A. In \emph{Handbook of High-Resolution Spectroscopy}; Quack,~M.,
  Merkt,~F., Eds.; John Wiley \& Sons, Ltd: Chichester, UK, 2011; Vol.~2; pp
  1153--1236\relax
\mciteBstWouldAddEndPuncttrue
\mciteSetBstMidEndSepPunct{\mcitedefaultmidpunct}
{\mcitedefaultendpunct}{\mcitedefaultseppunct}\relax
\EndOfBibitem
\bibitem[Diddams \latin{et~al.}(2020)Diddams, Kerry, and Thomas]{Diddams2020}
Diddams,~S.~A.; Kerry,~V.; Thomas,~U. Optical frequency combs: Coherently
  uniting the electromagnetic spectrum. \emph{Science} \textbf{2020},
  \emph{369}, eaay3676\relax
\mciteBstWouldAddEndPuncttrue
\mciteSetBstMidEndSepPunct{\mcitedefaultmidpunct}
{\mcitedefaultendpunct}{\mcitedefaultseppunct}\relax
\EndOfBibitem
\bibitem[Foltynowicz \latin{et~al.}(2011)Foltynowicz, Masasowski, Ban, Adler,
  Cossel, Briles, and Ye]{Foltynowicz2011}
Foltynowicz,~A.; Masasowski,~P.; Ban,~T.; Adler,~F.; Cossel,~K.~C.;
  Briles,~T.~C.; Ye,~J. Optical frequency comb spectroscopy. \emph{Faraday
  Discuss.} \textbf{2011}, \emph{150}, 23--31\relax
\mciteBstWouldAddEndPuncttrue
\mciteSetBstMidEndSepPunct{\mcitedefaultmidpunct}
{\mcitedefaultendpunct}{\mcitedefaultseppunct}\relax
\EndOfBibitem
\bibitem[Gambetta \latin{et~al.}(2016)Gambetta, Cassinerio, Gatti, Laporta, and
  Galzerano]{Gambetta2016}
Gambetta,~A.; Cassinerio,~M.; Gatti,~D.; Laporta,~P.; Galzerano,~G. Scanning
  micro-resonator direct-comb absolute spectroscopy. \emph{Scientific Reports}
  \textbf{2016}, \emph{6}, 35541\relax
\mciteBstWouldAddEndPuncttrue
\mciteSetBstMidEndSepPunct{\mcitedefaultmidpunct}
{\mcitedefaultendpunct}{\mcitedefaultseppunct}\relax
\EndOfBibitem
\bibitem[Muraviev \latin{et~al.}(2020)Muraviev, Konnov, and
  Vodopyanov]{Muraviev2020}
Muraviev,~A.~V.; Konnov,~D.; Vodopyanov,~K.~L. Broadband high-resolution
  molecular spectroscopy with interleaved mid-infrared frequency combs.
  \emph{Scientific Reports} \textbf{2020}, \emph{10}, 18700\relax
\mciteBstWouldAddEndPuncttrue
\mciteSetBstMidEndSepPunct{\mcitedefaultmidpunct}
{\mcitedefaultendpunct}{\mcitedefaultseppunct}\relax
\EndOfBibitem
\bibitem[Pelczer and Szalma(1991)Pelczer, and Szalma]{Pelczer1991}
Pelczer,~I.; Szalma,~S. Multidimensional NMR and Data Processing.
  \emph{Chemical Reviews} \textbf{1991}, \emph{91}, 1507--1524\relax
\mciteBstWouldAddEndPuncttrue
\mciteSetBstMidEndSepPunct{\mcitedefaultmidpunct}
{\mcitedefaultendpunct}{\mcitedefaultseppunct}\relax
\EndOfBibitem
\bibitem[Hoch \latin{et~al.}(2014)Hoch, Maciejewski, Mobli, Schuyler, and
  Stern]{Hoch2014}
Hoch,~J.~C.; Maciejewski,~M.~W.; Mobli,~M.; Schuyler,~A.~D.; Stern,~A.~S.
  Nonuniform Sampling and Maximum Entropy Reconstruction in Multidimensional
  NMR. \emph{Accounts of Chemical Research Acc. Chem. Res.} \textbf{2014},
  \emph{47}, 708--717\relax
\mciteBstWouldAddEndPuncttrue
\mciteSetBstMidEndSepPunct{\mcitedefaultmidpunct}
{\mcitedefaultendpunct}{\mcitedefaultseppunct}\relax
\EndOfBibitem
\bibitem[Diels and Rudolph(2006)Diels, and Rudolph]{Diels2006}
Diels,~J.-C.; Rudolph,~W. \emph{Ultrashort Laser Pulse Phenomena}, 2nd ed.;
  Elsevier Inc.: London, 2006\relax
\mciteBstWouldAddEndPuncttrue
\mciteSetBstMidEndSepPunct{\mcitedefaultmidpunct}
{\mcitedefaultendpunct}{\mcitedefaultseppunct}\relax
\EndOfBibitem
\bibitem[Sal\/en \latin{et~al.}(2019)Sal\/en, Basini, Bonetti, Hebling,
  Krasilnikov, Nikitin, Shamuilov, Tibai, Zhaunerchyk, and
  Goryashko]{Salen2019}
Sal\/en,~P.; Basini,~M.; Bonetti,~S.; Hebling,~J.; Krasilnikov,~M.;
  Nikitin,~A.~Y.; Shamuilov,~G.; Tibai,~Z.; Zhaunerchyk,~V.; Goryashko,~V.
  Matter manipulation with extreme terahertz light: Progress in the enabling
  THz technology. \emph{Physics Reports Matter manipulation with extreme
  terahertz light: Progress in the enabling THz technology} \textbf{2019},
  \emph{836-837}, 1--74\relax
\mciteBstWouldAddEndPuncttrue
\mciteSetBstMidEndSepPunct{\mcitedefaultmidpunct}
{\mcitedefaultendpunct}{\mcitedefaultseppunct}\relax
\EndOfBibitem
\bibitem[Stert \latin{et~al.}(1999)Stert, Radloff, Schulz, and
  Hertel]{Stert1999}
Stert,~V.; Radloff,~W.; Schulz,~C.~P.; Hertel,~I.~V. Ultrafast photoelectron
  spectroscopy: Femtosecond pump-probe coincidence detection of ammonia cluster
  ions and electrons. \emph{European Physical Journal D} \textbf{1999},
  \emph{5}, 97--106\relax
\mciteBstWouldAddEndPuncttrue
\mciteSetBstMidEndSepPunct{\mcitedefaultmidpunct}
{\mcitedefaultendpunct}{\mcitedefaultseppunct}\relax
\EndOfBibitem
\bibitem[Samoylova \latin{et~al.}(2009)Samoylova, Radloff, Ritze, and
  Schultz]{Samoylova2009}
Samoylova,~E.; Radloff,~W.; Ritze,~H.~H.; Schultz,~T. Observation of Proton
  Transfer in 2-Aminopyridine Dimer by Electron and Mass Spectroscopy.
  \emph{Journal of Physical Chemistry A} \textbf{2009}, \emph{113},
  8195--8201\relax
\mciteBstWouldAddEndPuncttrue
\mciteSetBstMidEndSepPunct{\mcitedefaultmidpunct}
{\mcitedefaultendpunct}{\mcitedefaultseppunct}\relax
\EndOfBibitem
\bibitem[Stolow \latin{et~al.}(2004)Stolow, Bragg, and Neumark]{Stolow2004}
Stolow,~A.; Bragg,~A.~E.; Neumark,~D.~M. Femtosecond time-resolved
  photoelectron spectroscopy. \emph{Chemical Reviews} \textbf{2004},
  \emph{104}, 1719--1757\relax
\mciteBstWouldAddEndPuncttrue
\mciteSetBstMidEndSepPunct{\mcitedefaultmidpunct}
{\mcitedefaultendpunct}{\mcitedefaultseppunct}\relax
\EndOfBibitem
\bibitem[Smith \latin{et~al.}(2010)Smith, Kalcic, Safran, Stemmer, Dantus, and
  Reida]{Smith2010}
Smith,~S.~A.; Kalcic,~C.~L.; Safran,~K.~A.; Stemmer,~P.~M.; Dantus,~M.;
  Reida,~G.~E. Enhanced characterization of singly protonated phosphopeptide
  ions by femtosecond laser-induced ionization/dissociation tandem mass
  spectrometry (fs-LID-MS/MS). \emph{Journal of the American Society for Mass
  Spectrometry} \textbf{2010}, \emph{21}, 2031--2040\relax
\mciteBstWouldAddEndPuncttrue
\mciteSetBstMidEndSepPunct{\mcitedefaultmidpunct}
{\mcitedefaultendpunct}{\mcitedefaultseppunct}\relax
\EndOfBibitem
\bibitem[Hartland \latin{et~al.}(1991)Hartland, Connell, and
  Felker]{Hartland1991}
Hartland,~G.~V.; Connell,~L.~L.; Felker,~P.~M. Theory of Rotational Coherence
  Spectroscopy as Implemented by Picosecond Fluorescence Depletion Schemes.
  \emph{Journal of Chemical Physics} \textbf{1991}, \emph{94}, 7649--7666\relax
\mciteBstWouldAddEndPuncttrue
\mciteSetBstMidEndSepPunct{\mcitedefaultmidpunct}
{\mcitedefaultendpunct}{\mcitedefaultseppunct}\relax
\EndOfBibitem
\bibitem[Gliserin \latin{et~al.}(2015)Gliserin, Walbran, Krausz, and
  Baum]{Gliserin2015}
Gliserin,~A.; Walbran,~M.; Krausz,~F.; Baum,~P. Sub-phonon-period compression
  of electron pulses for atomic diffraction. \emph{Nature Communications}
  \textbf{2015}, \emph{6}, 8723\relax
\mciteBstWouldAddEndPuncttrue
\mciteSetBstMidEndSepPunct{\mcitedefaultmidpunct}
{\mcitedefaultendpunct}{\mcitedefaultseppunct}\relax
\EndOfBibitem
\bibitem[Yang \latin{et~al.}(2015)Yang, Beck, Uiterwaal, and
  Centurion]{Yang2015}
Yang,~J.; Beck,~J.; Uiterwaal,~C.~J.; Centurion,~M. Imaging of alignment and
  structural changes of carbon disulfide molecules using ultrafast electron
  diffraction. \emph{Nature Communications} \textbf{2015}, \emph{6}, 8172\relax
\mciteBstWouldAddEndPuncttrue
\mciteSetBstMidEndSepPunct{\mcitedefaultmidpunct}
{\mcitedefaultendpunct}{\mcitedefaultseppunct}\relax
\EndOfBibitem
\bibitem[Yang and Centurion(2015)Yang, and Centurion]{Yang2015b}
Yang,~J.; Centurion,~M. Gas-phase electron diffraction from laser-aligned
  molecules. \emph{Structural Chemistry} \textbf{2015}, \emph{26},
  1513--1520\relax
\mciteBstWouldAddEndPuncttrue
\mciteSetBstMidEndSepPunct{\mcitedefaultmidpunct}
{\mcitedefaultendpunct}{\mcitedefaultseppunct}\relax
\EndOfBibitem
\bibitem[K\"upper \latin{et~al.}(2014)K\"upper, Stern, Holmegaard, Filsinger,
  Rouz\/ee, Rudenko, Johnsson, Martin, Adolph, Aquila, Bajt, Barty, Bostedt,
  Bozek, Caleman, Coffee, Coppola, Delmas, Epp, Erk, Foucar, Gorkhover,
  Gumprecht, Hartmann, Hartmann, Hauser, Holl, H\"omke, Kimmel, Krasniqi,
  K\"uhnel, Maurer, Messerschmidt, Moshammer, Reich, Rudek, Santra,
  Schlichting, Schmidt, Schorb, Schulz, Soltau, Spence, Starodub, Str\"uder,
  Th{\o}gersen, Vrakking, Weidenspointner, White, Wunderer, Meijer, Ullrich,
  Stapelfeldt, Rolles, and Chapman]{Kupper2014}
K\"upper,~J. \latin{et~al.}  X-Ray Diffraction from Isolated and Strongly
  Aligned Gas-Phase Molecules with a Free-Electron Laser. \emph{Physical Review
  Letters} \textbf{2014}, \emph{112}, 083002\relax
\mciteBstWouldAddEndPuncttrue
\mciteSetBstMidEndSepPunct{\mcitedefaultmidpunct}
{\mcitedefaultendpunct}{\mcitedefaultseppunct}\relax
\EndOfBibitem
\end{mcitethebibliography}

\end{document}